\pdfoutput=1
\documentclass[12pt,a4paper]{article}

\usepackage{ifthen} 
\newboolean{pdflatex}
\setboolean{pdflatex}{true} 

\newboolean{articletitles}
\setboolean{articletitles}{true} 

\newboolean{uprightparticles}
\setboolean{uprightparticles}{true} 

\newboolean{inbibliography}
\setboolean{inbibliography}{false} 

\newcommand{\upsmm}     {\decay{\PUpsilon}{\mumu}}
\newcommand{\ups}       {\PUpsilon}

\newcommand{\ones}      {\ensuremath{\PUpsilon(1\mathrm{S})}\xspace}
\newcommand{\twos}      {\ensuremath{\PUpsilon(2\mathrm{S})}\xspace}
\newcommand{\threes}    {\ensuremath{\PUpsilon(3\mathrm{S})}\xspace}

\def\Quark     {{\ensuremath{\PQ}}\xspace}
\def\Quarkbar  {{\ensuremath{\overline \Quark}}\xspace}
\def\QQbar     {{\ensuremath{\Quark\Quarkbar}}\xspace}
\newcommand{\bit}{\begin{itemize}}
\newcommand{\bce}{\begin{center}}
\newcommand{\eit}{\end{itemize}}
\newcommand{\ece}{\end{center}}

\newcommand{\pt}  {\ensuremath{p_{\mathrm{T}}}\xspace}

\def\sPlot{\mbox{\em sPlot}}

\newcommand{\pty}  {\ensuremath{p^{\ups}_{\mathrm{T}}}\xspace}

\newcommand{\yy}   {\ensuremath{y^{\ups}}\xspace}

\newcommand{\pz}{\ensuremath{\phantom{-}}}

\newcommand{\MYCIRCLE} {\ensuremath{\textrm{\ding{108}}}}
\newcommand{\MYSQUARE} {\ensuremath{\textrm{\ding{110}}}}
\newcommand{\MYDIAMOND}{\ensuremath{\textrm{\ding{117}}}}

\def\paperauthors{LHCb collaboration} 
\def\paperasciititle{Measurement of the  Upsilon(nS) polarization in
  proton-proton collisions at sqrt{s}=7 and 8TeV} 
\def\papertitle{Measurement of the~$\ups\mathrm{(nS)}$~polarizations in
  $\proton\proton$~collisions at~\mbox{$\sqrt{s}=7$} and~\mbox{$8\tev$}
} 
\def\paperkeywords{{High Energy Physics}, {LHCb}} 
\def\papercopyright{CERN on behalf of the LHCb collaboration}
\def\paperlicence{CC-BY-4.0}
\def\paperlicenceurl{https://creativecommons.org/licenses/by/4.0/}

\usepackage[top=1in, bottom=1.25in, left=1in, right=1in]{geometry}

\usepackage{microtype}
\usepackage{lineno}  
\usepackage{xspace} 
\usepackage{caption} 

\usepackage{graphicx}  
\usepackage{color}
\usepackage{colortbl}
\graphicspath{{./figs/}} 

\usepackage{amsmath} 
\usepackage{amssymb}
\usepackage{amsfonts}
\usepackage{upgreek} 

\newcommand*\patchAmsMathEnvironmentForLineno[1]{%
\expandafter\let\csname old#1\expandafter\endcsname\csname #1\endcsname
\expandafter\let\csname oldend#1\expandafter\endcsname\csname
end#1\endcsname
 \renewenvironment{#1}%
   {\linenomath\csname old#1\endcsname}%
   {\csname oldend#1\endcsname\endlinenomath}%
}
\newcommand*\patchBothAmsMathEnvironmentsForLineno[1]{%
  \patchAmsMathEnvironmentForLineno{#1}%
  \patchAmsMathEnvironmentForLineno{#1*}%
}
\AtBeginDocument{%
\patchBothAmsMathEnvironmentsForLineno{equation}%
\patchBothAmsMathEnvironmentsForLineno{align}%
\patchBothAmsMathEnvironmentsForLineno{flalign}%
\patchBothAmsMathEnvironmentsForLineno{alignat}%
\patchBothAmsMathEnvironmentsForLineno{gather}%
\patchBothAmsMathEnvironmentsForLineno{multline}%
\patchBothAmsMathEnvironmentsForLineno{eqnarray}%
}

\usepackage{hyperxmp}

\usepackage[pdftex,
            pdfauthor={\paperauthors},
            pdftitle={\paperasciititle},
            pdfkeywords={\paperkeywords},
            pdfcopyright={Copyright (C) \papercopyright},
            pdflicenseurl={\paperlicenceurl}]{hyperref}

\usepackage[all]{hypcap} 

\usepackage{xspace} 
\usepackage{upgreek}

\def\lhcb {\mbox{LHCb}\xspace}

\def\MagUp {\mbox{\em Mag\kern -0.05em Up}\xspace}

\ifthenelse{\boolean{uprightparticles}}%
{

 \def\Peta        {\ensuremath{\upeta}\xspace}

 \def\Pmu         {\ensuremath{\upmu}\xspace}

 \def\Pchi        {\ensuremath{\upchi}\xspace}                 
 \def\Ppsi        {\ensuremath{\uppsi}\xspace}

 \def\PDelta      {\ensuremath{\Delta}\xspace}                 
 \def\PXi      {\ensuremath{\Xi}\xspace}                 
 \def\PLambda      {\ensuremath{\Lambda}\xspace}                 
 \def\PSigma      {\ensuremath{\Sigma}\xspace}                 
 \def\POmega      {\ensuremath{\Omega}\xspace}                 
 \def\PUpsilon      {\ensuremath{\Upsilon}\xspace}                 
 
 \def\PB      {\ensuremath{\mathrm{B}}\xspace}                 
                  
 \def\PD      {\ensuremath{\mathrm{D}}\xspace}

 \def\PJ      {\ensuremath{\mathrm{J}}\xspace}                 
 \def\PK      {\ensuremath{\mathrm{K}}\xspace}

 \def\PQ      {\ensuremath{\mathrm{Q}}\xspace}

 \def\Pb      {\ensuremath{\mathrm{b}}\xspace}                 
 \def\Pc      {\ensuremath{\mathrm{c}}\xspace}

 \def\Pi      {\ensuremath{\mathrm{i}}\xspace}

 \def\Pp      {\ensuremath{\mathrm{p}}\xspace}

}
{

 \def\Peta        {\ensuremath{\eta}\xspace}

 \def\Pmu         {\ensuremath{\mu}\xspace}

 \def\Pchi        {\ensuremath{\chi}\xspace}                 
 \def\Ppsi        {\ensuremath{\psi}\xspace}                 
                  
 \mathchardef\PDelta="7101
 \mathchardef\PXi="7104
 \mathchardef\PLambda="7103
 \mathchardef\PSigma="7106
 \mathchardef\POmega="710A
 \mathchardef\PUpsilon="7107
                  
 \def\PB      {\ensuremath{B}\xspace}                 
                  
 \def\PD      {\ensuremath{D}\xspace}

 \def\PJ      {\ensuremath{J}\xspace}                 
 \def\PK      {\ensuremath{K}\xspace}

 \def\PQ      {\ensuremath{Q}\xspace}

 \def\Pb      {\ensuremath{b}\xspace}                 
 \def\Pc      {\ensuremath{c}\xspace}

 \def\Pi      {\ensuremath{i}\xspace}

 \def\Pp      {\ensuremath{p}\xspace}

}

\makeatletter
\ifcase \@ptsize \relax
  \newcommand{\miniscule}{\@setfontsize\miniscule{4}{5}}
\or
  \newcommand{\miniscule}{\@setfontsize\miniscule{5}{6}}
\or
  \newcommand{\miniscule}{\@setfontsize\miniscule{5}{6}}
\fi
\makeatother

\DeclareRobustCommand{\optbar}[1]{\shortstack{{\miniscule (\rule[.5ex]{1.25em}{.18mm})}
  \\ [-.7ex] $#1$}}

\def\mup        {{\ensuremath{\Pmu^+}}\xspace}
\def\mun        {{\ensuremath{\Pmu^-}}\xspace} 
\def\mumu       {{\ensuremath{\Pmu^+\Pmu^-}}\xspace}

\def\cquark    {{\ensuremath{\Pc}}\xspace}

\def\bquark    {{\ensuremath{\Pb}}\xspace}

  \def\Kbar    {{\kern 0.2em\overline{\kern -0.2em \PK}{}}\xspace}

\def\KorKbar    {\kern 0.18em\optbar{\kern -0.18em K}{}\xspace}

  \def\Dbar    {{\kern 0.2em\overline{\kern -0.2em \PD}{}}\xspace}

\def\DorDbar    {\kern 0.18em\optbar{\kern -0.18em D}{}\xspace}

\def\Bbar    {{\ensuremath{\kern 0.18em\overline{\kern -0.18em \PB}{}}}\xspace}

\def\BorBbar    {\kern 0.18em\optbar{\kern -0.18em B}{}\xspace}

\def\jpsi     {{\ensuremath{{\PJ\mskip -3mu/\mskip -2mu\Ppsi\mskip 2mu}}}\xspace}

  \def\Y#1S{\ensuremath{\PUpsilon{(#1S)}}\xspace}

\def\proton      {{\ensuremath{\Pp}}\xspace}
\def\antiproton  {{\ensuremath{\overline \proton}}\xspace}

\def\Lbar        {{\ensuremath{\kern 0.1em\overline{\kern -0.1em\PLambda}}}\xspace}
\def\LorLbar    {\kern 0.18em\optbar{\kern -0.18em \PLambda}{}\xspace}

\newcommand{\decay}[2]{\ensuremath{#1\!\to #2}\xspace}         

\def\to                 {\ensuremath{\rightarrow}\xspace}

\def\AT#1     {\ensuremath{A_{\mathrm{T}}^{#1}}\xspace}           

\def\C#1      {\ensuremath{\mathcal{C}_{#1}}\xspace}                       
\def\Cp#1     {\ensuremath{\mathcal{C}_{#1}^{'}}\xspace}                    
\def\Ceff#1   {\ensuremath{\mathcal{C}_{#1}^{\mathrm{(eff)}}}\xspace}        
\def\Cpeff#1  {\ensuremath{\mathcal{C}_{#1}^{'\mathrm{(eff)}}}\xspace}       
\def\Ope#1    {\ensuremath{\mathcal{O}_{#1}}\xspace}                       
\def\Opep#1   {\ensuremath{\mathcal{O}_{#1}^{'}}\xspace}                    

\newcommand{\tev}{\ifthenelse{\boolean{inbibliography}}{\ensuremath{~T\kern -0.05em eV}}{\ensuremath{\mathrm{\,Te\kern -0.1em V}}}\xspace}
\newcommand{\gev}{\ensuremath{\mathrm{\,Ge\kern -0.1em V}}\xspace}
\newcommand{\mev}{\ensuremath{\mathrm{\,Me\kern -0.1em V}}\xspace}
\newcommand{\kev}{\ensuremath{\mathrm{\,ke\kern -0.1em V}}\xspace}
\newcommand{\ev}{\ensuremath{\mathrm{\,e\kern -0.1em V}}\xspace}
\newcommand{\gevc}{\ensuremath{{\mathrm{\,Ge\kern -0.1em V\!/}c}}\xspace}
\newcommand{\mevc}{\ensuremath{{\mathrm{\,Me\kern -0.1em V\!/}c}}\xspace}
\newcommand{\gevcc}{\ensuremath{{\mathrm{\,Ge\kern -0.1em V\!/}c^2}}\xspace}
\newcommand{\gevgevcccc}{\ensuremath{{\mathrm{\,Ge\kern -0.1em V^2\!/}c^4}}\xspace}
\newcommand{\mevcc}{\ensuremath{{\mathrm{\,Me\kern -0.1em V\!/}c^2}}\xspace}

\newcommand{\bibgev}{\ifthenelse{\boolean{inbibliography}}{\ensuremath{~G\kern -0.05em eV}}{\ensuremath{\mathrm{\,Ge\kern -0.1em V}}}\xspace}
\newcommand{\bibgevc}{\ifthenelse{\boolean{inbibliography}}{\ensuremath{~G\kern -0.05em eV}/c}{\ensuremath{\mathrm{\,Ge\kern -0.1em V}/c}}\xspace}

\def\mum  {\ensuremath{{\,\upmu\mathrm{m}}}\xspace}

\def\invfb   {\ensuremath{\mbox{\,fb}^{-1}}\xspace}

\def\deriv {\ensuremath{\mathrm{d}}}

\def\gsim{{~\raise.15em\hbox{$>$}\kern-.85em
          \lower.35em\hbox{$\sim$}~}\xspace}
\def\lsim{{~\raise.15em\hbox{$<$}\kern-.85em
          \lower.35em\hbox{$\sim$}~}\xspace}

\def\sPlot{\mbox{\em sPlot}\xspace}

\def\sqs   {\ensuremath{\protect\sqrt{s}}\xspace}

\def\ptot       {\mbox{$p$}\xspace}
\def\pt         {\mbox{$p_{\mathrm{ T}}$}\xspace}

\def\evtgen     {\mbox{\textsc{EvtGen}}\xspace}

\def\geant      {\mbox{\textsc{Geant4}}\xspace}

\def\photos     {\mbox{\textsc{Photos}}\xspace}

\def\pythia     {\mbox{\textsc{Pythia}}\xspace}

\def\tell1  {TELL1\xspace}
\def\ukl1   {UKL1\xspace}

\newcommand{\eg}{\mbox{\itshape e.g.}\xspace}

\usepackage{cite} 
\usepackage{mciteplus}

\usepackage{longtable} 
\usepackage{mathtools}
\usepackage{rotating}  
\usepackage{graphpap}  
\usepackage{multirow}  
\usepackage{mathrsfs}  
\usepackage{pifont}    
\usepackage{cancel}    
\usepackage{afterpage} 
\usepackage{lscape}
\usepackage{bm}
\usepackage{wasysym}

\begin{document}

\renewcommand{\thefootnote}{\fnsymbol{footnote}}
\setcounter{footnote}{1}


\begin{titlepage}
\pagenumbering{roman}

\vspace*{-1.5cm}
\centerline{\large EUROPEAN ORGANIZATION FOR NUCLEAR RESEARCH (CERN)}
\vspace*{1.5cm}
\noindent
\begin{tabular*}{\linewidth}{lc@{\extracolsep{\fill}}r@{\extracolsep{0pt}}}
\ifthenelse{\boolean{pdflatex}}
{\vspace*{-2.7cm}\mbox{\!\!\!\includegraphics[width=.14\textwidth]{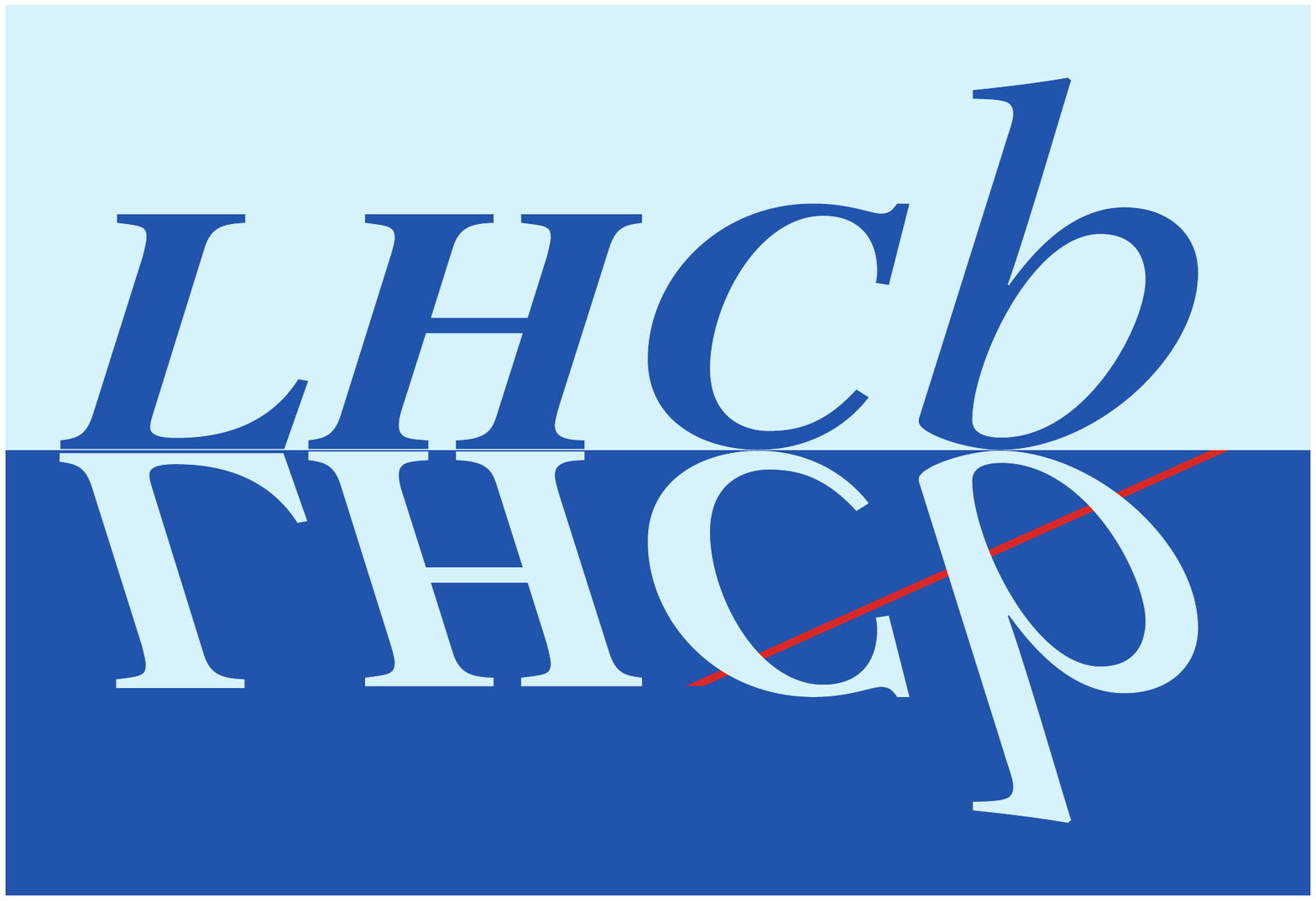}} & &}%
{\vspace*{-1.2cm}\mbox{\!\!\!\includegraphics[width=.12\textwidth]{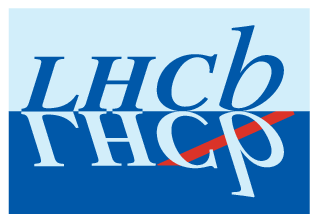}} & &}%
\\
 & & CERN-EP-2017-210    \\  
 & & LHCb-PAPER-2017-028 \\  
 & & August 15, 2017 \\ 
 & & \\
\end{tabular*}

\vspace*{4.0cm}

{\normalfont\bfseries\boldmath\huge
\begin{center}
  \papertitle 
\end{center}
}

\vspace*{2.0cm}

\begin{center}
\paperauthors\footnote{Authors are listed at the end of this paper.}
\end{center}

\vspace{\fill}

\begin{abstract}
  \noindent
  The polarization of the~\ones, \twos and \threes~mesons,
  produced in \proton\proton collisions at
  centre\nobreakdash-of\nobreakdash-mass energies 
  $\sqs=7$ and~\mbox{8\tev},
  is measured using data samples collected by the~\lhcb
  experiment, corresponding to integrated
  luminosities of 1 and 2\invfb, respectively.
  The~measurements are performed in three 
  polarization frames, using $\ups\to\mumu$~decays
  in the~kinematic region of the~transverse momentum 
  \mbox{$\pty<30~\gevc$} and rapidity \mbox{$2.2<\yy<4.5$}.
  No large polarization is observed.
\end{abstract}

\vspace*{2.0cm}

\begin{center}
  Published in JHEP\,12\, (2017)\,110. 
\end{center}

\vspace{\fill}

{\footnotesize 
\centerline{\copyright~\papercopyright, licence \href{\paperlicenceurl}{\paperlicence}.}}
\vspace*{2mm}

\end{titlepage}


\newpage
\setcounter{page}{2}
\mbox{~}

%
%
%
%

\cleardoublepage


\renewcommand{\thefootnote}{\arabic{footnote}}
\setcounter{footnote}{0}



\pagestyle{plain} 
\setcounter{page}{1}
\pagenumbering{arabic}


%

\section{Introduction}\label{sec:Introduction}

Studies of heavy quarkonium production
play an~important role in
the~development of quantum
chromodynamics\,(QCD)~\cite{Braaten_Russ_2014,Polar_perspective_2014}.
According to the~current theoretical framework,
nonrelativistic
QCD\,(NRQCD)~\cite{CaswellLepage1986PL,Bodwin_Braaten_Lepage},
the~production of heavy quarkonium factorizes into two steps,
separated by different time and energy scales.
In~the~first step, a~heavy quark\nobreakdash-antiquark pair, \QQbar, 
is created in a~short time, of order $\hbar/(2m_{\Quark}c^2)$, 
where $m_{\Quark}$ is
the~heavy\nobreakdash-quark mass.
In~the~second step, the~\QQbar pair, being produced
in a~colour\nobreakdash-singlet  
or colour\nobreakdash-octet~\cite{Braaten:1994vv} 
state, evolves nonperturbatively 
into a~quarkonium state.
The~nonperturbative transitions 
from the~initially produced \QQbar pairs to 
the~observable colourless quarkonium states are
described by long\nobreakdash-distance 
matrix elements. 
According to the~NRQCD factorization
approach~\cite{Bodwin_Braaten_Lepage},
these matrix elements are universal constants 
which are independent of
the~short\nobreakdash-time production processes,
and need to be determined from data.
Calculations based on the~colour\nobreakdash-singlet model~\cite{Kartvelishvili:1978id,Baier_Ruckl_1981,Berger_Jones_1981}
show good agreement~\cite{Campbell:2007ws,
  Gong_Wang_2008,
  CSM_ACLMT_2008,
  lansberg:2008gk,
  Han:2014kxa,
  Gong:2013qka}
with
the~experimental data~\cite{Brambilla:2010cs,
CDFpolar2002,
LHCb-PAPER-2011-036,
LHCb-PAPER-2013-016,
LHCb-PAPER-2013-066,
LHCb-PAPER-2015-045,
Khachatryan:2010zg,
Aad:2012dlq} 
for 
production cross\nobreakdash-sections 
and the~shapes of the~transverse momentum spectra.   
However, this~approach
fails to describe  the~spin\nobreakdash-alignment\,(usually 
labelled as polarization) of S\nobreakdash-wave
charmonium states~\cite{LHCb-PAPER-2013-008,LHCb-PAPER-2013-067}.
Leading\nobreakdash-order colour\nobreakdash-singlet calculations
predict a~transverse polarization for the~S\nobreakdash-wave quarkonia,
while next\nobreakdash-to\nobreakdash-leading\nobreakdash-order\,(NLO)
calculations predict a~longitudinal polarization for 
these states~\cite{Gong_Wang_2008}.
An~analysis using NLO~NRQCD calculations 
of the~short\nobreakdash-distance coefficients~\cite{Gong:2013qka}
concludes that the~\ones and \twos~bottomonium states 
should have  a~very small transverse polarization, 
almost independent of the~transverse momentum, while
the~\threes~meson should    show a~large
transverse polarization at high transverse momenta.
For~the~\threes~meson this~analysis neglects the~contributions from 
the~cascade decays of higher excited bottomonium states.
Accounting for these contributions, 
\eg from the~radiative
$\Pchi_{\bquark}\mathrm{(nP)}\to\ups\gamma$~decays\footnote{Throughout 
  this paper the~symbol \ups~represents generically 
  \ones, \twos and \threes~mesons.}~\cite{Abazov:2012gh,
  Aad:2011ih, 
  LHCb-PAPER-2012-015, 
  LHCb-PAPER-2014-031,
  LHCb-PAPER-2014-040}, 
is important for a~correct interpretation 
of results~\cite{Feng:2015wka}.
The~measured fractions of \ups~mesons originating from 
$\Pchi_{\bquark}$~decays are large, 
around $30-40\,\%$\cite{LHCb-PAPER-2014-031},
for \ups~mesons with high transverse momenta, $\pty\gtrsim20\gevc$.

Experimentally, polarization of \ups~mesons produced in 
high\nobreakdash-energy hadron\nobreakdash-hadron interactions 
was studied by the~CDF collaboration in $\proton\antiproton$~collisions
at~\mbox{$\sqs=1.8$} and $1.96\tev$~\cite{CDFpolar2002,CDFpolar2012}.
It~was found that the~angular distributions of muons 
from~\mbox{$\PUpsilon\to\mumu$}~decays for 
all three \ups~states are nearly isotropic
in the~central rapidity region~\mbox{$\left|\yy\right|<0.6$}
and~\mbox{$\pty<40\gevc$}~\cite{CDFpolar2012}.
This~result 
is inconsistent with the~measurement performed by
the~D0 collaboration,
where a~significant $\pty$\nobreakdash-dependent
longitudinal polarization 
was observed for \ones~mesons
produced in $\proton\antiproton$~collisions
at~\mbox{$\sqrt{s}=1.96\tev$}, 
for~\mbox{$\left|\yy\right|<1.8$}
and~\mbox{$\pty<20~\gevc$}~\cite{D0polar2008}.
At~the~LHC the~\ups polarization was measured
by the~CMS  collaboration~\cite{CMSpolar2013}
using $\proton\proton$~collision data at~\mbox{$\sqs=7\tev$},
for the~rapidity ranges~\mbox{$\left|\yy\right|<0.6$}
and
\mbox{$0.6<\left|\yy\right|<1.2$},
and for~\mbox{$10<\pty<50~\gevc$}.
No~evidence of large transverse or longitudinal
polarization was found for any of the~three \ups~mesons in
the~considered kinematic region.

The~spin\nobreakdash-alignment state of 
\ups~mesons
is measured through the~analysis of the~angular distribution of
muons from the~decay $\ups\to\mumu$~\cite{Oakes1966,LamTung1978,Faccioli_clarification}
\begin{equation}
 \dfrac{1}{\sigma}\dfrac{\deriv\sigma}{\deriv\Omega} =
 \dfrac{3}{4\pi}~\dfrac{1}{3 + \uplambda_{\theta}}
 \left(1 + \uplambda_{\theta}\cos^2\theta 
         + \uplambda_{\theta\phi}\sin2\theta\cos\phi 
         + \uplambda_{\phi}\sin^2\theta\cos2\phi \right),
\label{eq:MainAngDistrib}
\end{equation}
where the angular quantities $\Omega = \left(\cos\theta, \phi \right)$
denote the~direction of the~positive muon in the~rest frame of the \ups~meson
with respect to the~chosen reference frame.
The~spin\nobreakdash-alignment parameters $\pmb{\uplambda}\equiv
\left(\uplambda_{\theta}, 
  \uplambda_{\theta\phi}, 
  \uplambda_{\phi}\right)$ 
are directly related to the~spin\nobreakdash-1 density\nobreakdash-matrix
elements~\cite{Oakes1966,Beneke_Kramer_Vanttinen_1998}.
Alternatively, these parameters are often denoted 
as~\mbox{$\left(\uplambda,\upmu,\upnu/2\right)$}.
The~case~\mbox{$\pmb{\uplambda}=(0,0,0)$}
    corresponds to unpolarized \ups~mesons and for
the chosen reference frame transverse\,(longitudinal) polarization corresponds
to $\uplambda_{\theta} > 0\,(<0)$.

The~parameters $\pmb{\uplambda}$ depend on 
the~choice for the~reference system
in the~rest frame of the $\ups$~meson.
The~following three frames are widely used
in polarization analyses:
helicity\,(HX), 
Collins\nobreakdash-Soper\,(CS) 
and
Gottfried\nobreakdash-Jackson\,(GJ).
In~the~HX frame~\cite{Jacob_Wick:1959},
the~$z$~axis is defined as 
the~direction of the~$\ups$ momentum in
the~centre\nobreakdash-of\nobreakdash-mass 
frame of the~colliding LHC protons,
that is~\mbox{$\hat{z}\equiv-\left(\vec{p}_{1}+\vec{p}_{2}\right)/\left|\vec{p}_{1}+\vec{p}_{2}\right|$},
where $\vec{p}_{1}$ and $\vec{p}_{2}$
are the~three\nobreakdash-momenta of 
the~colliding protons
in the~rest frame of the~\ups~meson.\footnote{A~beam proton
  travelling in the~positive\,(negative) direction of the~$z$~axis
  of the~coordinate system of the~LHCb detector~\cite{Alves:2008zz} 
  is designated as the~first\,(second).} 
In~the~CS frame~\cite{Collins_Soper},
the~$z$ axis is defined such  that it~bisects 
the~angle between $\vec{p}_{1}$ and $-\vec{p}_{2}$
in the~rest frame of the~\ups~meson.
In~the~GJ frame~\cite{Gottfried_Jackson},
the~$z$ axis is defined as
the~direction of $\vec{p}_{1}$
in the~rest frame of the~\ups~meson.
In~all of these frames, 
the~$y$~axis is normal 
to the~production plane of 
the~\ups~mesons 
with the~direction 
along the~vector product
$\vec{p}_{1} \times \vec{p}_{2}$
defined in the~rest frame of 
the~\ups~meson.\footnote{This~definition is adopted from 
Ref.~\cite{Faccioli_clarification},
and is opposite to the~Madison convention~\cite{Madison1971},
$\vec{p}_{1} \times \vec{p}_{\ups}$,
where $\vec{p}_{1}$ and $\vec{p}_{\ups}$
are the~three\nobreakdash-momenta of 
the~first beam proton
and the~\ups meson in 
the~centre\nobreakdash-of\nobreakdash-mass 
frame of the~two colliding protons.
The~two conventions differ by 
the~sign of the~$\uplambda_{\theta\phi}$~parameter
while 
keeping the~same values for 
$\uplambda_{\theta}$ and
$\uplambda_{\phi}$~\cite{Faccioli_clarification}.
}   
The~$x$~axis is defined 
to complete a~right\nobreakdash-handed
coordinate system.

Since the~HX, CS and GJ~reference frames are related by rotations
around the~$y$~axis, the~three polarization
parameters measured in one frame can be translated into
another~\cite{Telegdi1986,Faccioli_clarification}.
The~frame\nobreakdash-transformation relations
imply the~existence of quantities $\mathcal{F}$ that are invariant 
under these rotations~\cite{Faccioli_2010_a,Faccioli_2010_b}.
These~quantities are defined in terms of 
$\uplambda_{\theta}$ and $\uplambda_{\phi}$~as 
\begin{equation*}
  \mathcal{F}\left(c_1,c_2,c_3\right) \equiv 
  \dfrac{ \phantom{c_2}\left(3+\uplambda_\theta\right) + c_1 \left(1-\uplambda_\phi\right)} 
        {         c_2  \left(3+\uplambda_\theta\right) + c_3 \left(1-\uplambda_\phi\right)}   
\end{equation*} 
for arbitrary numbers~$c_1$, $c_2$  and~$c_3$.
In~particular, the~frame\nobreakdash-invariant polarization parameter $\tilde\uplambda$ is 
defined for the~specific choice of constants $c_1$, $c_2$ 
and $c_3$~\cite{Teryaev:SPIN2005,Faccioli_2010_a,Faccioli_2010_b,Faccioli1,Faccioli_2011_c}
\begin{equation}
  \label{eq:lamtilde}
  \tilde\uplambda\equiv\mathcal{F}(-3,0,1) = 
  \dfrac{\lambda_\theta+3\lambda_\phi}{1-\lambda_\phi}.
\end{equation} 

This paper presents 
a~full angular analysis performed on 
\ups~mesons produced in $\proton\proton$~collisions 
at~\mbox{$\sqs=7$} and~\mbox{$8\tev$} 
corresponding to integrated luminosities of
1 and 2\invfb, respectively.
The~polarization parameters 
are measured in the~HX, CS and GJ~frames
in the~region
\mbox{$2.2<\yy<4.5$}
as functions of~\pty and~\yy
for \mbox{$\pty<20~\gevc$}
and as functions of~\pty
for \mbox{$\pty<30~\gevc$}. The~latter 
range is   referred to as the~wide~\pty range in 
the~following.

\section{LHCb detector and simulation}\label{sec:Detector}
The \lhcb detector~\cite{Alves:2008zz,LHCb-DP-2014-002} is 
a~single\nobreakdash-arm forward
spectrometer covering the pseudorapidity range~\mbox{$2<\eta <5$},
designed for the~study of particles containing \bquark or \cquark
quarks. The~detector includes a~high\nobreakdash-precision tracking system
consisting of a~silicon\nobreakdash-strip vertex detector surrounding
the~$\proton\proton$~interaction region,
a~large\nobreakdash-area silicon\nobreakdash-strip detector located
upstream of a~dipole magnet with a~bending power of about
$4{\mathrm{\,Tm}}$, and three stations of silicon\nobreakdash-strip 
detectors and
straw drift tubes placed downstream of the~magnet.
The~tracking system provides a measurement of momentum,
\ptot, of charged particles with a~relative uncertainty that varies
from 0.5\,\% at low momentum to 1.0\,\% at~200\gevc.
The~minimum distance of a~track to a~primary vertex, the~impact
parameter, is measured with a~resolution of~\mbox{$(15+29/\pt)\mum$},
where \pt is
in\,\gevc.
Different types of charged hadrons are distinguished using information
from two ring\nobreakdash-imaging Cherenkov detectors. 
Photons, electrons and hadrons are identified by a~calorimeter system
consisting of scintillating\nobreakdash-pad and preshower detectors, 
an~electromagnetic calorimeter and a~hadronic calorimeter. 
Muons are identified by a~system composed of alternating layers of iron and
multiwire proportional chambers~\cite{LHCb-DP-2012-002}.
The~online event selection is performed by a
trigger~\cite{LHCb-DP-2012-004}, 
which consists of a hardware stage, based on information from 
the~calorimeter and muon systems, followed by a~software stage, 
which applies a~full event reconstruction.

The~hardware trigger stage is implemented by
requiring the~presence of two muon candidates with 
the~product of their~\pt exceeding 
\mbox{$1.7\,(2.6)\,(\!\gevc)^2$} for data
collected at~\mbox{$\sqs=7\,(8)\tev$}.
In~the~subsequent software stage, the~trigger requires the presence of
two well\nobreakdash-reconstructed tracks with hits in the~muon system,
\mbox{$\pt>0.5\gevc$} and \mbox{$\ptot>6\gevc$}.
Only events with a~pair of tracks identified as oppositely charged
muons with a~common vertex and a~mass~\mbox{$m_{\mumu}>4.7\gevcc$}
are retained for further analysis.

In Monte Carlo simulation, $\proton\proton$
collisions are generated using
\mbox{$\pythia\,6$}~\cite{Sjostrand:2006za} with 
a~specific \lhcb configuration~\cite{LHCb-PROC-2010-056}. 
Decays of hadrons
are described by \evtgen~\cite{Lange:2001uf} with final\nobreakdash-state
radiation generated using \photos~\cite{Golonka:2005pn}. 
The~interaction of the~generated particles with the~detector, and its
response, are implemented using the~\geant
toolkit~\cite{Allison:2006ve, *Agostinelli:2002hh} as described in
Ref.~\cite{LHCb-PROC-2011-006}.
The~bottomonium production is simulated according
to the~leading order colour\nobreakdash-singlet and 
colour\nobreakdash-octet 
mechanisms~\cite{LHCb-PROC-2010-056,LHCb-2007-042}, and 
the~bottomonium states are generated without polarization.

\section{Data selection}\label{sec:evsel}
\begin{figure}[t]
  \setlength{\unitlength}{1mm}
  \centering
  \begin{picture}(150,60)
    \put( 75,  0){
      \includegraphics*[width=75mm,height=60mm,%
      ]{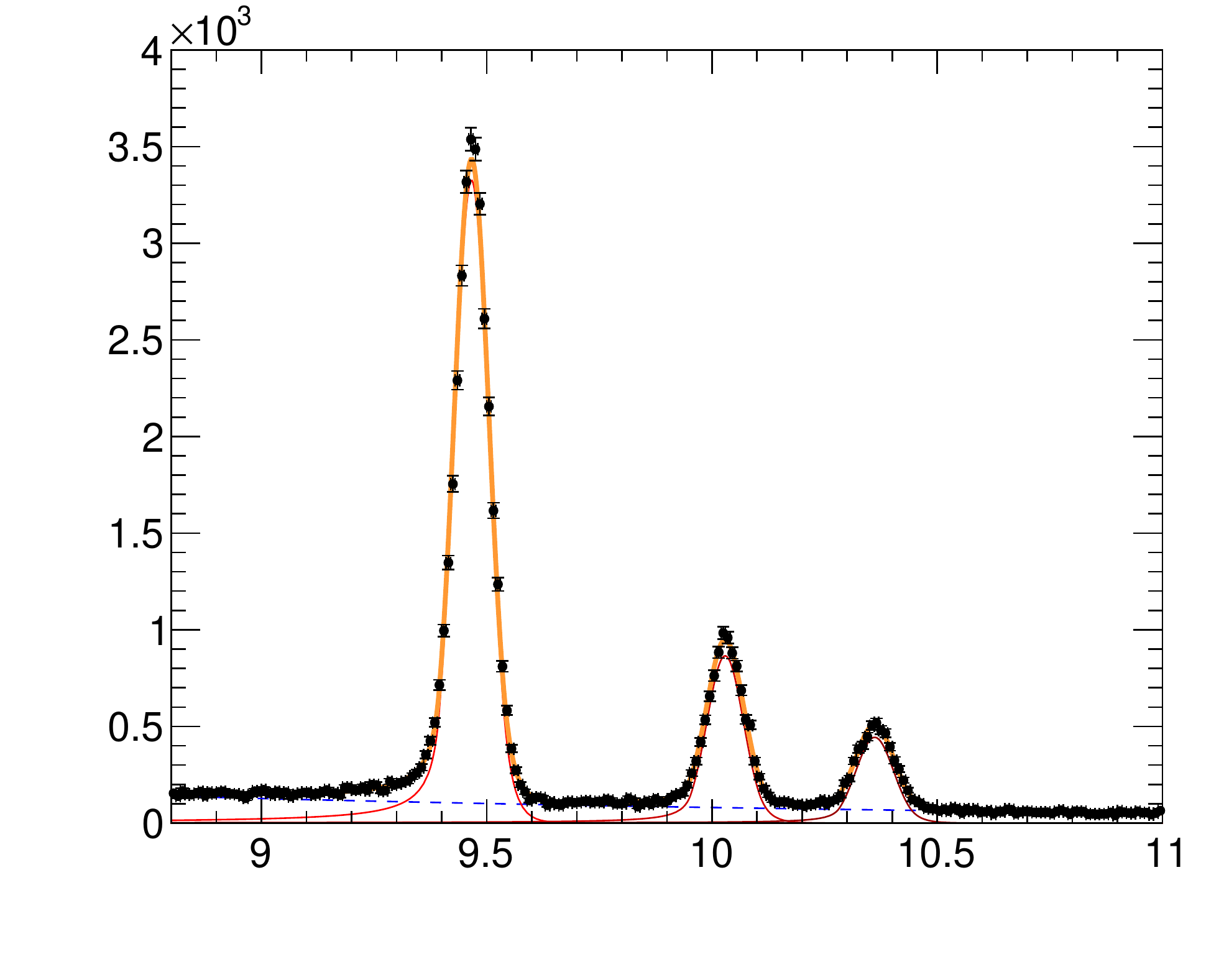}
    }
    \put(  0,  0){
      \includegraphics*[width=75mm,height=60mm,%
      ]{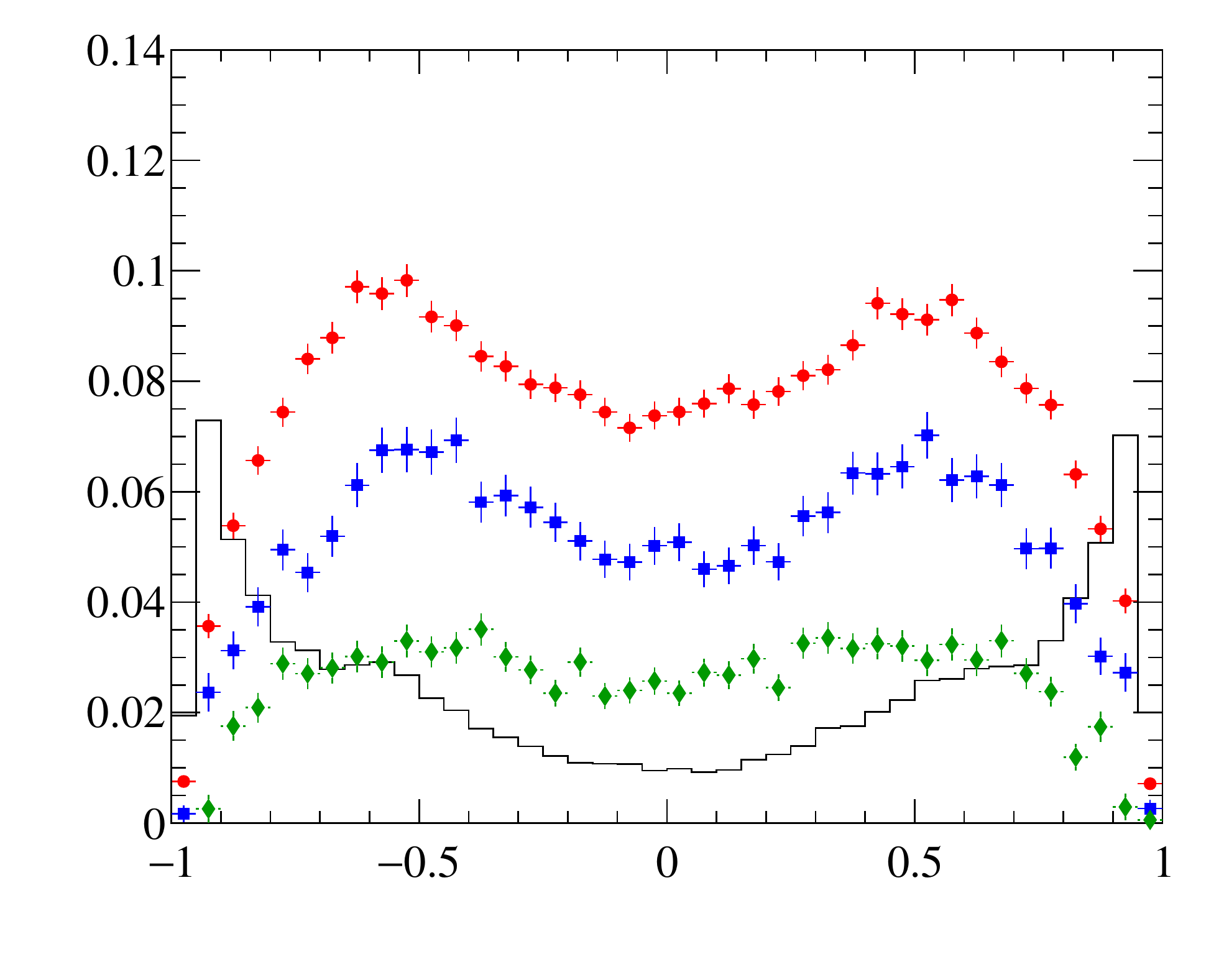}
    }
    \put(108, 45.5) { \small { {$\begin{array}{l}\lhcb~\sqs=7\tev
            \\ 6<\pty<8\gevc \\ 2.2<\yy<3.0\end{array}$}}}
    \put ( 25, 48.0) { \footnotesize $\begin{array}{cl} 
        {\color{red}\MYCIRCLE}           & \!\!\!\!\ones\times3 \\
        {\color{blue}\MYSQUARE}          & \!\!\!\!\twos\times2 \\ 
        {\color[rgb]{0,0.6,0}\MYDIAMOND} & \!\!\!\!\threes 
      \end{array}$ }
    \put ( 46,46.0) { \footnotesize $\begin{array}{cl} 
        \framebox[1.5\width]{\phantom{o}} & \!\!\!\!\mathrm{background} 
      \end{array}$ }
    \put(57, 51) { \small $\lhcb$}
    \put( 78, 16.0){\begin{sideways}\small Candidates/(10\mevcc)\end{sideways}}
    \put( -2, 29.5){\begin{sideways}
        $\frac{1}{N}\frac{\deriv N}{\deriv \cos\theta_{\mathrm{GJ}}}~\left[\frac{1}{0.05}\right]$
      \end{sideways}}
    \put(112,  2) { $m_{\mumu}$}  \put(132,2) {
      $\left[\!\gevcc\right]$}
    \put( 37,  2) { $\cos\theta_{\,\text{GJ}}$}
    \put(15,50){ a)}
    \put(90,50){ b)}
  \end{picture}
  \caption { \small
    (a)~Distributions of $\cos\theta_{\mathrm{GJ}}$
    for \ones\,(red circles), 
    \twos\,(blue squares) and
    \threes\,(green diamonds) signal candidates 
    and
    the~background component\,(histogram)
    in the~region 
    \mbox{$6<\pty<8\gevc$}, 
    \mbox{$2.2<\yy<3.0$}
    for data  accumulated at~\mbox{$\sqrt{s}=7\,\mathrm{TeV}$}.
    To~improve visibility,
    the~distributions for~the \ones and~\twos 
    signals are scaled by factors of~3 and~2, respectively.
    (b)~Dimuon mass distribution 
    in the~region \mbox{$6<\pty<8\gevc$}, \mbox{$2.2<\yy<3.0$}
    for data  accumulated at~\mbox{$\sqrt{s}=7\,\mathrm{TeV}$}.
    The~thick dark yellow solid curve shows the~result of the~fit,
    as described in the~text.
    The~three peaks, shown with thin red solid lines,
    correspond to the~\ones, 
    \twos and \threes~signals\,(left to right).
    The~background component is indicated with a~dashed blue line.
  }
  \label{fig:figure01}
\end{figure}

The~selection of \ups candidates is based on the~same criteria
as used in the~previous \lhcb analyses~\mbox{\cite{LHCb-PAPER-2011-036,
    LHCb-PAPER-2013-016,LHCb-PAPER-2013-066,LHCb-PAPER-2015-045}}.
The~\ups candidates are formed from pairs of oppositely charged tracks.
Each track is required to have a~good reconstruction
quality~\cite{LHCb-DP-2013-002}
and to be identified as a~muon~\cite{LHCb-DP-2013-001}.
Each~muon is then required to satisfy
\mbox{$1<\pt<25\gevc$},
\mbox{$10<\ptot<400\gevc$}
and have pseudorapidity within~\mbox{$2.2<\Peta<4.5$}.
The~two muons are required to be consistent with originating 
from a~common vertex.
%
The~consistency of the~dimuon vertex with a primary vertex
is ensured via the~quality requirement of 
a~global fit, performed for each dimuon
candidate using the~primary vertex position 
as a~constraint~\cite{Hulsbergen:2005pu}.
This~requirement also reduces the~background from genuine muons
coming from decays of
long\nobreakdash-lived charm and beauty hadrons. 
The~mass of the~muon pair is required to be
in the~range~\mbox{$8.8 < m_{\mumu} < 11.0\,\gevcc$}.
A~large~fraction of the~combinatorial background is characterized 
by a~large absolute value of the~cosine of the~polar angle,
\mbox{$\cos\theta_{\text{GJ}}$},
of the~$\mup$~lepton in the~GJ~frame.
The~distributions of~\mbox{$\cos\theta_{\text{GJ}}$}
for signal and for background components
in a~typical bin, \mbox{$6<\pty<8\gevc$} and~\mbox{$2.2<\yy<3.0$},
are shown
in Fig.~\ref{fig:figure01}a.
The~components   are determined using the~\sPlot technique, 
described below.
To~reduce this background,
a~requirement~\mbox{$\left|\cos\theta_{\text{GJ}}\right|<0.8$}
is applied.

Dimuon mass distributions of the~\mbox{\upsmm}~candidates 
selected in the~region~\mbox{$6<\pty<8\gevc$} 
and~\mbox{$2.2<\yy<3.0$}
for data collected at~\mbox{$\sqs=7\tev$}
are shown in~Fig.~\ref{fig:figure01}b.
In~each $(\pty,\yy)$~bin, the dimuon mass distribution is parametrized
by a~sum of three double\nobreakdash-sided Crystal Ball
functions~\cite{Skwarnicki:1986xj,LHCb-PAPER-2011-013},
describing
the~three \ups~meson signals, and 
an~exponential function for
the~combinatorial background.
A~double\nobreakdash-sided Crystal Ball
function is defined
as a~Gaussian function with power\nobreakdash-law tails on both the~low\nobreakdash- and
high\nobreakdash-mass sides.
The~peak position and
the~resolution parameters  
of the~Crystal~Ball function
describing the~mass distribution of the~\ones~meson
are free parameters for 
the~unbinned extended maximum likelihood fit.
The~peak position parameters of the~\twos and \threes
signal components are fixed using
the~known values of the~mass differences
of the~\twos and \threes~mesons to that of
the~\ones~meson~\cite{PDG2017},
while the~resolution parameters are fixed to the~value of 
the~resolution parameter of the~\ones~signal, scaled by
the~ratio of the~masses of the~\twos and \threes~mesons
to  that of
the~\ones~meson. 
The~power\nobreakdash-law tail parameters 
of the~three Crystal Ball  functions
are common for the~three \ups~meson signals
and left free in the~fit.

\section{Polarization fit}\label{sec:polarfit}

The polarization parameters are determined from unbinned maximum likelihood fits~\cite{Xie_2009}
to  the~two\nobreakdash-dimensional $\left(\cos\theta,\phi\right)$~angular distribution
of the~\mup~lepton from the~\mbox{$\ups\to\mumu$}~decay,
described by Eq.~\ref{eq:MainAngDistrib}, following 
the~approach of Refs.~\cite{LHCb-PAPER-2013-008,LHCb-PAPER-2013-067}.
The~projections of the~two\nobreakdash-dimensional $\left(\cos\theta,\phi\right)$~angular 
distribution in the~HX~frame are shown in Fig.~\ref{fig:figurexx} for
data accumulated at~\mbox{$\sqs=7\tev$}
in the~kinematic region \mbox{$6<\pty<8\gevc$}, \mbox{$2.2<\yy<3.0$}.

\input{figurexx}

The~logarithm of the likelihood function for each \ups~state,
in each $\left(\pty,~\yy\right)$ bin, is defined as~\cite{LHCb-PAPER-2013-067}
\begin{eqnarray}
  \log \mathcal{L}^{\ups}(\pmb{\uplambda}) & =
  & s_{w} \sum_{i} w_i^{\ups} 
  \log \left[\dfrac{\mathcal{P}(\Omega_i \vert \pmb{\uplambda})\;
      \varepsilon(\Omega_i)}
    {\mathcal{N}( \pmb{\uplambda})}\right] \nonumber \\ 
  & = & s_{w} \sum_{i} w_i^{\ups} 
  \log \left[\dfrac{\mathcal{P}(\Omega_i \vert \pmb{\uplambda}) }
    {\mathcal{N}( \pmb{\uplambda} )}\right] \label{eq:UpsLikelihood_2} \\
  & + & s_{w} \sum_{i} w_{i}^{\ups} 
  \log \left[\varepsilon ( \Omega_i)\right]\;,
  \nonumber 
\end{eqnarray}
where 
$\mathcal{P}(\Omega_i \vert \pmb{\uplambda} ) \equiv
1+\uplambda_\theta \cos^2 \theta_{i}
+ \uplambda_{\theta \phi}\sin 2\theta_{i} \cos \phi_{i}
+ \uplambda_\phi\sin^2 \theta_{i} \cos 2\phi_{i}$,
\mbox{$\varepsilon(\Omega_i)$}~is the~total efficiency
for the~$i^{\mathrm{th}}$ \ups~candidate
and $\mathcal{N}(\pmb{\uplambda})$
is the~normalization integral defined below.
The~weights $w_i^{\ups}$ are determined
from the~fit of the~dimuon mass distribution
using the~$sPlot$
technique~\cite{Pivk:2004ty},
which projects out the~corresponding signal component
from the~combined signal plus background densities.
The~sum in Eq.~\ref{eq:UpsLikelihood_2} 
runs over all selected \ups~candidates.
The~constant scale factor
\mbox{$s_{w}\equiv\sum_{i} w_i^{\ups}/\sum_{i}\left(w_i^{\ups}\right)^2$}
accounts for statistical fluctuations in the~background 
subtraction~\cite{LHCb-PAPER-2013-002,LHCb-PAPER-2013-067,Eadie:100342}
and was
validated using pseudoexperiments.
Numerically, it increases by approximately 5\% the~uncertainties in the~polarization parameters.
The~last term in Eq.~\ref{eq:UpsLikelihood_2}
is ignored in the~fit as it has no dependence on
the~polarization parameters.
The~normalization factor $\mathcal{N}(\pmb{\uplambda})$ is defined as
\begin{equation}\label{eq:NormFactor_all}
  \mathcal{N}(\pmb{\uplambda})
  \equiv \int \deriv\Omega~ \mathcal{P}(\Omega \vert  \pmb{\uplambda})\,\varepsilon( \Omega )
\end{equation}
and is calculated using simulated events. 
In~the~simulation, where the~\ups~mesons are generated unpolarized,
the~two\nobreakdash-dimensional $\left(\cos\theta,\phi\right)$ distribution
of selected candidates is proportional to the~total efficiency 
$\varepsilon(\Omega)$, so
$\mathcal{N}( \pmb{\uplambda} )$
is evaluated by summing 
$\mathcal{P}(\Omega_i \vert \pmb{\uplambda} )$ 
over the~selected \ups~candidates in the~simulated sample
\begin{equation}
  \mathcal{N}( \pmb{\uplambda} )
  \propto  \sum_{i}
  \varepsilon^{\mumu}
  \mathcal{P}(\Omega_i \vert \pmb{\uplambda} ). \label{eq:NormFactor_4}
\end{equation}
For~simulated events no muon identification requirement is applied
when selecting the~\ups~candidates.
Instead, a~muon\nobreakdash-pair identification efficiency factor $\varepsilon^{\mumu}$
is applied for all selected simulated $\ups\to\mumu$~decays.
This~factor is calculated on a~per\nobreak-event basis as 
\begin{equation}
  \varepsilon^{\mumu} = \varepsilon^{\Pmu\mathrm{ID}}(\mup)\,\varepsilon^{\Pmu\mathrm{ID}}(\mun),
\end{equation}
where $\varepsilon^{\Pmu\mathrm{ID}}$ is 
the~single\nobreakdash-muon identification efficiency, 
measured in data, using large samples of prompt \jpsi~mesons
decaying to muon pairs.
Given that the~reconstruction and selection efficiencies 
are taken from  simulation,
the~efficiency $\varepsilon^{\mumu}$ is further corrected 
to account for small differences between data and simulation 
in the~track reconstruction efficiency~\cite{LHCb-DP-2013-002,LHCb-DP-2013-001}
and in the~\pty~and \yy~spectra~\cite{Sjostrand:2006za,LHCb-PROC-2010-056}.

\section{Systematic uncertainties}\label{sec:systematic}

The~sources of systematic uncertainty studied in this analysis are
summarized in Table~\ref{tab:sum_syst_y1s_2011_2012}.
They are considered for 
the~polarization parameters 
$\uplambda_{\theta}$, $\uplambda_{\theta\phi}$,
$\uplambda_{\phi}$ and 
for the~frame\nobreakdash-invariant 
parameter $\tilde{\uplambda}$
in the~HX, CS and GJ~frames for each \mbox{$\left(\pty,~\yy\right)$}~bin.

The~systematic uncertainty related to the~signal determination
procedure is studied by varying the~mass model describing 
the~shape of the~dimuon mass distributions.
For~the~signal parametrizations, the~power\nobreakdash-law tail 
parameters of the~double\nobreakdash-sided Crystal~Ball 
functions are fixed to the values obtained 
in the~simulation, and the~constraints
for the~mean values of the~Crystal~Ball
functions  describing the \twos and \threes~signals
are removed. 
The~variation of the background parametrization is done by
replacing the~exponential function with the~product of
an~exponential function and a~polynomial function.
Mass~fit ranges are also varied.
The~maximum differences in each parameter
$\pmb{\uplambda}$ and $\tilde{\uplambda}$
with respect to the~nominal fit results 
are taken as systematic uncertainties.
For~all $\left(\pty,\yy\right)$~bins
these uncertainties are around 10\,\% of
the~corresponding  statistical uncertainties.

\begin{table}[t]
  \centering
  \caption{ \small 
    Ranges of the~absolute 
    systematic uncertainties of the parameters $\pmb{\uplambda}$ and~$\tilde{\uplambda}$.
    The ranges indicate variations depending on the~\mbox{$\left(\pty,\yy\right)$}~bin 
    and frame.
  } \label{tab:sum_syst_y1s_2011_2012}
  \vspace*{1.5mm}
  \begin{tabular*}{0.99\textwidth}{@{\hspace{1mm}}l@{\extracolsep{\fill}}cccc@{\hspace{1mm}}}
    Source 
    & $\upsigma_{\uplambda_{\theta}}~\left[10^{-3}\right]$ 
    & $\upsigma_{\uplambda_{\theta\phi}}~\left[10^{-3}\right]$ 
    & $\upsigma_{\uplambda_{\phi}}~\left[10^{-3}\right]$  
    & $\upsigma_{\tilde{\uplambda}}~\left[10^{-3}\right]$
    \\
    \hline
    \\[-3mm]
    \multicolumn{5}{c}{ \ones }
    \\[1mm] 
    \hline 
    \\[-3mm] 
    Dimuon mass fit                     & $\phantom{0}1.0-12\phantom{0}$ & $\phantom{0}0.2-10\phantom{0}$
                                        & $\phantom{0}0.1-7$\phantom{00} & $\phantom{0}1.8-20\phantom{0}$
    \\
    Efficiency calculation              &   &   &    &              \\
    \quad muon identification           & $\phantom{0}0.2-10\phantom{0}$  & $\phantom{0}0.1-7\phantom{00}$
                                        & $\phantom{0}0.1-6\phantom{00}$  & $\phantom{0}0.2-17\phantom{0}$
                                        \\
    \quad correction factors for $\varepsilon^{\mumu}$   
                                        & $\phantom{0}0.7-12\phantom{0}$  & $\phantom{0}0.4-5\phantom{00}$ 
                                        & $\phantom{0}0.1-4\phantom{00}$  & $\phantom{0}2.1-14\phantom{0}$ \\ 

    \quad trigger                       & $\phantom{0}0.1-18\phantom{0}$ & $\phantom{0}0.1-8\phantom{00}$
                                        & $\phantom{0}0.1-5\phantom{00}$ & $\phantom{0}0.3-19\phantom{0}$  \\ 
    Finite size of simulated samples    & $\phantom{0}6.0-82\phantom{0}$ & $\phantom{0}1.3-29\phantom{0}$
                                        & $\phantom{0}0.9-35\phantom{0}$ & $\phantom{0}6.9-95\phantom{0}$ \\
    \hline
    \\[-3mm]
    \multicolumn{5}{c}{ \twos }
    \\[1mm]
    \hline 
    \\[-3mm]
    Dimuon mass fit                     & $\phantom{0}0.6-37\phantom{0}$ & $\phantom{0}0.2-19\phantom{0}$
                                        & $\phantom{0}0.3-16\phantom{0}$ & $\phantom{0}4.6-53\phantom{0}$  \\
    Efficiency calculation              &    &   &   &   \\
    \quad muon identification           & $\phantom{0}0.2-11\phantom{0}$ & $\phantom{0}0.1-6\phantom{00}$
                                        & $\phantom{0}0.1-5\phantom{00}$ & $\phantom{0}0.2-13\phantom{0}$  \\
    \quad correction factors for $\varepsilon^{\mumu}$   
                                        & $\phantom{0}0.7-12\phantom{0}$ & $\phantom{0}0.3-5\phantom{00}$
                                        & $\phantom{0}0.1-5\phantom{00}$ & $\phantom{0}2.1-13\phantom{0}$ \\ 
    \quad trigger                       & $\phantom{0}0.1-17\phantom{0}$ & $\phantom{0}0.1-7\phantom{00}$
                                        & $\phantom{0}0.1-5\phantom{00}$ & $\phantom{0}0.3-18\phantom{0}$  \\
    Finite size of simulated samples    & $\phantom{0}9.8-210$           & $\phantom{0}2.5-98\phantom{0}$
                                        & $\phantom{0}1.5-120$           & $\phantom{0.}14-320$ \\ 
    \hline
    \\[-3mm]
    \multicolumn{5}{c}{ \threes }
    \\[1mm]
    \hline 
    \\[-3mm]
    Dimuon mass fit model               & $\phantom{0}1.4-72\phantom{0}$  & $\phantom{0}0.2-24\phantom{0}$
                                        & $\phantom{0}0.5-21\phantom{0}$  & $\phantom{0}7.2-86\phantom{0}$  \\ 
    Efficiency calculation              &    &   &   &   \\ 
    \quad muon identification           & $\phantom{0}0.2-12\phantom{0}$  & $\phantom{0}0.1-7\phantom{00}$
                                        & $\phantom{0}0.1-5\phantom{00}$  & $\phantom{0}0.3-22\phantom{0}$  \\
    \quad correction factors for $\varepsilon^{\mumu}$   
                                        & $\phantom{0}0.6-14\phantom{0}$  & $\phantom{0}0.3-6\phantom{00}$  
                                        & $\phantom{0}0.1-5\phantom{00}$  & $\phantom{0}2.1-18\phantom{0}$   \\
    \quad trigger                       & $\phantom{0}0.2-17\phantom{0}$  & $\phantom{0}0.1-8\phantom{00}$
                                        & $\phantom{0}0.1-4\phantom{00}$  & $\phantom{0}0.3-19\phantom{0}$  \\
    Finite size of simulated samples    & $\phantom{0.}12-280$            & $\phantom{0}3.5-100$
                                        & $\phantom{0}2.1-110$            & $\phantom{0.}16-350$
  \end{tabular*}   
\end{table}

For several sources of systematic uncertainty pseudoexperiments are used, whereby
an~ensemble of pseudodata samples is generated, 
each with a~random value of the~appropriate parameter taken from 
a~Gaussian distribution. 
The~fit is then performed for each sample, 
and the~observed  variations in the~fit 
parameters are used to assign  
the~corresponding systematic uncertainties.

The~single\nobreakdash-muon identification efficiency,
$\varepsilon^{\Pmu\mathrm{ID}}$,  
is determined from large samples 
of~\mbox{$\jpsi\to\mumu$}~decays.
The~efficiency $\varepsilon^{\Pmu\mathrm{ID}}$
is measured as a~function of muon transverse momentum and pseudorapidity.
The~systematic uncertainty in
the~$\pmb{\uplambda}$ and 
$\tilde\uplambda$ parameters 
related to the~muon identification is obtained from the~uncertainties 
of the~single particle identification efficiency $\varepsilon^{\Pmu\mathrm{ID}}$
using pseudoexperiments. 
This~uncertainty is around 2\,\% of 
the~statistical uncertainty 
for data in low\nobreakdash-\pty~bins
and rises to 8\,\% of the~statistical uncertainty 
in~high\nobreakdash-\pty~bins.
The~uncertainties in the~correction factors 
for the~muon\nobreakdash-pair identification efficiency $\varepsilon^{\mumu}$, 
related to small differences
in the~tracking and muon reconstruction efficiencies
between data and simulation,
are propagated to 
the~determination of the~polarization parameters 
using pseudoexperiments.
These~uncertainties are 20\,\% of 
the~statistical uncertainty for 
low\nobreakdash-\pty~bins
and decrease to 10\,\% of 
the~statistical uncertainty 
for high\nobreakdash-\pty~bins.

In  this analysis the~efficiency of  the~trigger is taken from simulation.
The~systematic uncertainty associated with a~possible small difference
in the~trigger 
efficiency between data and simulation is assessed by
studying the performance of the~dimuon trigger, described in Sect.~\ref{sec:Detector}, 
for events selected using the~single\nobreakdash-muon 
high\nobreakdash-\pt 
trigger~\cite{LHCb-DP-2012-004}.
The~fractions of \ones~signal candidates selected
using both trigger requirements are compared for 
the~data and simulation in $\left(\pty,~\yy\right)$ bins
and found to agree
within 2\,\%~\cite{LHCb-PAPER-2013-066,LHCb-PAPER-2015-045}.
The~corresponding systematic uncertainties in the polarization parameters
are obtained using pseudoexperiments
and found to be between 2\,\% and 4\,\% of the~statistical uncertainty.

Good agreement between the~data and simulated
samples is observed for all variables used 
to select the~\ups~candidates~\cite{LHCb-PAPER-2013-066,LHCb-PAPER-2015-045}.
The~discrepancies in the~corresponding integrated normalized distributions 
do not exceed 1\,\% and therefore no~systematic uncertainty 
related to possible mismodelling is assigned
to the polarization parameters.

The~finite size of the~simulated samples
introduces a~systematic uncertainty
related to the~normalization factors 
$\mathcal{N}(\pmb{\uplambda})$
of Eq.~\ref{eq:NormFactor_4}.
This~uncertainty is also propagated to
the~final uncertainty of the polarization results
using pseudoexperiments.
This~systematic uncertainty is dominant
for most of the~\mbox{$\left(\pty,\yy\right)$}~bins,
varying between 30\,\% and 70\,\% of the~statistical uncertainty.

The~total systematic uncertainty for each polarization
parameter is calculated as the~quadratic sum of the systematic
uncertainties from all the~considered sources, 
assuming no correlations.
The~systematic uncertainties of the $\pmb{\uplambda}$ and
$\tilde{\uplambda}$~parameters in the~different frames
are comparable.
For~the~majority of the~\mbox{$\left(\pty,\yy\right)$}~bins
the~total systematic uncertainty 
is much smaller than the~statistical uncertainty.
For~some high-\pty bins the~systematic
and statistical uncertainties are comparable.

\section{Results}\label{sec:results}

\begin{figure}[p]
  \setlength{\unitlength}{1mm}
  \centering
  \begin{picture}(150,180)
    \put( 0,  0){ 
      \includegraphics*[width=150mm,height=180mm,%
      ]{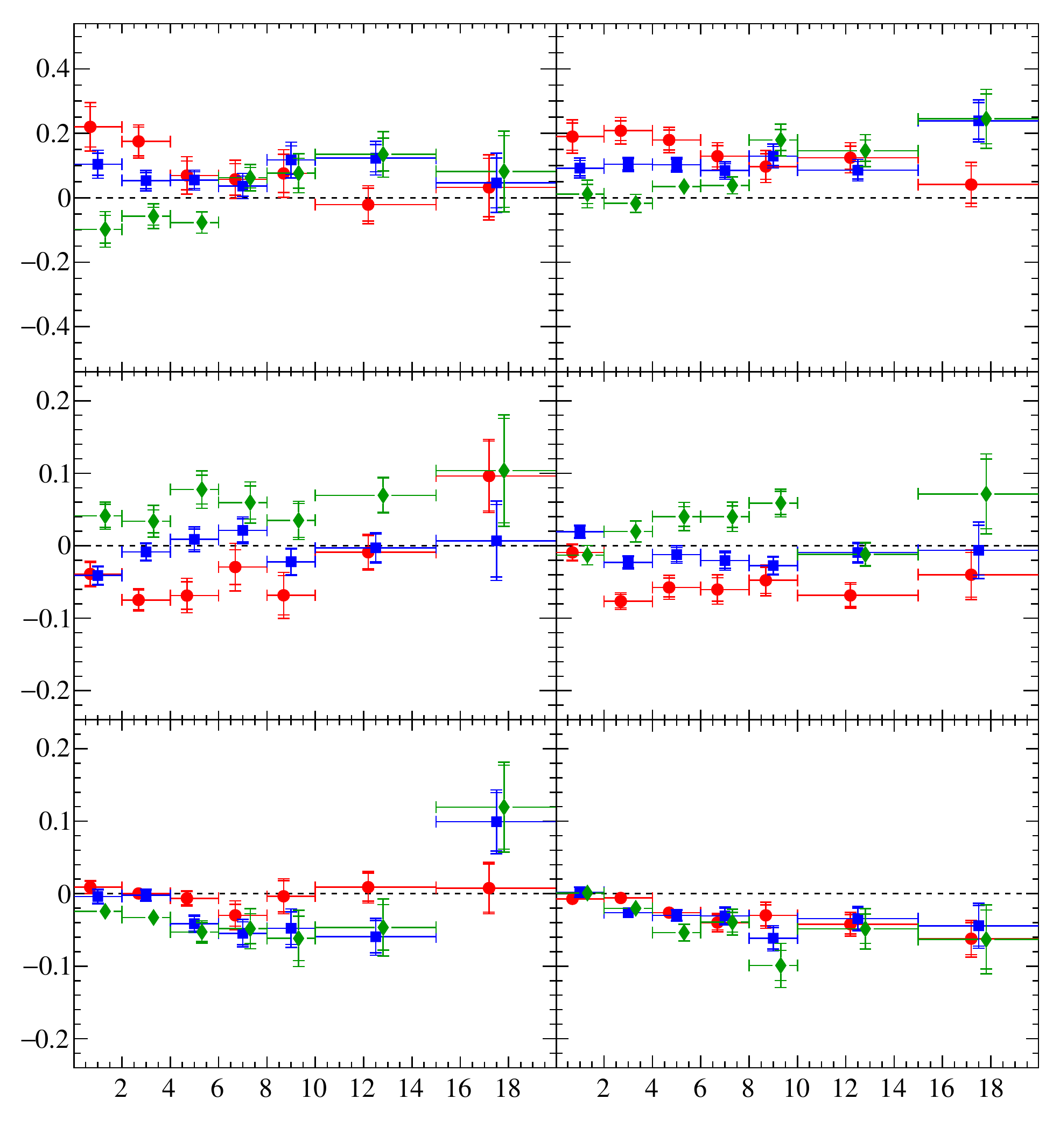}
    }
    \put(15,169) {\small $\begin{array}{l}\text{LHCb} \\ \sqrt{s}=7\tev \end{array}$}
    \put(83,169) {\small $\begin{array}{l}\text{LHCb} \\ \sqrt{s}=8\tev \end{array}$}
    \put(125,128) {\small $\begin{array}{l} \ones \\ \mathrm{HX~frame} \end{array}$}
    \put (43,130){ \small $\begin{array}{cl} 
        {\color{red}\MYCIRCLE}                & 2.2<\yy<3.0 \\
        {\color{blue}\MYSQUARE}          & 3.0<\yy<3.5 \\
        {\color[rgb]{0,0.6,0}\MYDIAMOND} & 3.5<\yy<4.5
      \end{array}$}
    \put( -2,148) { \begin{sideways} $\uplambda_{\theta}$     \end{sideways}}
    \put( -2, 93) { \begin{sideways} $\uplambda_{\theta\phi}$ \end{sideways}}
    \put( -2, 38) { \begin{sideways} $\uplambda_{\phi}$       \end{sideways}}
    \put(111, 2) { $\pty$} \put(135, 2){ $\left[\!\gevc\right]$}
    \put( 43, 2) { $\pty$} \put( 65, 2){ $\left[\!\gevc\right]$}
  \end{picture}
  \caption { \small
    The~polarization parameters 
    (top)\,$\uplambda_{\theta}$,
    (middle)\,$\uplambda_{\theta\phi}$ and 
    (bottom)\,$\uplambda_{\phi}$,
    measured in the~HX frame 
    for the~\ones~state
    in different
    bins of \pty and three rapidity ranges, 
    for data collected 
    at~(left)\,\mbox{$\sqs=7\,\mathrm{TeV}$}
    and~(right)\,\mbox{$\sqs=8\,\mathrm{TeV}$}.
    The~results for 
    the~rapidity ranges 
    \mbox{$2.2<\yy<3.0$}, 
    \mbox{$3.0<\yy<3.5$} and
    \mbox{$3.5<\yy<4.5$} are shown with 
    red circles, 
    blue squares and green diamonds, respectively.
    The~vertical inner error bars indicate the~statistical uncertainty, 
    whilst the~outer error bars indicate the~sum of 
    the~statistical and systematic uncertainties added in quadrature. 
    The~horizontal error bars indicate the~bin width.
    Some data points are displaced from 
    the~bin centers to improve visibility.
  }
  \label{fig:figure02hx}
\end{figure}

\begin{figure}[t]
  \setlength{\unitlength}{1mm}
  \centering
  \begin{picture}(150,180)
    \put( 0,  0){ 
      \includegraphics*[width=150mm,height=180mm,%
      ]{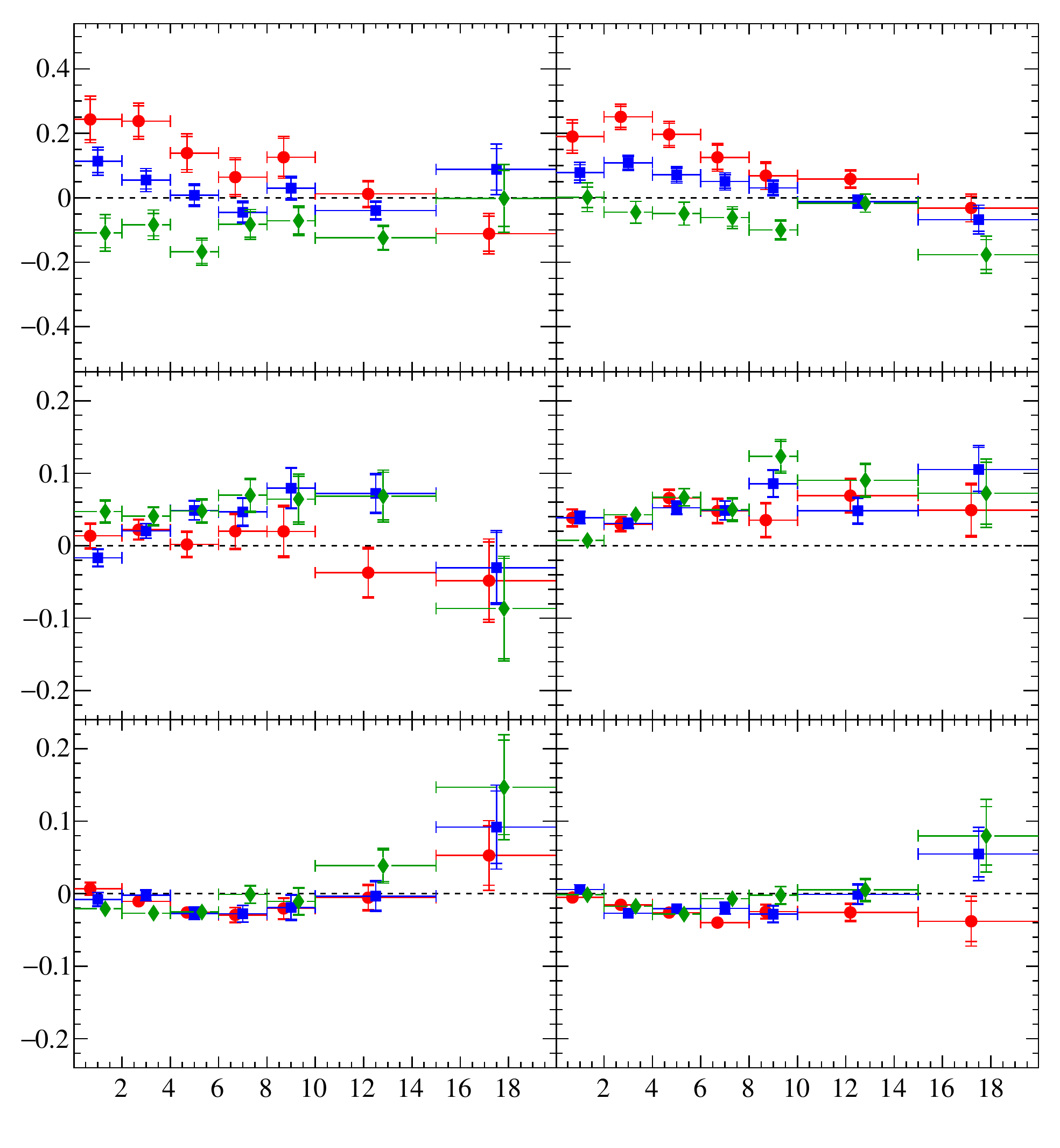}
    }
    \put(15,169) {\small $\begin{array}{l}\text{LHCb} \\ \sqrt{s}=7\tev \end{array}$}
    \put(83,169) {\small $\begin{array}{l}\text{LHCb} \\ \sqrt{s}=8\tev \end{array}$}
    \put(125,128) {\small $\begin{array}{l} \ones \\ \mathrm{CS~frame} \end{array}$}
    \put (43,130){ \small $\begin{array}{cl} 
        {\color{red}\MYCIRCLE}                & 2.2<\yy<3.0 \\
        {\color{blue}\MYSQUARE}          & 3.0<\yy<3.5 \\
        {\color[rgb]{0,0.6,0}\MYDIAMOND} & 3.5<\yy<4.5
      \end{array}$}
    \put( -2,148) { \begin{sideways} $\uplambda_{\theta}$     \end{sideways}}
    \put( -2, 93) { \begin{sideways} $\uplambda_{\theta\phi}$ \end{sideways}}
    \put( -2, 38) { \begin{sideways} $\uplambda_{\phi}$       \end{sideways}}
    \put(111, 2) { $\pty$} \put(135, 2){ $\left[\!\gevc\right]$}
    \put( 43, 2) { $\pty$} \put( 65, 2){ $\left[\!\gevc\right]$}
  \end{picture}
  \caption { \small
    The~polarization parameters 
    (top)\,$\uplambda_{\theta}$,
    (middle)\,$\uplambda_{\theta\phi}$ and 
    (bottom)\,$\uplambda_{\phi}$,
    measured in the~CS frame 
    for the~\ones~state
    in different
    bins of \pty and three rapidity ranges, 
    for data collected 
    at~(left)\,\mbox{$\sqs=7\,\mathrm{TeV}$}
    and~(right)\,\mbox{$\sqs=8\,\mathrm{TeV}$}.
    The~results for 
    the~rapidity ranges 
    \mbox{$2.2<\yy<3.0$}, 
    \mbox{$3.0<\yy<3.5$} and
    \mbox{$3.5<\yy<4.5$} are shown with 
    red circles, 
    blue squares and green diamonds, respectively.
    The~vertical inner error bars indicate the~statistical uncertainty,
    whilst the~outer error bars indicate the~sum of 
    the~statistical and systematic uncertainties added in quadrature. 
    The~horizontal error bars indicate the~bin width.
    Some data points are displaced from 
    the~bin centers to improve visibility.
  }
  \label{fig:figure02cs}
\end{figure}

\begin{figure}[t]
  \setlength{\unitlength}{1mm}
  \centering
  \begin{picture}(150,180)
    \put( 0,  0){ 
      \includegraphics*[width=150mm,height=180mm,%
      ]{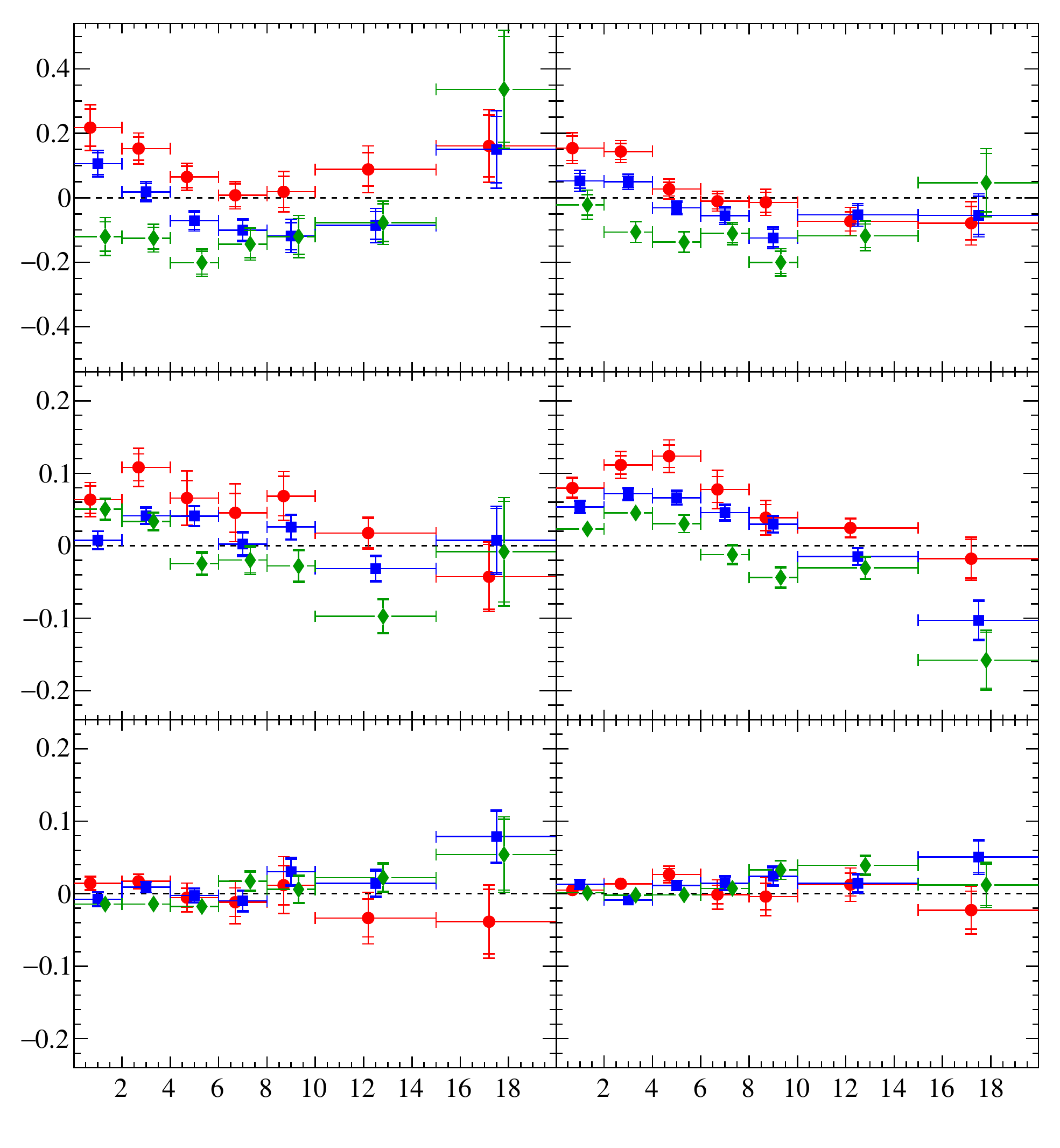}
    }
    \put(15,169) {\small $\begin{array}{l}\text{LHCb} \\ \sqrt{s}=7\tev \end{array}$}
    \put(83,169) {\small $\begin{array}{l}\text{LHCb} \\ \sqrt{s}=8\tev \end{array}$}
    \put(125,128) {\small $\begin{array}{l} \ones \\ \mathrm{GJ~frame} \end{array}$}
    \put (43,130){ \small $\begin{array}{cl} 
        {\color{red}\MYCIRCLE}                & 2.2<\yy<3.0 \\
        {\color{blue}\MYSQUARE}          & 3.0<\yy<3.5 \\
        {\color[rgb]{0,0.6,0}\MYDIAMOND} & 3.5<\yy<4.5
      \end{array}$}
    \put( -2,148) { \begin{sideways} $\uplambda_{\theta}$     \end{sideways}}
    \put( -2, 93) { \begin{sideways} $\uplambda_{\theta\phi}$ \end{sideways}}
    \put( -2, 38) { \begin{sideways} $\uplambda_{\phi}$       \end{sideways}}
    \put(111, 2) { $\pty$} \put(135, 2){ $\left[\!\gevc\right]$}
    \put( 43, 2) { $\pty$} \put( 65, 2){ $\left[\!\gevc\right]$}
  \end{picture}
  \caption { \small
    The~polarization parameters 
    (top)\,$\uplambda_{\theta}$,
    (middle)\,$\uplambda_{\theta\phi}$ and 
    (bottom)\,$\uplambda_{\phi}$,
    measured in the~GJ frame 
    for the \ones~state
    in different
    bins of \pty and three rapidity ranges, 
    for data collected 
    at~(left)\,\mbox{$\sqs=7\,\mathrm{TeV}$}
    and~(right)\,\mbox{$\sqs=8\,\mathrm{TeV}$}.
    The~results for 
    the~rapidity ranges 
    \mbox{$2.2<\yy<3.0$}, 
    \mbox{$3.0<\yy<3.5$} and
    \mbox{$3.5<\yy<4.5$} are shown with 
    red circles, 
    blue squares and green diamonds, respectively.
    The~vertical inner error bars indicate the~statistical uncertainty, 
    whilst the~outer error bars indicate the~sum of 
    the~statistical and systematic uncertainties added in quadrature. 
    The~horizontal error bars indicate the~bin width.
    Some data points are displaced from 
    the~bin centers to improve visibility.
  }
  \label{fig:figure02gj}
\end{figure}

\begin{figure}[t]
  \setlength{\unitlength}{1mm}
  \centering
  \begin{picture}(150,180)
    \put( 0,  0){ 
      \includegraphics*[width=150mm,height=180mm,%
      ]{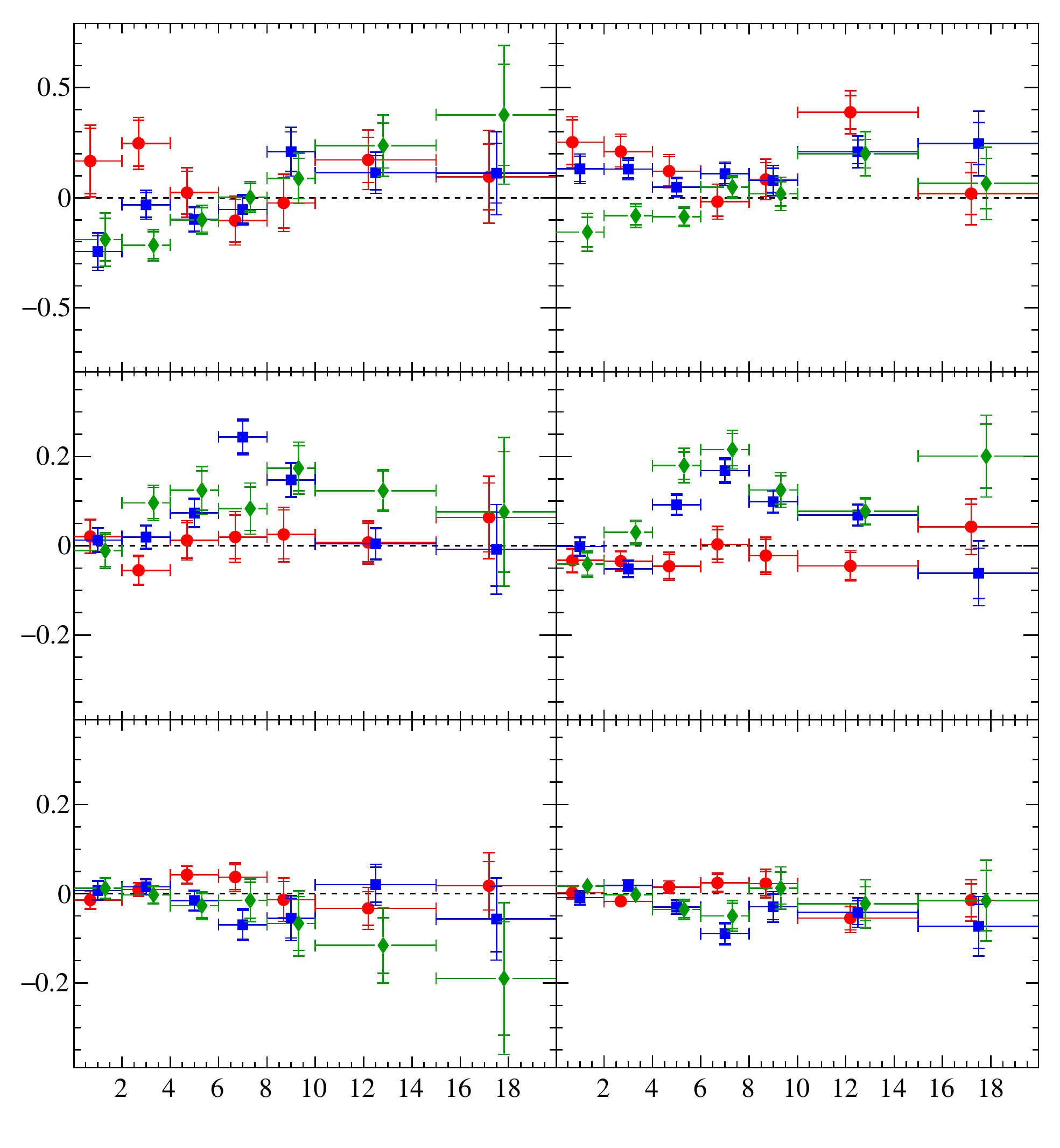}
    }
    \put(15,169) {\small $\begin{array}{l}\text{LHCb} \\ \sqrt{s}=7\tev \end{array}$}
    \put(83,169) {\small $\begin{array}{l}\text{LHCb} \\ \sqrt{s}=8\tev \end{array}$}
    \put(125,128) {\small $\begin{array}{l} \twos \\ \mathrm{HX~frame} \end{array}$}
    \put (43,130){ \small $\begin{array}{cl} 
        {\color{red}\MYCIRCLE}                & 2.2<\yy<3.0 \\
        {\color{blue}\MYSQUARE}          & 3.0<\yy<3.5 \\
        {\color[rgb]{0,0.6,0}\MYDIAMOND} & 3.5<\yy<4.5
      \end{array}$}
    \put( -2,148) { \begin{sideways} $\uplambda_{\theta}$     \end{sideways}}
    \put( -2, 93) { \begin{sideways} $\uplambda_{\theta\phi}$ \end{sideways}}
    \put( -2, 38) { \begin{sideways} $\uplambda_{\phi}$       \end{sideways}}
    \put(111, 2) { $\pty$} \put(135, 2){ $\left[\!\gevc\right]$}
    \put( 43, 2) { $\pty$} \put( 65, 2){ $\left[\!\gevc\right]$}
  \end{picture}
  \caption { \small 
    The~polarization parameters 
    (top)\,$\uplambda_{\theta}$,
    (middle)\,$\uplambda_{\theta\phi}$ and 
    (bottom)\,$\uplambda_{\phi}$,
    measured in the~HX frame 
    for the \twos~state
    in different
    bins of \pty and three rapidity ranges, 
    for data collected 
    at~(left)\,\mbox{$\sqs=7\,\mathrm{TeV}$}
    and~(right)\,\mbox{$\sqs=8\,\mathrm{TeV}$}.
    The~results for 
    the~rapidity ranges 
    \mbox{$2.2<\yy<3.0$}, 
    \mbox{$3.0<\yy<3.5$} and
    \mbox{$3.5<\yy<4.5$} are shown with 
    red circles, 
    blue squares and green diamonds, respectively.
    The~vertical inner error bars indicate the~statistical uncertainty, 
    whilst the~outer error bars indicate the~sum of 
    the~statistical and systematic uncertainties added in quadrature. 
    The~horizontal error bars indicate the~bin width.
    Some data points are displaced from 
    the~bin centers to improve visibility.
 }
  \label{fig:figure03hx}
\end{figure}

\begin{figure}[t]
  \setlength{\unitlength}{1mm}
  \centering
  \begin{picture}(150,180)
    \put( 0,  0){ 
      \includegraphics*[width=150mm,height=180mm,%
      ]{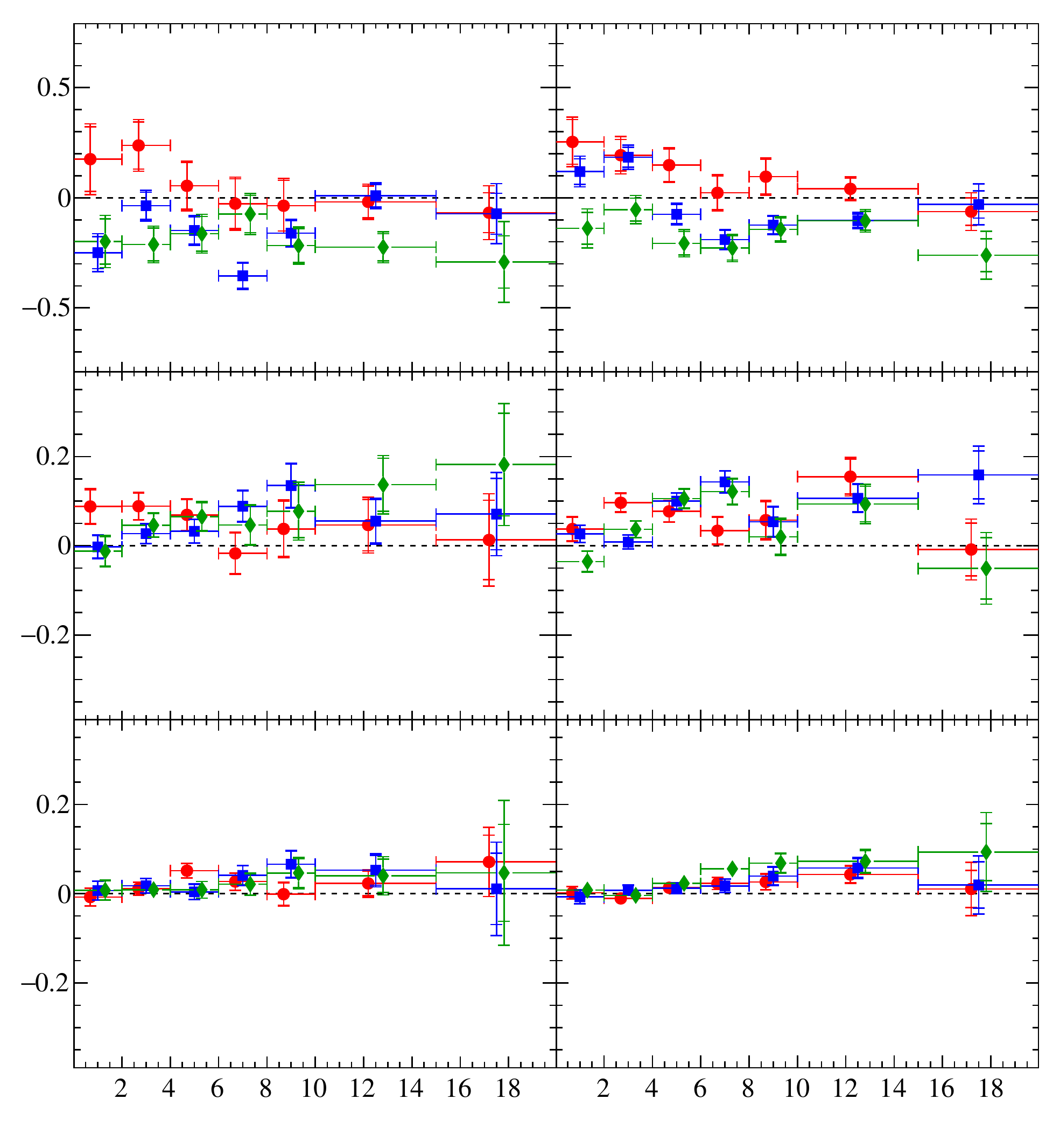}
    }
    \put(15,169) {\small $\begin{array}{l}\text{LHCb} \\ \sqrt{s}=7\tev \end{array}$}
    \put(83,169) {\small $\begin{array}{l}\text{LHCb} \\ \sqrt{s}=8\tev \end{array}$}
    \put(125,128) {\small $\begin{array}{l} \twos \\ \mathrm{CS~frame} \end{array}$}
    \put (43,165){ \small $\begin{array}{cl} 
        {\color{red}\MYCIRCLE}                & 2.2<\yy<3.0 \\
        {\color{blue}\MYSQUARE}          & 3.0<\yy<3.5 \\
        {\color[rgb]{0,0.6,0}\MYDIAMOND} & 3.5<\yy<4.5
      \end{array}$}
    \put( -2,148) { \begin{sideways} $\uplambda_{\theta}$     \end{sideways}}
    \put( -2, 93) { \begin{sideways} $\uplambda_{\theta\phi}$ \end{sideways}}
    \put( -2, 38) { \begin{sideways} $\uplambda_{\phi}$       \end{sideways}}
    \put(111, 2) { $\pty$} \put(135, 2){ $\left[\!\gevc\right]$}
    \put( 43, 2) { $\pty$} \put( 65, 2){ $\left[\!\gevc\right]$}
  \end{picture}
  \caption { \small
    The~polarization parameters 
    (top)\,$\uplambda_{\theta}$,
    (middle)\,$\uplambda_{\theta\phi}$ and 
    (bottom)\,$\uplambda_{\phi}$,
    measured in the~CS frame 
    for the \twos~state
    in different
    bins of \pty and three rapidity ranges, 
    for data collected 
    at~(left)\,\mbox{$\sqs=7\,\mathrm{TeV}$}
    and~(right)\,\mbox{$\sqs=8\,\mathrm{TeV}$}.
    The~results for 
    the~rapidity ranges 
    \mbox{$2.2<\yy<3.0$}, 
    \mbox{$3.0<\yy<3.5$} and
    \mbox{$3.5<\yy<4.5$} are shown with 
    red circles, 
    blue squares and green diamonds, respectively.
    The~vertical inner error bars indicate the~statistical uncertainty, 
    whilst the~outer error bars indicate the~sum of 
    the~statistical and systematic uncertainties added in quadrature. 
    The~horizontal error bars indicate the~bin width.
    Some data points are displaced from 
    the~bin centers to improve visibility.
  }
  \label{fig:figure03cs}
\end{figure}

\begin{figure}[t]
  \setlength{\unitlength}{1mm}
  \centering
  \begin{picture}(150,180)
    \put( 0,  0){ 
      \includegraphics*[width=150mm,height=180mm,%
      ]{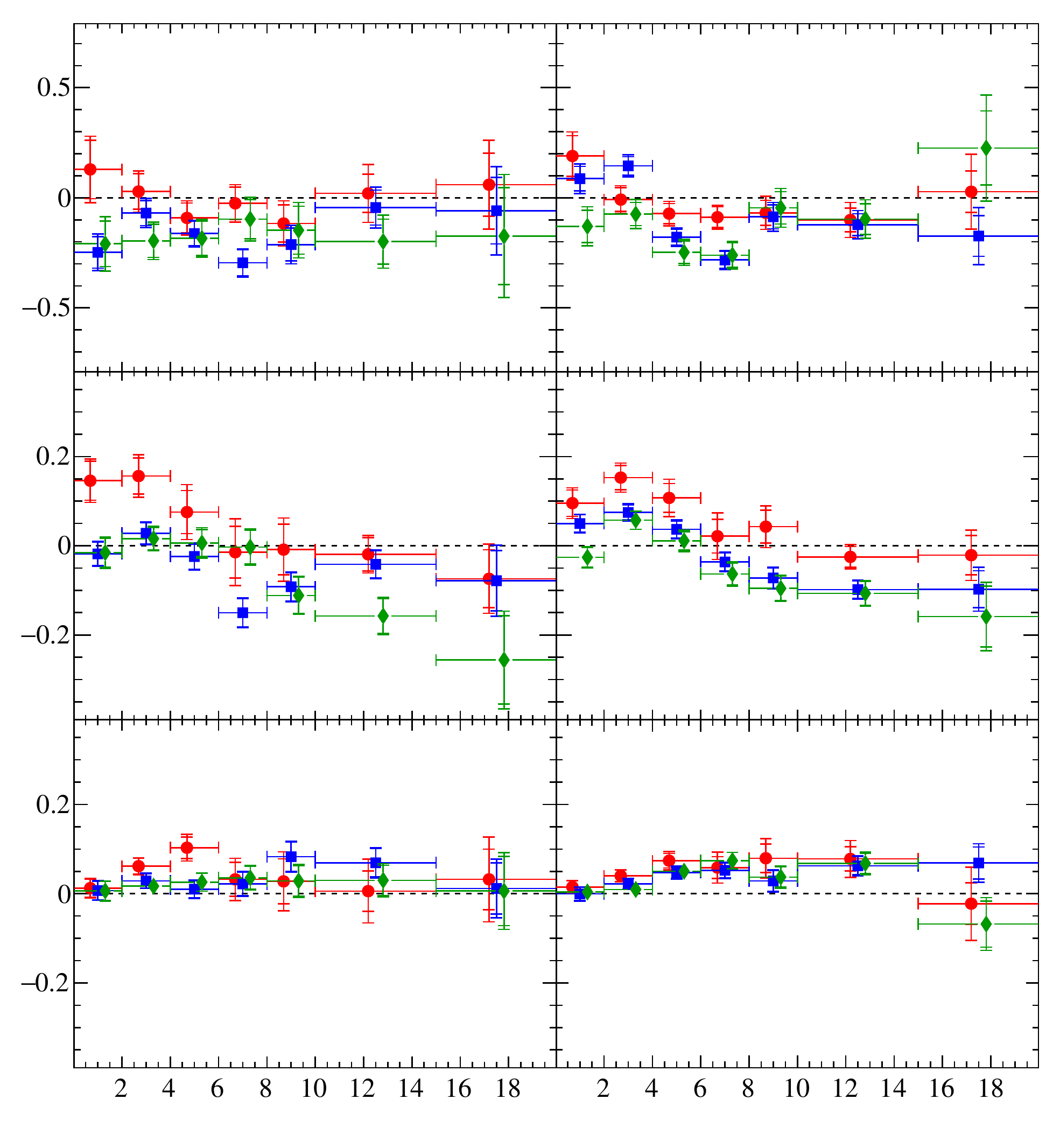}
    }
    \put(15,169) {\small $\begin{array}{l}\text{LHCb} \\ \sqrt{s}=7\tev \end{array}$}
    \put(83,169) {\small $\begin{array}{l}\text{LHCb} \\ \sqrt{s}=8\tev \end{array}$}
    \put(125,128) {\small $\begin{array}{l} \twos \\ \mathrm{GJ~frame} \end{array}$}
    \put (43,165){ \small $\begin{array}{cl} 
        {\color{red}\MYCIRCLE}                & 2.2<\yy<3.0 \\
        {\color{blue}\MYSQUARE}          & 3.0<\yy<3.5 \\
        {\color[rgb]{0,0.6,0}\MYDIAMOND} & 3.5<\yy<4.5
      \end{array}$}
    \put( -2,148) { \begin{sideways} $\uplambda_{\theta}$     \end{sideways}}
    \put( -2, 93) { \begin{sideways} $\uplambda_{\theta\phi}$ \end{sideways}}
    \put( -2, 38) { \begin{sideways} $\uplambda_{\phi}$       \end{sideways}}
    \put(111, 2) { $\pty$} \put(135, 2){ $\left[\!\gevc\right]$}
    \put( 43, 2) { $\pty$} \put( 65, 2){ $\left[\!\gevc\right]$}
  \end{picture}
  \caption { \small
    The~polarization parameters 
    (top)\,$\uplambda_{\theta}$,
    (middle)\,$\uplambda_{\theta\phi}$ and 
    (bottom)\,$\uplambda_{\phi}$,
    measured in the~GJ frame 
    for the \twos~state
    in different
    bins of \pty and three rapidity ranges, 
    for data collected 
    at~(left)\,\mbox{$\sqs=7\,\mathrm{TeV}$}
    and~(right)\,\mbox{$\sqs=8\,\mathrm{TeV}$}.
    The~results for 
    the~rapidity ranges 
    \mbox{$2.2<\yy<3.0$}, 
    \mbox{$3.0<\yy<3.5$} and
    \mbox{$3.5<\yy<4.5$} are shown with 
    red circles, 
    blue squares and green diamonds, respectively.
    The~vertical inner error bars indicate the~statistical uncertainty, 
    whilst the~outer error bars indicate the~sum of 
    the~statistical and systematic uncertainties added in quadrature. 
    The~horizontal error bars indicate the~bin width.
    Some data points are displaced from 
    the~bin centers to improve visibility.
  }
  \label{fig:figure03gj}
\end{figure}

\begin{figure}[t]
  \setlength{\unitlength}{1mm}
  \centering
  \begin{picture}(150,180)
    \put( 0,  0){ 
      \includegraphics*[width=150mm,height=180mm,%
      ]{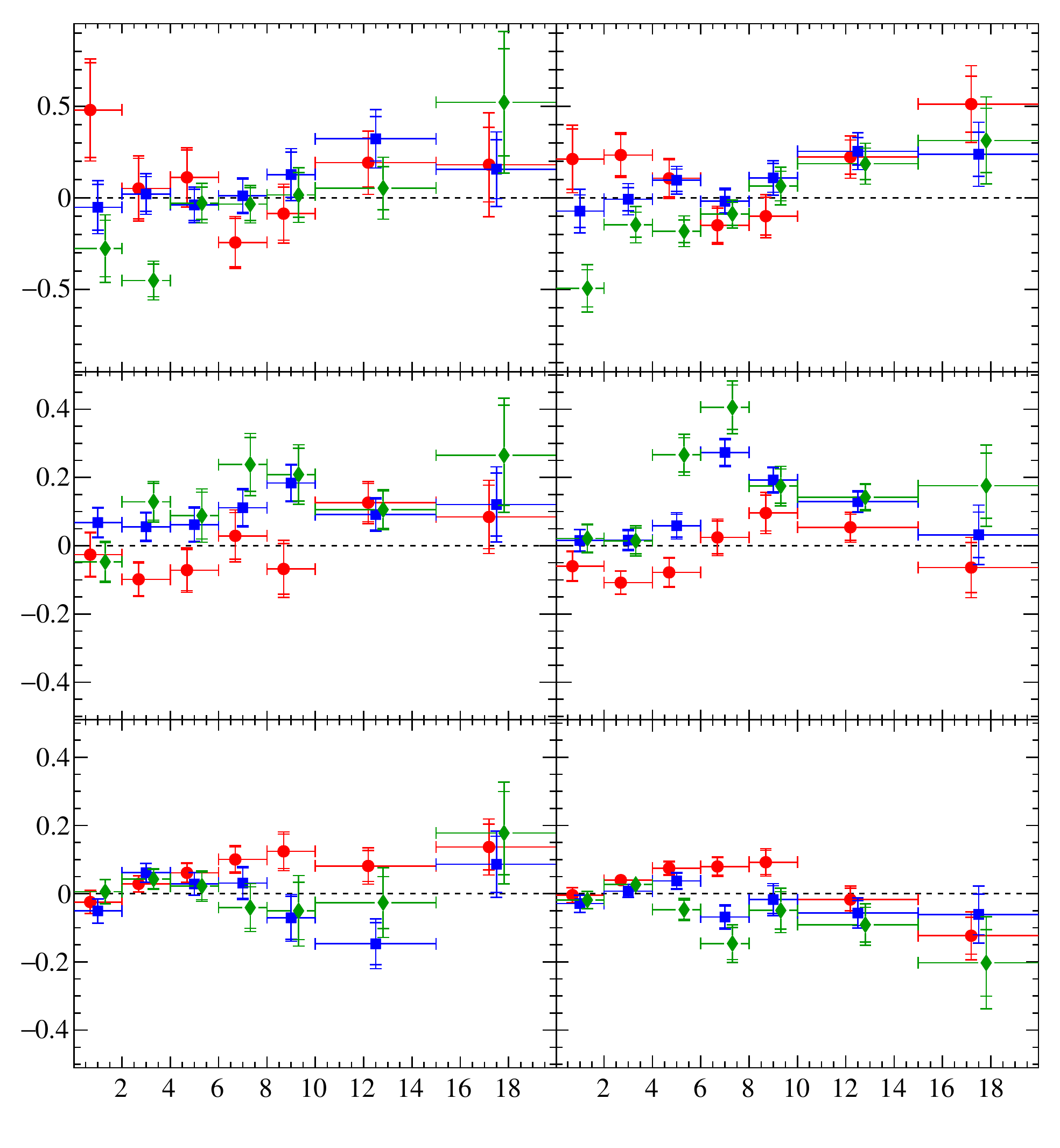}
    }
    \put(15,169) {\small $\begin{array}{l}\text{LHCb} \\ \sqrt{s}=7\tev \end{array}$}
    \put(83,169) {\small $\begin{array}{l}\text{LHCb} \\ \sqrt{s}=8\tev \end{array}$}
    \put(125,128) {\small $\begin{array}{l} \threes \\ \mathrm{HX~frame} \end{array}$}
    \put (43,130){ \small $\begin{array}{cl} 
        {\color{red}\MYCIRCLE}                & 2.2<\yy<3.0 \\
        {\color{blue}\MYSQUARE}          & 3.0<\yy<3.5 \\
        {\color[rgb]{0,0.6,0}\MYDIAMOND} & 3.5<\yy<4.5
      \end{array}$}
    \put( -2,148) { \begin{sideways} $\uplambda_{\theta}$     \end{sideways}}
    \put( -2, 93) { \begin{sideways} $\uplambda_{\theta\phi}$ \end{sideways}}
    \put( -2, 38) { \begin{sideways} $\uplambda_{\phi}$       \end{sideways}}
    \put(111, 2) { $\pty$} \put(135, 2){ $\left[\!\gevc\right]$}
    \put( 43, 2) { $\pty$} \put( 65, 2){ $\left[\!\gevc\right]$}
  \end{picture}
  \caption { \small
    The~polarization parameters 
    (top)\,$\uplambda_{\theta}$,
    (middle)\,$\uplambda_{\theta\phi}$ and 
    (bottom)\,$\uplambda_{\phi}$,
    measured in the~HX frame 
    for the \threes~state
    in different
    bins of \pty and three rapidity ranges, 
    for data collected 
    at~(left)\,\mbox{$\sqs=7\,\mathrm{TeV}$}
    and~(right)\,\mbox{$\sqs=8\,\mathrm{TeV}$}.
    The~results for 
    the~rapidity ranges 
    \mbox{$2.2<\yy<3.0$}, 
    \mbox{$3.0<\yy<3.5$} and
    \mbox{$3.5<\yy<4.5$} are shown with 
    red circles, 
    blue squares and green diamonds, respectively.
    The~vertical inner error bars indicate the~statistical uncertainty, 
    whilst the~outer error bars indicate the~sum of 
    the~statistical and systematic uncertainties added in quadrature. 
    The~horizontal error bars indicate the~bin width.
    Some data points are displaced from 
    the~bin centers to improve visibility.
  }
  \label{fig:figure04hx}
\end{figure}

\begin{figure}[t]
  \setlength{\unitlength}{1mm}
  \centering
  \begin{picture}(150,180)
    \put( 0,  0){ 
      \includegraphics*[width=150mm,height=180mm,%
      ]{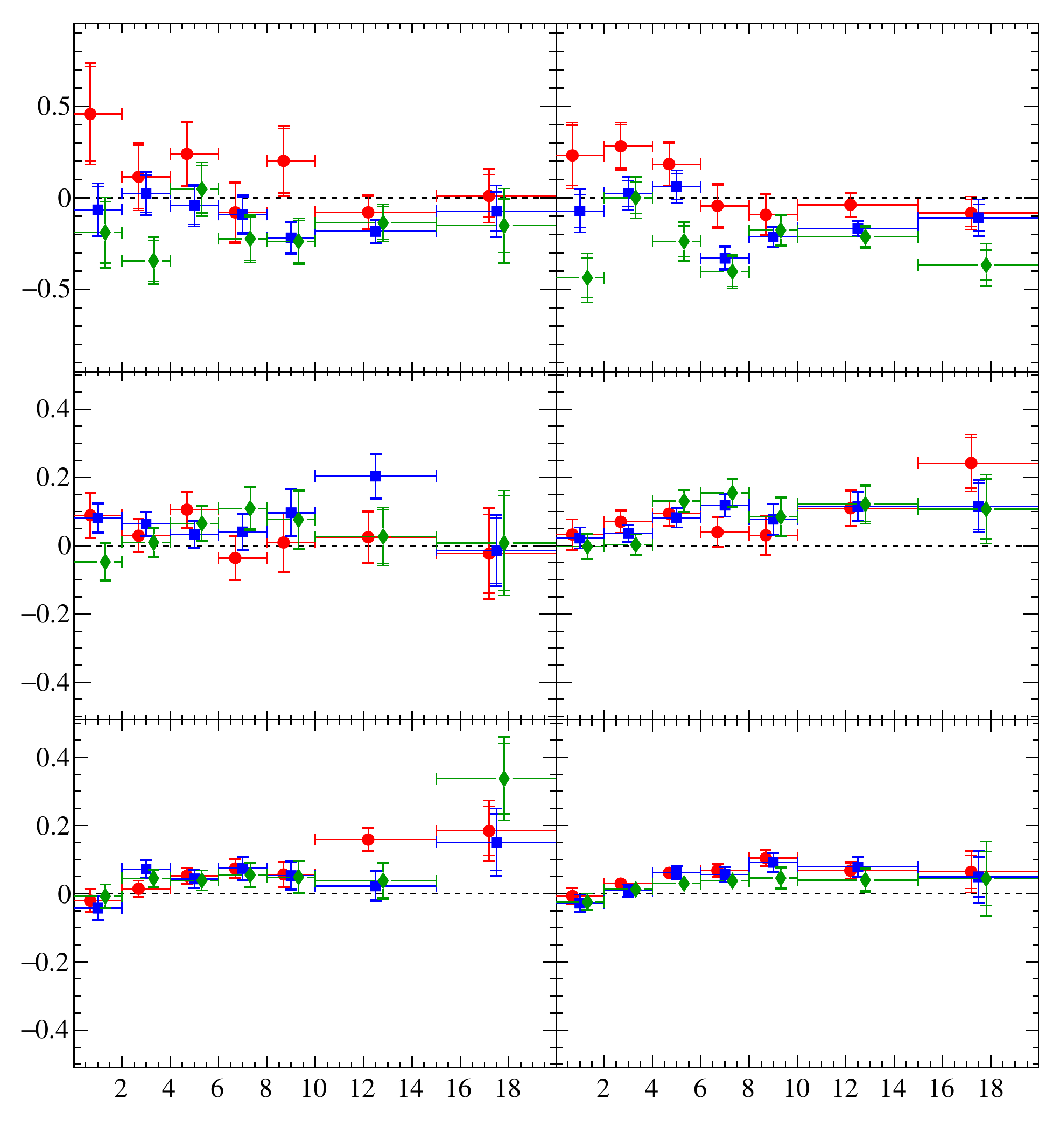}
    }
    \put(15,169) {\small $\begin{array}{l}\text{LHCb} \\ \sqrt{s}=7\tev \end{array}$}
    \put(83,169) {\small $\begin{array}{l}\text{LHCb} \\ \sqrt{s}=8\tev \end{array}$}
    \put(125,128) {\small $\begin{array}{l} \threes \\ \mathrm{CS~frame} \end{array}$}
    \put (43,130){ \small $\begin{array}{cl} 
        {\color{red}\MYCIRCLE}                & 2.2<\yy<3.0 \\
        {\color{blue}\MYSQUARE}          & 3.0<\yy<3.5 \\
        {\color[rgb]{0,0.6,0}\MYDIAMOND} & 3.5<\yy<4.5
      \end{array}$}
    \put( -2,148) { \begin{sideways} $\uplambda_{\theta}$     \end{sideways}}
    \put( -2, 93) { \begin{sideways} $\uplambda_{\theta\phi}$ \end{sideways}}
    \put( -2, 38) { \begin{sideways} $\uplambda_{\phi}$       \end{sideways}}
    \put(111, 2) { $\pty$} \put(135, 2){ $\left[\!\gevc\right]$}
    \put( 43, 2) { $\pty$} \put( 65, 2){ $\left[\!\gevc\right]$}
  \end{picture}
  \caption { \small
    The~polarization parameters 
    $\uplambda_{\theta}$\,(top),
    $\uplambda_{\theta\phi}$\,(middle) and 
    $\uplambda_{\phi}$\,(bottom),
    measured in the~CS frame 
    for the \threes~state
    in different
    bins of \pty and three rapidity ranges, 
    for data collected 
    at~\mbox{$\sqs=7\,\mathrm{TeV}$}\,(left)
    and~\mbox{$\sqs=8\,\mathrm{TeV}$}\,(right).
    The~results for 
    the~rapidity ranges 
    \mbox{$2.2<\yy<3.0$}, 
    \mbox{$3.0<\yy<3.5$} and
    \mbox{$3.5<\yy<4.5$} are shown with 
    red circles, 
    blue squares and green diamonds, respectively.
    The~vertical inner error bars indicate the~statistical uncertainty, 
    whilst the~outer error bars indicate the~sum of 
    the~statistical and systematic uncertainties added in quadrature. 
    The~horizontal error bars indicate the~bin width.
    Some data points are displaced from 
    the~bin centers to improve visibility.
  }
  \label{fig:figure04cs}
\end{figure}

\begin{figure}[t]
  \setlength{\unitlength}{1mm}
  \centering
  \begin{picture}(150,180)
    \put( 0,  0){ 
      \includegraphics*[width=150mm,height=180mm,%
      ]{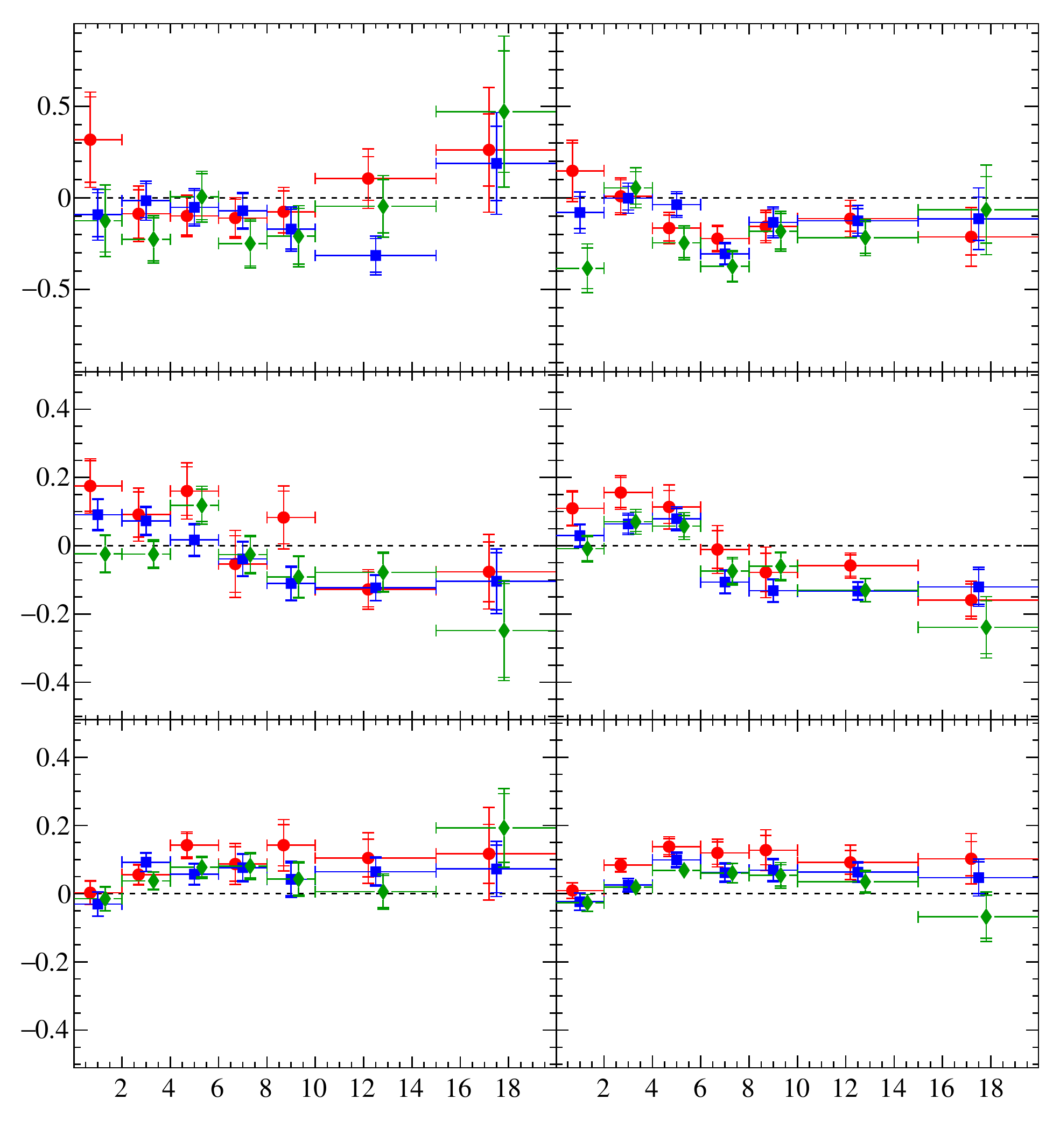}
    }
    \put(15,169) {\small $\begin{array}{l}\text{LHCb} \\ \sqrt{s}=7\tev \end{array}$}
    \put(83,169) {\small $\begin{array}{l}\text{LHCb} \\ \sqrt{s}=8\tev \end{array}$}
    \put(125,128) {\small $\begin{array}{l} \threes \\ \mathrm{GJ~frame} \end{array}$}
    \put (43,129){ \small $\begin{array}{cl} 
        {\color{red}\MYCIRCLE}                & 2.2<\yy<3.0 \\
        {\color{blue}\MYSQUARE}          & 3.0<\yy<3.5 \\
        {\color[rgb]{0,0.6,0}\MYDIAMOND} & 3.5<\yy<4.5
      \end{array}$}
    \put( -2,148) { \begin{sideways} $\uplambda_{\theta}$     \end{sideways}}
    \put( -2, 93) { \begin{sideways} $\uplambda_{\theta\phi}$ \end{sideways}}
    \put( -2, 38) { \begin{sideways} $\uplambda_{\phi}$       \end{sideways}}
    \put(111, 2) { $\pty$} \put(135, 2){ $\left[\!\gevc\right]$}
    \put( 43, 2) { $\pty$} \put( 65, 2){ $\left[\!\gevc\right]$}
  \end{picture}
  \caption { \small
    The~polarization parameters 
    $\uplambda_{\theta}$\,(top),
    $\uplambda_{\theta\phi}$\,(middle) and 
    $\uplambda_{\phi}$\,(bottom),
    measured in the~GJ frame 
    for the \threes~state
    in different
    bins of \pty and three rapidity ranges, 
    for data collected 
    at~\mbox{$\sqs=7\,\mathrm{TeV}$}\,(left)
    and~\mbox{$\sqs=8\,\mathrm{TeV}$}\,(right).
    The~results for 
    the~rapidity ranges 
    \mbox{$2.2<\yy<3.0$}, 
    \mbox{$3.0<\yy<3.5$} and
    \mbox{$3.5<\yy<4.5$} are shown with 
    red circles, 
    blue squares and green diamonds, respectively.
    The~vertical inner error bars indicate the~statistical uncertainty, 
    whilst the~outer error bars indicate the~sum of 
    the~statistical and systematic uncertainties added in quadrature. 
    The~horizontal error bars indicate the~bin width.
    Some data points are displaced from 
    the~bin centers to improve visibility.
  }
  \label{fig:figure04gj}
\end{figure}

\begin{figure}[t]
  \setlength{\unitlength}{1mm}
  \centering
  \begin{picture}(150,180)
    \put( 0,  0){ 
      \includegraphics*[width=150mm,height=180mm,%
      ]{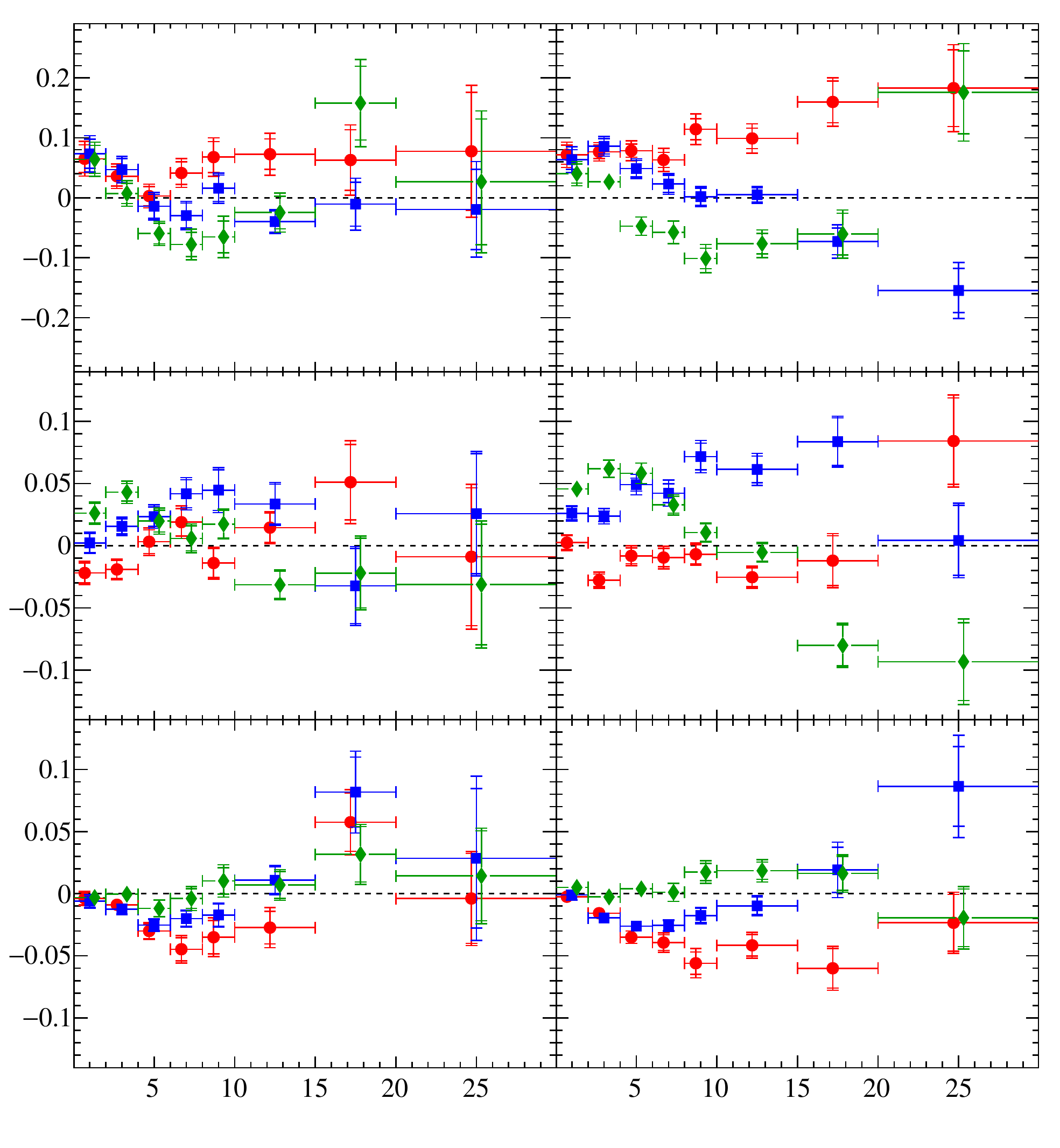}
    }
    \put(15,169) {\small $\begin{array}{l}\text{LHCb} \\ \sqrt{s}=7\tev \end{array}$}
    \put(83,169) {\small $\begin{array}{l}\text{LHCb} \\ \sqrt{s}=8\tev \end{array}$}
    \put(131,126) {\small $\begin{array}{l} \ones \end{array}$}
    \put(50,129) {\small $\begin{array}{cl}
        {\color{red}\MYCIRCLE}                & \mathrm{HX~frame} \\
        {\color{blue}\MYSQUARE}          & \mathrm{CS~frame} \\
        {\color[rgb]{0,0.6,0}\MYDIAMOND} & \mathrm{GJ~frame}
      \end{array}$}
    \put( -2,148) { \begin{sideways} $\uplambda_{\theta}$ \end{sideways}}
    \put( -2, 93) { \begin{sideways} $\uplambda_{\theta\phi}$ \end{sideways}}
    \put( -2, 38) { \begin{sideways} $\uplambda_{\phi}$ \end{sideways}}
    \put(111, 2) { $\pty$} \put(135, 2){ $\left[\!\gevc\right]$}
    \put( 43, 2) { $\pty$} \put( 65, 2){ $\left[\!\gevc\right]$}
  \end{picture}
  \caption { \small
    The~polarization parameters 
    (top)\,$\uplambda_{\theta}$,
    (middle)\,$\uplambda_{\theta\phi}$ and 
    (bottom)\,$\uplambda_{\phi}$,
    for \ones~mesons 
    as a~function of \pty,
    for the~rapidity range $2.2<\yy<4.5$,
    for data collected 
    at~(left)\,\mbox{$\sqs=7\,\mathrm{TeV}$}
    and~(right)\,\mbox{$\sqs=8\,\mathrm{TeV}$}.
    The~results for 
    the~HX, CS and GJ~frames 
    are shown with     
    red circles, 
    blue squares and green diamonds, respectively.
    The~inner error bars indicate the~statistical uncertainty, 
    whilst the~outer error bars indicate the~sum of 
    the~statistical and systematic uncertainties added in quadrature. 
    Some data points are displaced from 
    the~bin centers to improve visibility.
  }
  \label{fig:figure05}
\end{figure}

\begin{figure}[t]
  \setlength{\unitlength}{1mm}
  \centering
  \begin{picture}(150,180)
    \put( 0,  0){ 
      \includegraphics*[width=150mm,height=180mm,%
      ]{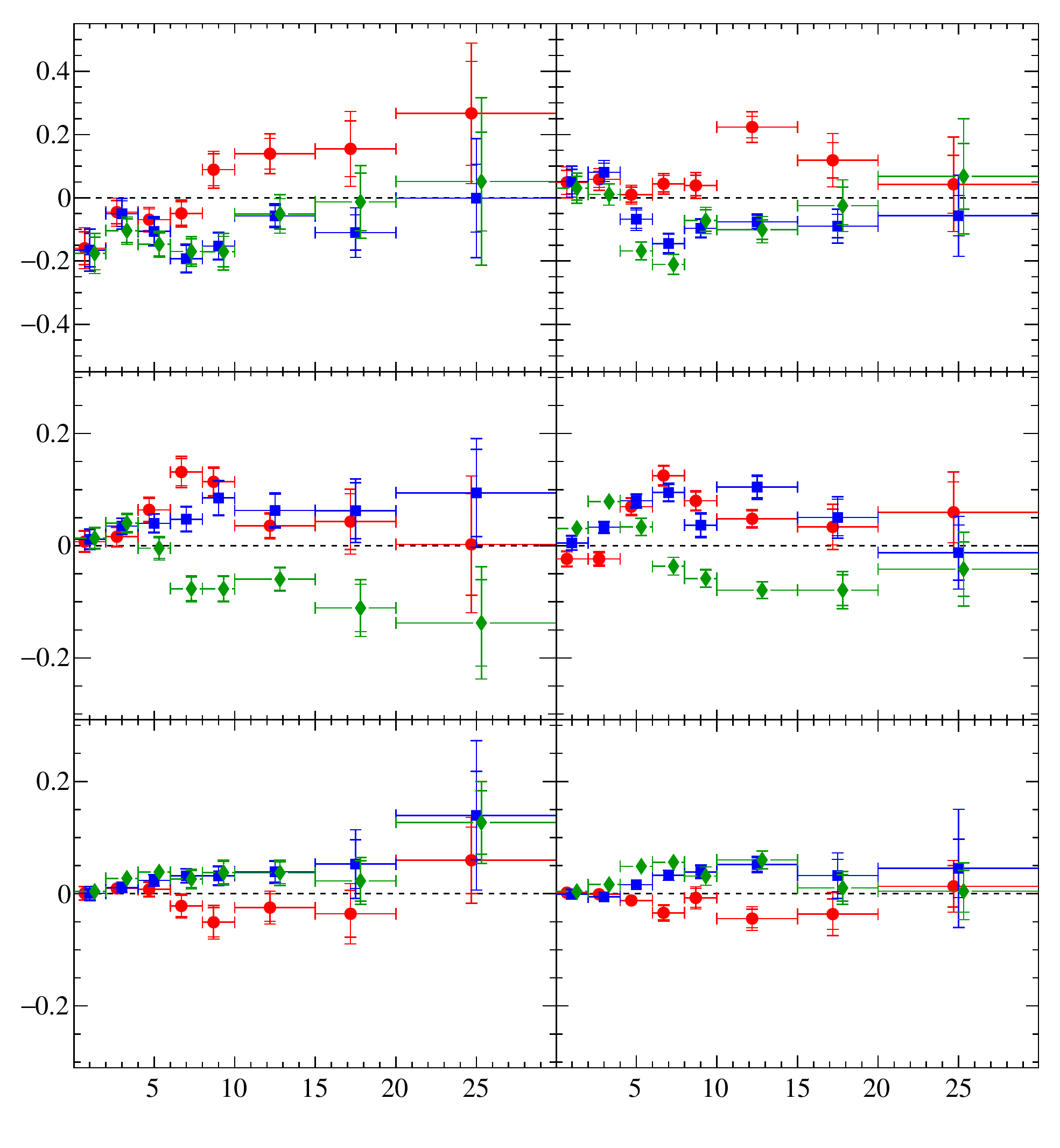}
    }
    \put(15,169) {\small $\begin{array}{l}\text{LHCb} \\ \sqrt{s}=7\tev \end{array}$}
    \put(83,169) {\small $\begin{array}{l}\text{LHCb} \\ \sqrt{s}=8\tev \end{array}$}
    \put(131,126) {\small $\begin{array}{l} \twos \end{array}$}
    \put(50,129) {\small $\begin{array}{cl}
        {\color{red}\MYCIRCLE}                & \mathrm{HX~frame} \\
        {\color{blue}\MYSQUARE}          & \mathrm{CS~frame} \\
        {\color[rgb]{0,0.6,0}\MYDIAMOND} & \mathrm{GJ~frame}
      \end{array}$}
    \put( -2,148) { \begin{sideways} $\uplambda_{\theta}$ \end{sideways}}
    \put( -2, 93) { \begin{sideways} $\uplambda_{\theta\phi}$ \end{sideways}}
    \put( -2, 38) { \begin{sideways} $\uplambda_{\phi}$ \end{sideways}}
    \put(111, 2) { $\pty$} \put(135, 2){ $\left[\!\gevc\right]$}
    \put( 43, 2) { $\pty$} \put( 65, 2){ $\left[\!\gevc\right]$}
  \end{picture}
  \caption { \small
    The~polarization parameters 
    (top)\,$\uplambda_{\theta}$,
    (middle)\,$\uplambda_{\theta\phi}$ and 
    (bottom)\,$\uplambda_{\phi}$,
    for \twos~mesons 
    as a~function of \pty,
    for the~rapidity range $2.2<\yy<4.5$,
    for data collected 
    at~(left)\,\mbox{$\sqs=7\,\mathrm{TeV}$}
    and~(right)\,\mbox{$\sqs=8\,\mathrm{TeV}$}.
    The~results for 
    the~HX, CS and GJ~frames 
    are shown with     
    red circles, 
    blue squares and green diamonds, respectively.
    The~inner error bars indicate the~statistical uncertainty, 
    whilst the~outer error bars indicate the~sum of 
    the~statistical and systematic uncertainties added in quadrature. 
    Some data points are displaced from 
    the~bin centers to improve visibility.
  }
  \label{fig:figure06}
\end{figure}

\begin{figure}[t]
  \setlength{\unitlength}{1mm}
  \centering
  \begin{picture}(150,180)
    \put( 0,  0){ 
      \includegraphics*[width=150mm,height=180mm,%
      ]{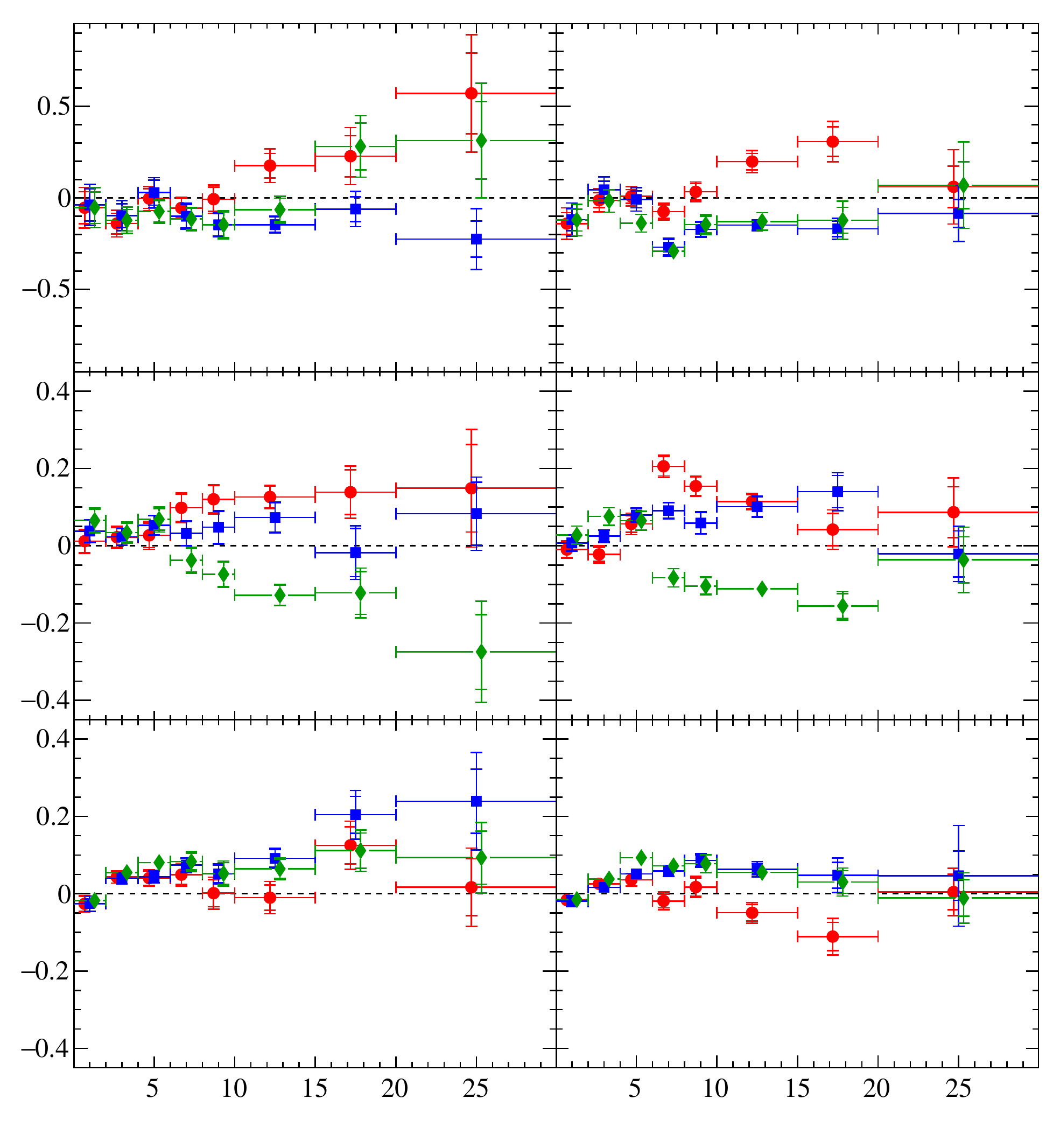}
    }
    \put(15,169) {\small $\begin{array}{l}\text{LHCb} \\ \sqrt{s}=7\tev \end{array}$}
    \put(83,169) {\small $\begin{array}{l}\text{LHCb} \\ \sqrt{s}=8\tev \end{array}$}
    \put(131,126) {\small $\begin{array}{l} \threes \end{array}$}
    \put(50,129) {\small $\begin{array}{cl}
        {\color{red}\MYCIRCLE}                & \mathrm{HX~frame} \\
        {\color{blue}\MYSQUARE}          & \mathrm{CS~frame} \\
        {\color[rgb]{0,0.6,0}\MYDIAMOND} & \mathrm{GJ~frame}
      \end{array}$}
    \put( -2,148) { \begin{sideways} $\uplambda_{\theta}$ \end{sideways}}
    \put( -2, 93) { \begin{sideways} $\uplambda_{\theta\phi}$ \end{sideways}}
    \put( -2, 38) { \begin{sideways} $\uplambda_{\phi}$ \end{sideways}}
    \put(111, 2) { $\pty$} \put(135, 2){ $\left[\!\gevc\right]$}
    \put( 43, 2) { $\pty$} \put( 65, 2){ $\left[\!\gevc\right]$}
  \end{picture}
  \caption { \small
    The~polarization parameters 
    (top)\,$\uplambda_{\theta}$,
    (middle)\,$\uplambda_{\theta\phi}$ and 
    (bottom)\,$\uplambda_{\phi}$,
    for \threes~mesons 
    as a~function of \pty,
    for the~rapidity range $2.2<\yy<4.5$,
    for data collected 
    at~(left)\,\mbox{$\sqs=7\,\mathrm{TeV}$}
    and~(right)\,\mbox{$\sqs=8\,\mathrm{TeV}$}.
    The~results for 
    the~HX, CS and GJ~frames 
    are shown with     
    red circles, 
    blue squares and green diamonds, respectively.
    The~inner error bars indicate the~statistical uncertainty, 
    whilst the~outer error bars indicate the~sum of 
    the~statistical and systematic uncertainties added in quadrature. 
    Some data points are displaced from 
    the~bin centers to improve visibility.
  }
  \label{fig:figure07}
\end{figure}

The~polarization parameters
$\uplambda_{\theta}$, $\uplambda_{\theta\phi}$,
$\uplambda_{\phi}$ for the \ups~mesons, measured
in the HX, CS and GJ~frames
for different $\left(\pty,~\yy\right)$ bins,
for data collected at~\mbox{$\sqs=7\tev$} 
and $8\tev$, are shown in 
   Figs.~\ref{fig:figure02hx},\,\ref{fig:figure02cs}
   and~\ref{fig:figure02gj} for the~\ones~meson,
in Figs.~\ref{fig:figure03hx},\,\ref{fig:figure03cs}
   and~\ref{fig:figure03gj} for the~\twos~meson and
in Figs.~\ref{fig:figure04hx},\,\ref{fig:figure04cs}
and~\ref{fig:figure04gj} for the~\threes~meson.
The~parameters $\pmb{\uplambda}$ do not show significant variations
as a function of \yy,
in accordance with expectations.
Figures~\ref{fig:figure05},\,\ref{fig:figure06} and~\ref{fig:figure07} show 
the polarization parameters measured in the~full considered 
rapidity range $2.2<\yy<4.5$ and
the~wide region of transverse momentum up to 30\gevc.
All~polarization parameters are listed
in Appendix~\ref{sec:upsonespol} for the~\ones~mesons,
Appendix~\ref{sec:upstwospol} for the~\twos~mesons
and
Appendix~\ref{sec:upsthreespol} for the~\threes~mesons.
The~correlation coefficients between the different polarization parameters
are, in general, small, especially between the~$\uplambda_{\theta}$ and
$\uplambda_{\phi}$~parameters.
The~smallest correlation coefficients are obtained in the~CS~frame.

The~values of the~parameter $\uplambda_{\theta}$
measured in the~HX, CS and GJ frames do not show large  
transverse or longitudinal polarization
over the~considered kinematic region.
The~values of the~parameters $\uplambda_{\theta\phi}$ and
$\uplambda_{\phi}$ are small in all polarization 
frames, over all $\left(\pty,~\yy\right)$ bins.
The~\ups polarization results are in good agreement with
those obtained by the~CMS collaboration~\cite{CMSpolar2013}.
The~polarization results obtained for the~two centre\nobreakdash-of\nobreakdash-mass
energies, \mbox{$\sqrt{s}=7\tev$} and~\mbox{$8\tev$}, 
are similar and show a~good agreement.

In~the~rest frame of the~\ups meson,
the~spin\nobreakdash-1 density matrix is 
proportional to~\cite{Teryaev:2011zza,Teryaev:SPIN2011}
\begin{equation*}
\left(\begin{matrix}
    \dfrac{1-\uplambda_{\theta}}{2} & \uplambda_{\theta\phi} & 0 \\ 
    \uplambda_{\theta\phi}          &  \dfrac{1+\uplambda_{\theta}-2\uplambda_{\phi}}{2} & 0 \\ 
      0                             & 0    & \dfrac{1+\uplambda_{\theta}+2\uplambda_{\phi}}{2}  
\end{matrix}\right).
\end{equation*}
The~positivity of the~density matrix
imposes constraints on the~$\pmb{\uplambda}$~parameters
as follows~\mbox{\cite{Lam:1980uc,Palestini:2010xu,Faccioli_2010_b,Faccioli_2011_c,Teryaev:2011zza,Teryaev:SPIN2011}}
\begin{eqnarray*}
0 & \le & \mathcal{C}_1 = 1 - \left| \uplambda_{\theta} \right| \\
0 & \le & \mathcal{C}_2 = 1 + \uplambda_{\theta} - 2\left| \uplambda_{\phi} \right| \\
0 & \le & \mathcal{C}_3 = \left( 1  - \uplambda_{\theta} \right)\left( 1+\uplambda_{\theta} - 2\uplambda_{\phi}\right) - 4 \uplambda_{\theta\phi}^2 \\ 
0 & \le & \mathcal{C}_4 = \left( 1 - \uplambda_{\theta} \right) \left( 1 + \uplambda_{\theta} + 2 \uplambda_{\phi}\right)  \\
0 & \le & \mathcal{C}_5 = \left( 1 + \uplambda_{\theta} \right)^2 - 4\uplambda_{\phi}^2 \\
0 & \le & \mathcal{C}_6 = \left(1+\uplambda_{\theta}+2\uplambda_{\phi}\right)\left( \left(1-\uplambda_{\theta}\right)\left(1+\uplambda_{\theta}-2\uplambda_{\phi}\right) - 4\uplambda_{\theta\phi}^2\right).
\end{eqnarray*}
The~measured values of the~$\pmb{\uplambda}$~parameters satisfy
these positivity constraints $\mathcal{C}_i$
in all frames over all   $\left(\pty,\yy\right)$~bins.

\begin{figure}[t]
  \setlength{\unitlength}{1mm}
  \centering
  \begin{picture}(150,180)
    \put( 0,  0){ 
      \includegraphics*[width=150mm,height=180mm,%
      ]{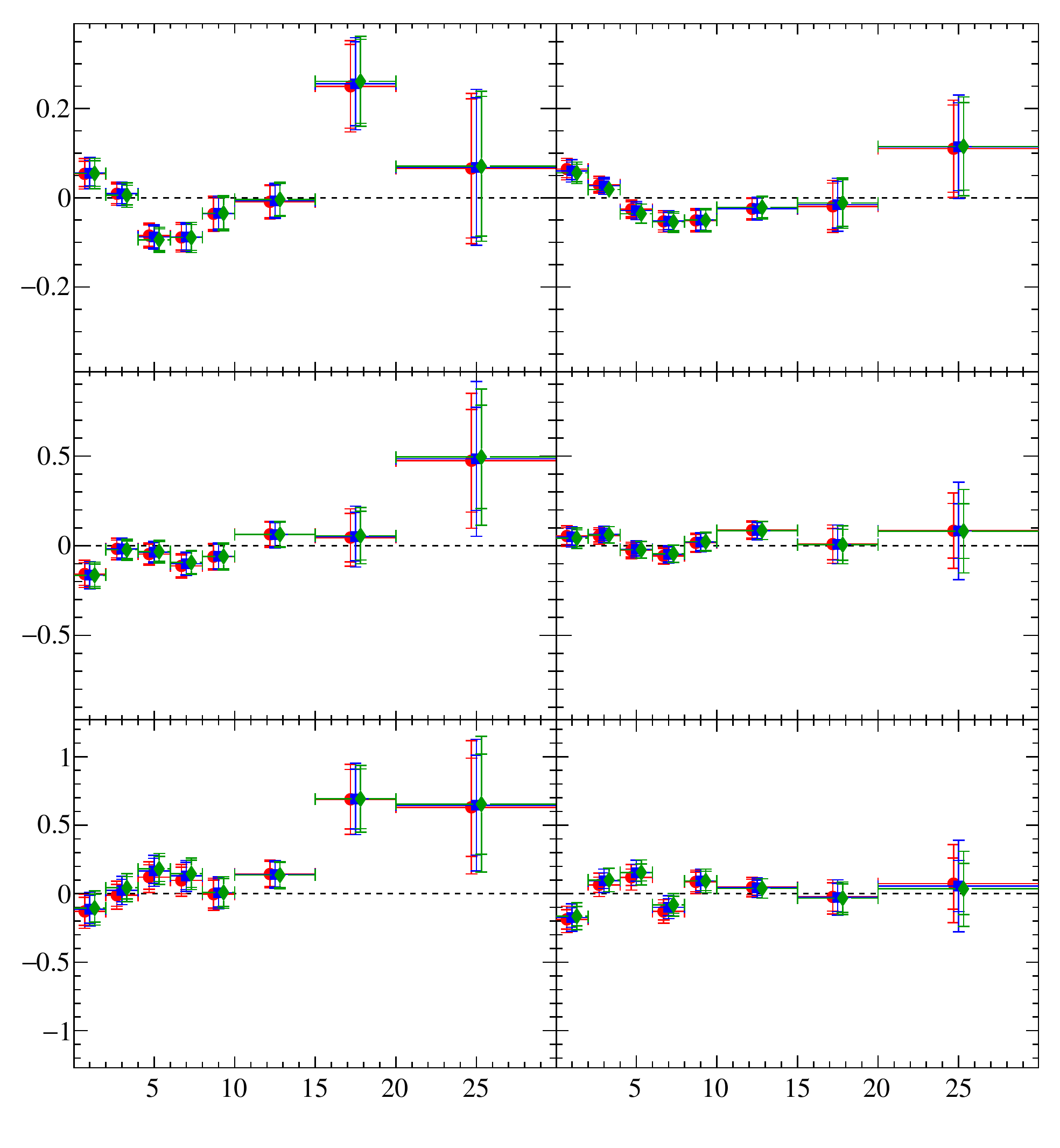}
    }
    \put(15,169) {\small $\begin{array}{l}\text{LHCb} \\ \sqrt{s}=7\tev \end{array}$}
    \put(83,169) {\small $\begin{array}{l}\text{LHCb} \\ \sqrt{s}=8\tev \end{array}$}
    \put(131,126) {\small $\begin{array}{l} \ones   \end{array}$}
    \put(131, 71) {\small $\begin{array}{l} \twos   \end{array}$}
    \put(131, 16) {\small $\begin{array}{l} \threes \end{array}$}
    \put(50,129) {\small $\begin{array}{cl}
        {\color{red}\MYCIRCLE}                & \mathrm{HX~frame} \\
        {\color{blue}\MYSQUARE}          & \mathrm{CS~frame} \\
        {\color[rgb]{0,0.6,0}\MYDIAMOND} & \mathrm{GJ~frame}
      \end{array}$}
    \put( -2,148) { \begin{sideways} $\tilde \uplambda$ \end{sideways}}
    \put( -2, 93) { \begin{sideways} $\tilde \uplambda$ \end{sideways}}
    \put( -2, 39) { \begin{sideways} $\tilde \uplambda$ \end{sideways}}
    \put(111, 2) { $\pty$} \put(135, 2){ $\left[\!\gevc\right]$}
    \put( 43, 2) { $\pty$} \put( 65, 2){ $\left[\!\gevc\right]$}
  \end{picture}
  \caption { \small
    The~polarization parameter $\tilde\uplambda$
    for 
    (top)\,\ones~mesons,
    (middle)\,\twos~mesons and 
    (bottom)\,\threes~mesons
    as a~function of \pty,
    for the~rapidity range $2.2<\yy<4.5$,
    for data collected 
    at~(left)\,\mbox{$\sqs=7\,\mathrm{TeV}$}
    and~(right)\,\mbox{$\sqs=8\,\mathrm{TeV}$}.
    The~results for 
    the~HX, CS and GJ~frames 
    are shown with     
    red circles, 
    blue squares and green diamonds, respectively.
    The~inner error bars indicate the~statistical uncertainty, 
    whilst the~outer error bars indicate the~sum of 
    the~statistical and systematic uncertainties added in quadrature. 
    Some data points are displaced from 
    the~bin centers to improve visibility.
  }
  \label{fig:figure08}
\end{figure}

The~frame\nobreakdash-invariant polarization parameter $\tilde\uplambda$
measured in the~HX, CS and GJ frames is shown in Fig.~\ref{fig:figure08}.
A~possible disagreement between the~values of $\tilde\uplambda$ measured in
the~different frames would indicate an~unaccounted systematic 
uncertainty, \eg related to limitations
of the~simulation. No~such disagreement is found.
The~rotation angles between
the~different frames depend
on the~transverse momentum of 
the~\ups~mesons
and vanish for small \pty,
resulting in a~degeneracy between
the~three frames~\cite{Telegdi1986,Faccioli_clarification}.
Due~to this degeneracy, the~polarization results
for different frames are very similar
for low\nobreakdash-\pty bins, see \eg Figs.~\ref{fig:figure05},
\ref{fig:figure06} and~\ref{fig:figure07}.

%

%

%

%

%

%

\section{Summary}\label{sec:summary}

A~polarization analysis is carried out
for the \ones, \twos and \threes mesons
in $\proton\proton$~collision data 
at~\mbox{$\sqs=7$} and $8\tev$ 
at~LHCb, corresponding to
integrated luminosities of~1~and 2\invfb, respectively.
The~analysis is performed in the~helicity,
Collins\nobreakdash-Soper and 
Gottfried\nobreakdash-Jackson frames
by studying the~angular distribution 
of the~$\mup$~lepton
in the~rest frame of the~$\ups$~meson, 
in~\mbox{$\ups\to\mumu$}~decays.
The~angular distribution parameters 
$\uplambda_{\theta}$,
$\uplambda_{\theta\phi}$ and 
$\uplambda_{\phi}$,
as well as the~frame\nobreakdash-invariant parameter
$\tilde{\uplambda}$, are measured as functions of
the~$\ups$~transverse momentum $\pty$ and rapidity $\yy$,
in the~regions $\pty<30~\gevc$ and $2.2<\yy<4.5$.

The~values of the~$\uplambda_{\theta}$ parameter
obtained for all the~\ups~mesons
show no large transverse or longitudinal
polarization in all frames
over the~accessible phase space domain.
The~values of the~$\uplambda_{\theta\phi}$ and
$\uplambda_{\phi}$ parameters are small
in all frames over the~accessible kinematic region.
The~values of the~frame\nobreakdash-invariant parameter
$\tilde{\uplambda}$ measured in the~HX, CS and 
GJ~frames are consistent.
The~polarization results corresponding to
$\sqrt{s}=7$~and \mbox{8\tev}~are in good agreement.
The~\ups polarization results  agree with 
the~results obtained by the~CMS collaboration~\cite{CMSpolar2013}.

\section*{Acknowledgements}


\noindent 
We would like to thank
S.~P.~Baranov,
K.-T.~Chao,
J.-P.~Lansberg, 
A.~K.~Likhoded,
H.-S.~Shao,
S.~R.~Slabospitsky and 
O.~V.~Teryaev
for interesting 
and stimulating discussions on quarkonium polarization.~We~express our 
gratitude to our colleagues in the~CERN
accelerator departments for the excellent performance of the~LHC.
We~thank the~technical and administrative staff at the LHCb
institutes. 
We~acknowledge support from CERN and from the~national agencies: 
CAPES, CNPq, FAPERJ and FINEP\,(Brazil); 
MOST and NSFC\,(China);
CNRS/IN2P3\,(France); 
BMBF, DFG and MPG\,(Germany); 
INFN\,(Italy); 
NWO\,(The~Netherlands); 
MNiSW and NCN\,(Poland); 
MEN/IFA\,(Romania); 
MinES and FASO\,(Russia); 
MinECo\,(Spain); 
SNSF and SER\,(Switzerland); 
NASU\,(Ukraine); 
STFC\,(United Kingdom); 
NSF\,(USA).
We~acknowledge the~computing resources that are provided by CERN, 
IN2P3\,(France), 
KIT and DESY\,(Germany), 
INFN\,(Italy), 
SURF\,(The~Netherlands), 
PIC\,(Spain), 
GridPP\,(United Kingdom), 
RRCKI and Yandex LLC\,(Russia), 
CSCS\,(Switzerland), 
IFIN\nobreakdash-HH\,(Romania), 
CBPF\,(Brazil), 
PL\nobreakdash-GRID\,(Poland) and 
OSC\,(USA). 
We~are indebted to the~communities behind the~multiple open 
source software packages on which we depend.
Individual groups or members have received support from 
AvH Foundation\,(Germany),
EPLANET, Marie Sk\l{}odowska\nobreakdash-Curie Actions and ERC\,(European Union), 
Conseil G\'{e}n\'{e}ral de Haute\nobreakdash-Savoie, Labex ENIGMASS and OCEVU, 
R\'{e}gion Auvergne\,(France), 
RFBR and Yandex LLC\,(Russia), 
GVA, XuntaGal and GENCAT\,(Spain), 
Herchel Smith Fund, 
The~Royal Society, Royal Commission for the~Exhibition of 1851 and the~Leverhulme Trust\,(United Kingdom).

\clearpage
{\noindent\normalfont\bfseries\Large Appendices}
\appendix


{\boldmath{\section{Polarization results for the \ones state}\label{sec:upsonespol}}}

Values of the polarization parameters $\pmb{\uplambda}$ and $\tilde\uplambda$
for the \ones~meson
are presented 
in Tables~\ref{tab:Y1S_Results_HX_Slices_7TeV} and~\ref{tab:Y1S_Results_HX_Slices_8TeV}
for the~HX~frame,
in Tables~\ref{tab:Y1S_Results_CS_Slices_7TeV} and~\ref{tab:Y1S_Results_CS_Slices_8TeV}
for the~CS~frame and
in Tables~\ref{tab:Y1S_Results_GJ_Slices_7TeV} and~\ref{tab:Y1S_Results_GJ_Slices_8TeV}
for the~GJ~frame for $\sqs=7$~and~8\tev, respectively.
The~polarization parameters $\pmb{\uplambda}$ 
measured in the~wide rapidity bin $2.2<\yy<4.5$ are presented 
in Tables~\ref{tab:Y1S_Results_HX_CS_GJ_2.2y4.5_7TeV} 
      and~\ref{tab:Y1S_Results_HX_CS_GJ_2.2y4.5_8TeV}, while 
the~parameters $\tilde\uplambda$ are listed 
in Table~\ref{tab:Y1S_lamT_Results_HX_CS_GJ_2.2y4.5}.

\begin{table}[htb] 
\begin{center} 
\caption{\small Values of
  $\uplambda_{\theta}$, $\uplambda_{\theta\phi}$, $\uplambda_{\phi}$
  and $\tilde \uplambda$
  measured in the HX frame for the \ones produced
  at $\sqrt{s}=7\,\,{\mathrm{TeV}}$.
  The first uncertainty is statistical and the second systematic.
}\label{tab:Y1S_Results_HX_Slices_7TeV}
\vspace*{2mm}
\begin{footnotesize}
\begin{tabular*}{0.99\textwidth}{@{\hspace{1mm}}c@{\extracolsep{\fill}}cccc@{\hspace{1mm}}}
 $\pty~\left[\!\gevc\right]$ 
 & $\uplambda$ 
 & $2.2<y<3.0$
 & $3.0<y<3.5$ 
 & $3.5<y<4.5$ 
\\[ 1.5mm]
\hline
\\[-2.5mm]
 \multirow{4}{*}{$0-2$}
 &  $\uplambda_{\theta}$
 &  $\pz0.220\pm0.063\pm0.042$
 &  $\pz0.104\pm0.034\pm0.027$
 &  $-0.098\pm0.043\pm0.035$
\\
 &  $\uplambda_{\theta\phi}$
 &  $-0.039\pm0.016\pm0.007$
 &  $-0.041\pm0.012\pm0.006$
 &  $\pz0.041\pm0.016\pm0.010$
\\
 &  $\uplambda_{\phi}$
 &  $\pz0.009\pm0.008\pm0.004$
 &  $-0.004\pm0.009\pm0.005$
 &  $-0.024\pm0.009\pm0.004$
\\
 &  $\tilde \uplambda$
 &  $\pz0.249\pm0.069\pm0.045$
 &  $\pz0.092\pm0.045\pm0.032$
 &  $-0.167\pm0.051\pm0.036$
\\[ 1.5mm]
\hline
\\[-2.5mm]
 \multirow{4}{*}{$2-4$}
 &  $\uplambda_{\theta}$
 &  $\pz0.175\pm0.045\pm0.025$
 &  $\pz0.053\pm0.025\pm0.020$
 &  $-0.057\pm0.028\pm0.027$
\\
 &  $\uplambda_{\theta\phi}$
 &  $-0.075\pm0.014\pm0.007$
 &  $-0.009\pm0.011\pm0.006$
 &  $\pz0.034\pm0.016\pm0.015$
\\
 &  $\uplambda_{\phi}$
 &  $\pz0.000\pm0.007\pm0.002$
 &  $-0.002\pm0.007\pm0.003$
 &  $-0.033\pm0.009\pm0.005$
\\
 &  $\tilde \uplambda$
 &  $\pz0.176\pm0.051\pm0.029$
 &  $\pz0.047\pm0.036\pm0.026$
 &  $-0.151\pm0.039\pm0.036$
\\[ 1.5mm]
\hline
\\[-2.5mm]
 \multirow{4}{*}{$4-6$}
 &  $\uplambda_{\theta}$
 &  $\pz0.069\pm0.045\pm0.037$
 &  $\pz0.055\pm0.026\pm0.017$
 &  $-0.077\pm0.025\pm0.021$
\\
 &  $\uplambda_{\theta\phi}$
 &  $-0.069\pm0.019\pm0.015$
 &  $\pz0.009\pm0.014\pm0.009$
 &  $\pz0.078\pm0.020\pm0.016$
\\
 &  $\uplambda_{\phi}$
 &  $-0.006\pm0.009\pm0.005$
 &  $-0.041\pm0.011\pm0.006$
 &  $-0.053\pm0.013\pm0.008$
\\
 &  $\tilde \uplambda$
 &  $\pz0.050\pm0.057\pm0.045$
 &  $-0.066\pm0.040\pm0.025$
 &  $-0.223\pm0.044\pm0.035$
\\[ 1.5mm]
\hline
\\[-2.5mm]
 \multirow{4}{*}{$6-8$}
 &  $\uplambda_{\theta}$
 &  $\pz0.057\pm0.049\pm0.034$
 &  $\pz0.036\pm0.031\pm0.024$
 &  $\pz0.062\pm0.032\pm0.026$
\\
 &  $\uplambda_{\theta\phi}$
 &  $-0.029\pm0.024\pm0.023$
 &  $\pz0.021\pm0.016\pm0.009$
 &  $\pz0.060\pm0.023\pm0.017$
\\
 &  $\uplambda_{\phi}$
 &  $-0.030\pm0.015\pm0.013$
 &  $-0.055\pm0.016\pm0.010$
 &  $-0.048\pm0.021\pm0.019$
\\
 &  $\tilde \uplambda$
 &  $-0.031\pm0.064\pm0.054$
 &  $-0.121\pm0.047\pm0.025$
 &  $-0.079\pm0.056\pm0.041$
\\[ 1.5mm]
\hline
\\[-2.5mm]
 \multirow{4}{*}{$8-10$}
 &  $\uplambda_{\theta}$
 &  $\pz0.076\pm0.059\pm0.044$
 &  $\pz0.117\pm0.044\pm0.034$
 &  $\pz0.076\pm0.047\pm0.039$
\\
 &  $\uplambda_{\theta\phi}$
 &  $-0.068\pm0.027\pm0.016$
 &  $-0.022\pm0.018\pm0.006$
 &  $\pz0.035\pm0.023\pm0.012$
\\
 &  $\uplambda_{\phi}$
 &  $-0.004\pm0.021\pm0.011$
 &  $-0.048\pm0.023\pm0.014$
 &  $-0.062\pm0.031\pm0.024$
\\
 &  $\tilde \uplambda$
 &  $\pz0.065\pm0.077\pm0.045$
 &  $-0.024\pm0.059\pm0.024$
 &  $-0.103\pm0.070\pm0.038$
\\[ 1.5mm]
\hline
\\[-2.5mm]
 \multirow{4}{*}{$10-15$}
 &  $\uplambda_{\theta}$
 &  $-0.021\pm0.051\pm0.029$
 &  $\pz0.123\pm0.042\pm0.030$
 &  $\pz0.135\pm0.051\pm0.048$
\\
 &  $\uplambda_{\theta\phi}$
 &  $-0.009\pm0.023\pm0.010$
 &  $-0.003\pm0.019\pm0.009$
 &  $\pz0.070\pm0.024\pm0.007$
\\
 &  $\uplambda_{\phi}$
 &  $\pz0.009\pm0.019\pm0.010$
 &  $-0.059\pm0.022\pm0.013$
 &  $-0.047\pm0.031\pm0.024$
\\
 &  $\tilde \uplambda$
 &  $\pz0.006\pm0.064\pm0.026$
 &  $-0.052\pm0.060\pm0.026$
 &  $-0.005\pm0.076\pm0.033$
\\[ 1.5mm]
\hline
\\[-2.5mm]
 \multirow{4}{*}{$15-20$}
 &  $\uplambda_{\theta}$
 &  $\pz0.032\pm0.091\pm0.045$
 &  $\pz0.046\pm0.077\pm0.049$
 &  $\pz0.082\pm0.111\pm0.058$
\\
 &  $\uplambda_{\theta\phi}$
 &  $\pz0.096\pm0.048\pm0.015$
 &  $\pz0.007\pm0.050\pm0.022$
 &  $\pz0.104\pm0.072\pm0.027$
\\
 &  $\uplambda_{\phi}$
 &  $\pz0.008\pm0.033\pm0.013$
 &  $\pz0.099\pm0.040\pm0.018$
 &  $\pz0.119\pm0.058\pm0.022$
\\
 &  $\tilde \uplambda$
 &  $\pz0.055\pm0.130\pm0.044$
 &  $\pz0.382\pm0.175\pm0.083$
 &  $\pz0.500\pm0.252\pm0.094$
\end{tabular*}
\end{footnotesize} 
\end{center} 
\end{table}  
%
%
%
%
%
\begin{table}[t] 
\begin{center} 
\caption{\small Values of
 $\uplambda_{\theta}$, $\uplambda_{\theta\phi}$, $\uplambda_{\phi}$
 and $\tilde \uplambda$
 measured in the HX frame for the \ones produced
 at $\sqrt{s}=8\,\,{\mathrm{TeV}}$.
 The first uncertainty is statistical and the second systematic.
}\label{tab:Y1S_Results_HX_Slices_8TeV}
\vspace*{2mm}
\begin{footnotesize}
\begin{tabular*}{0.99\textwidth}{@{\hspace{1mm}}c@{\extracolsep{\fill}}cccc@{\hspace{1mm}}}
 $\pty~\left[\!\gevc\right]$ 
 & $\uplambda$ 
 & $2.2<y<3.0$ 
 & $3.0<y<3.5$ 
 & $3.5<y<4.5$ 
\\[ 1.5mm]
\hline
\\[-2.5mm]
 \multirow{4}{*}{$0-2$}
 &  $\uplambda_{\theta}$
 &  $\pz0.190\pm0.042\pm0.030$
 &  $\pz0.092\pm0.023\pm0.020$
 &  $\pz0.012\pm0.030\pm0.031$
\\
 &  $\uplambda_{\theta\phi}$
 &  $-0.009\pm0.011\pm0.005$
 &  $\pz0.019\pm0.008\pm0.004$
 &  $-0.013\pm0.011\pm0.007$
\\
 &  $\uplambda_{\phi}$
 &  $-0.007\pm0.006\pm0.003$
 &  $\pz0.002\pm0.006\pm0.003$
 &  $\pz0.001\pm0.006\pm0.003$
\\
 &  $\tilde \uplambda$
 &  $\pz0.168\pm0.046\pm0.032$
 &  $\pz0.098\pm0.031\pm0.023$
 &  $\pz0.014\pm0.037\pm0.033$
\\[ 1.5mm]
\hline
\\[-2.5mm]
 \multirow{4}{*}{$2-4$}
 &  $\uplambda_{\theta}$
 &  $\pz0.208\pm0.030\pm0.028$
 &  $\pz0.104\pm0.017\pm0.014$
 &  $-0.017\pm0.019\pm0.021$
\\
 &  $\uplambda_{\theta\phi}$
 &  $-0.077\pm0.009\pm0.007$
 &  $-0.023\pm0.008\pm0.005$
 &  $\pz0.020\pm0.011\pm0.010$
\\
 &  $\uplambda_{\phi}$
 &  $-0.006\pm0.004\pm0.002$
 &  $-0.026\pm0.005\pm0.003$
 &  $-0.020\pm0.006\pm0.004$
\\
 &  $\tilde \uplambda$
 &  $\pz0.189\pm0.035\pm0.030$
 &  $\pz0.025\pm0.024\pm0.017$
 &  $-0.076\pm0.027\pm0.028$
\\[ 1.5mm]
\hline
\\[-2.5mm]
 \multirow{4}{*}{$4-6$}
 &  $\uplambda_{\theta}$
 &  $\pz0.179\pm0.031\pm0.025$
 &  $\pz0.102\pm0.018\pm0.014$
 &  $\pz0.035\pm0.018\pm0.017$
\\
 &  $\uplambda_{\theta\phi}$
 &  $-0.057\pm0.013\pm0.010$
 &  $-0.012\pm0.010\pm0.007$
 &  $\pz0.040\pm0.014\pm0.014$
\\
 &  $\uplambda_{\phi}$
 &  $-0.026\pm0.006\pm0.004$
 &  $-0.030\pm0.007\pm0.004$
 &  $-0.054\pm0.009\pm0.008$
\\
 &  $\tilde \uplambda$
 &  $\pz0.098\pm0.038\pm0.031$
 &  $\pz0.012\pm0.028\pm0.020$
 &  $-0.119\pm0.031\pm0.030$
\\[ 1.5mm]
\hline
\\[-2.5mm]
 \multirow{4}{*}{$6-8$}
 &  $\uplambda_{\theta}$
 &  $\pz0.129\pm0.033\pm0.025$
 &  $\pz0.084\pm0.021\pm0.017$
 &  $\pz0.038\pm0.021\pm0.017$
\\
 &  $\uplambda_{\theta\phi}$
 &  $-0.060\pm0.016\pm0.012$
 &  $-0.020\pm0.011\pm0.008$
 &  $\pz0.040\pm0.015\pm0.014$
\\
 &  $\uplambda_{\phi}$
 &  $-0.040\pm0.010\pm0.007$
 &  $-0.031\pm0.011\pm0.007$
 &  $-0.039\pm0.013\pm0.012$
\\
 &  $\tilde \uplambda$
 &  $\pz0.009\pm0.044\pm0.030$
 &  $-0.007\pm0.033\pm0.020$
 &  $-0.077\pm0.037\pm0.029$
\\[ 1.5mm]
\hline
\\[-2.5mm]
 \multirow{4}{*}{$8-10$}
 &  $\uplambda_{\theta}$
 &  $\pz0.097\pm0.040\pm0.030$
 &  $\pz0.129\pm0.030\pm0.022$
 &  $\pz0.180\pm0.033\pm0.037$
\\
 &  $\uplambda_{\theta\phi}$
 &  $-0.048\pm0.018\pm0.011$
 &  $-0.027\pm0.012\pm0.006$
 &  $\pz0.059\pm0.016\pm0.010$
\\
 &  $\uplambda_{\phi}$
 &  $-0.030\pm0.015\pm0.011$
 &  $-0.062\pm0.015\pm0.009$
 &  $-0.099\pm0.021\pm0.022$
\\
 &  $\tilde \uplambda$
 &  $\pz0.007\pm0.050\pm0.030$
 &  $-0.052\pm0.039\pm0.015$
 &  $-0.107\pm0.045\pm0.033$
\\[ 1.5mm]
\hline
\\[-2.5mm]
 \multirow{4}{*}{$10-15$}
 &  $\uplambda_{\theta}$
 &  $\pz0.124\pm0.036\pm0.030$
 &  $\pz0.086\pm0.027\pm0.020$
 &  $\pz0.146\pm0.033\pm0.037$
\\
 &  $\uplambda_{\theta\phi}$
 &  $-0.069\pm0.016\pm0.008$
 &  $-0.009\pm0.013\pm0.007$
 &  $-0.012\pm0.015\pm0.007$
\\
 &  $\uplambda_{\phi}$
 &  $-0.042\pm0.013\pm0.009$
 &  $-0.034\pm0.014\pm0.008$
 &  $-0.048\pm0.020\pm0.019$
\\
 &  $\tilde \uplambda$
 &  $-0.002\pm0.041\pm0.022$
 &  $-0.017\pm0.040\pm0.018$
 &  $\pz0.001\pm0.049\pm0.031$
\\[ 1.5mm]
\hline
\\[-2.5mm]
 \multirow{4}{*}{$15-20$}
 &  $\uplambda_{\theta}$
 &  $\pz0.041\pm0.058\pm0.036$
 &  $\pz0.239\pm0.057\pm0.032$
 &  $\pz0.245\pm0.077\pm0.049$
\\
 &  $\uplambda_{\theta\phi}$
 &  $-0.040\pm0.031\pm0.015$
 &  $-0.006\pm0.034\pm0.019$
 &  $\pz0.071\pm0.048\pm0.027$
\\
 &  $\uplambda_{\phi}$
 &  $-0.062\pm0.022\pm0.012$
 &  $-0.044\pm0.028\pm0.014$
 &  $-0.063\pm0.041\pm0.025$
\\
 &  $\tilde \uplambda$
 &  $-0.137\pm0.075\pm0.036$
 &  $\pz0.102\pm0.098\pm0.045$
 &  $\pz0.052\pm0.126\pm0.067$
\end{tabular*}
\end{footnotesize}
\end{center} 
\end{table}  
%
%
%
%
%
\begin{table}[t] 
\begin{center} 
\caption{\small Values of
 $\uplambda_{\theta}$, $\uplambda_{\theta\phi}$, $\uplambda_{\phi}$
 and $\tilde \uplambda$
 measured in the CS frame for the \ones produced
 at $\sqrt{s}=7\,\,{\mathrm{TeV}}$.
 The first uncertainty is statistical and the second systematic.
}\label{tab:Y1S_Results_CS_Slices_7TeV}
\vspace*{2mm}
\begin{footnotesize}
\begin{tabular*}{0.99\textwidth}{@{\hspace{1mm}}c@{\extracolsep{\fill}}cccc@{\hspace{1mm}}}
 $\pty~\left[\!\gevc\right]$ 
 & $\uplambda$ 
 & 2.2~$<y<$~3.0 
 & 3.0~$<y<$~3.5 
 & 3.5~$<y<$~4.5 
\\[ 1.5mm]
\hline
\\[-2.5mm]
 \multirow{4}{*}{$0-2$}
 &  $\uplambda_{\theta}$
 &  $\pz0.243\pm0.063\pm0.036$
 &  $\pz0.114\pm0.035\pm0.026$
 &  $-0.109\pm0.046\pm0.034$
\\
 &  $\uplambda_{\theta\phi}$
 &  $\pz0.014\pm0.016\pm0.008$
 &  $-0.017\pm0.012\pm0.004$
 &  $\pz0.047\pm0.014\pm0.007$
\\
 &  $\uplambda_{\phi}$
 &  $\pz0.007\pm0.008\pm0.003$
 &  $-0.008\pm0.009\pm0.003$
 &  $-0.021\pm0.009\pm0.004$
\\
 &  $\tilde \uplambda$
 &  $\pz0.265\pm0.069\pm0.038$
 &  $\pz0.089\pm0.046\pm0.028$
 &  $-0.168\pm0.052\pm0.036$
\\[ 1.5mm]
\hline
\\[-2.5mm]
 \multirow{4}{*}{$2-4$}
 &  $\uplambda_{\theta}$
 &  $\pz0.238\pm0.047\pm0.031$
 &  $\pz0.055\pm0.028\pm0.022$
 &  $-0.084\pm0.034\pm0.030$
\\
 &  $\uplambda_{\theta\phi}$
 &  $\pz0.022\pm0.013\pm0.006$
 &  $\pz0.020\pm0.010\pm0.003$
 &  $\pz0.041\pm0.012\pm0.006$
\\
 &  $\uplambda_{\phi}$
 &  $-0.011\pm0.006\pm0.003$
 &  $-0.002\pm0.007\pm0.003$
 &  $-0.027\pm0.007\pm0.003$
\\
 &  $\tilde \uplambda$
 &  $\pz0.204\pm0.052\pm0.033$
 &  $\pz0.048\pm0.037\pm0.024$
 &  $-0.160\pm0.040\pm0.033$
\\[ 1.5mm]
\hline
\\[-2.5mm]
 \multirow{4}{*}{$4-6$}
 &  $\uplambda_{\theta}$
 &  $\pz0.139\pm0.052\pm0.030$
 &  $\pz0.008\pm0.030\pm0.018$
 &  $-0.168\pm0.036\pm0.021$
\\
 &  $\uplambda_{\theta\phi}$
 &  $\pz0.002\pm0.017\pm0.008$
 &  $\pz0.049\pm0.013\pm0.005$
 &  $\pz0.048\pm0.015\pm0.008$
\\
 &  $\uplambda_{\phi}$
 &  $-0.026\pm0.007\pm0.003$
 &  $-0.027\pm0.008\pm0.003$
 &  $-0.025\pm0.008\pm0.003$
\\
 &  $\tilde \uplambda$
 &  $\pz0.059\pm0.058\pm0.033$
 &  $-0.072\pm0.041\pm0.021$
 &  $-0.238\pm0.045\pm0.024$
\\[ 1.5mm]
\hline
\\[-2.5mm]
 \multirow{4}{*}{$6-8$}
 &  $\uplambda_{\theta}$
 &  $\pz0.064\pm0.055\pm0.025$
 &  $-0.045\pm0.030\pm0.018$
 &  $-0.083\pm0.039\pm0.024$
\\
 &  $\uplambda_{\theta\phi}$
 &  $\pz0.020\pm0.023\pm0.010$
 &  $\pz0.047\pm0.019\pm0.007$
 &  $\pz0.070\pm0.022\pm0.009$
\\
 &  $\uplambda_{\phi}$
 &  $-0.029\pm0.010\pm0.004$
 &  $-0.028\pm0.011\pm0.004$
 &  $-0.001\pm0.012\pm0.005$
\\
 &  $\tilde \uplambda$
 &  $-0.023\pm0.065\pm0.031$
 &  $-0.125\pm0.047\pm0.022$
 &  $-0.086\pm0.057\pm0.032$
\\[ 1.5mm]
\hline
\\[-2.5mm]
 \multirow{4}{*}{$8-10$}
 &  $\uplambda_{\theta}$
 &  $\pz0.126\pm0.059\pm0.027$
 &  $\pz0.030\pm0.032\pm0.018$
 &  $-0.071\pm0.041\pm0.020$
\\
 &  $\uplambda_{\theta\phi}$
 &  $\pz0.020\pm0.034\pm0.012$
 &  $\pz0.079\pm0.027\pm0.009$
 &  $\pz0.064\pm0.032\pm0.014$
\\
 &  $\uplambda_{\phi}$
 &  $-0.020\pm0.014\pm0.005$
 &  $-0.019\pm0.016\pm0.007$
 &  $-0.011\pm0.018\pm0.007$
\\
 &  $\tilde \uplambda$
 &  $\pz0.063\pm0.078\pm0.035$
 &  $-0.027\pm0.059\pm0.026$
 &  $-0.102\pm0.070\pm0.029$
\\[ 1.5mm]
\hline
\\[-2.5mm]
 \multirow{4}{*}{$10-15$}
 &  $\uplambda_{\theta}$
 &  $\pz0.012\pm0.038\pm0.018$
 &  $-0.039\pm0.026\pm0.014$
 &  $-0.124\pm0.035\pm0.017$
\\
 &  $\uplambda_{\theta\phi}$
 &  $-0.037\pm0.033\pm0.012$
 &  $\pz0.072\pm0.026\pm0.010$
 &  $\pz0.069\pm0.033\pm0.015$
\\
 &  $\uplambda_{\phi}$
 &  $-0.005\pm0.017\pm0.008$
 &  $-0.003\pm0.019\pm0.009$
 &  $\pz0.039\pm0.022\pm0.010$
\\
 &  $\tilde \uplambda$
 &  $-0.004\pm0.063\pm0.022$
 &  $-0.049\pm0.060\pm0.024$
 &  $-0.009\pm0.077\pm0.035$
\\[ 1.5mm]
\hline
\\[-2.5mm]
 \multirow{4}{*}{$15-20$}
 &  $\uplambda_{\theta}$
 &  $-0.112\pm0.054\pm0.031$
 &  $\pz0.088\pm0.064\pm0.044$
 &  $-0.002\pm0.087\pm0.060$
\\
 &  $\uplambda_{\theta\phi}$
 &  $-0.048\pm0.053\pm0.021$
 &  $-0.030\pm0.048\pm0.016$
 &  $-0.087\pm0.069\pm0.021$
\\
 &  $\uplambda_{\phi}$
 &  $\pz0.053\pm0.041\pm0.025$
 &  $\pz0.092\pm0.050\pm0.030$
 &  $\pz0.147\pm0.065\pm0.032$
\\
 &  $\tilde \uplambda$
 &  $\pz0.049\pm0.129\pm0.061$
 &  $\pz0.401\pm0.176\pm0.079$
 &  $\pz0.514\pm0.253\pm0.090$
\end{tabular*}
\end{footnotesize}
\end{center} 
\end{table}  
%
%
%
%
%
\begin{table}[t] 
\begin{center} 
\caption{\small Values of
 $\uplambda_{\theta}$, $\uplambda_{\theta\phi}$, $\uplambda_{\phi}$
 and $\tilde \uplambda$
 measured in the CS frame for the \ones produced
 at $\sqrt{s}=8\,\,{\mathrm{TeV}}$.
 The first uncertainty is statistical and the second systematic.
}\label{tab:Y1S_Results_CS_Slices_8TeV}
\vspace*{2mm}
\begin{footnotesize}
\begin{tabular*}{0.99\textwidth}{@{\hspace{1mm}}c@{\extracolsep{\fill}}cccc@{\hspace{1mm}}}
 $\pty~\left[\!\gevc\right]$ 
 & $\uplambda$ 
 & $2.2<y<3.0$ 
 & $3.0<y<3.5$
 & $3.5<y<4.5$ 
\\[ 1.5mm]
\hline
\\[-2.5mm]
 \multirow{4}{*}{$0-2$}
 &  $\uplambda_{\theta}$
 &  $\pz0.190\pm0.042\pm0.030$
 &  $\pz0.078\pm0.024\pm0.021$
 &  $\pz0.002\pm0.032\pm0.031$
\\
 &  $\uplambda_{\theta\phi}$
 &  $\pz0.038\pm0.011\pm0.006$
 &  $\pz0.039\pm0.008\pm0.004$
 &  $\pz0.007\pm0.010\pm0.005$
\\
 &  $\uplambda_{\phi}$
 &  $-0.005\pm0.006\pm0.002$
 &  $\pz0.006\pm0.006\pm0.003$
 &  $-0.001\pm0.006\pm0.003$
\\
 &  $\tilde \uplambda$
 &  $\pz0.173\pm0.046\pm0.031$
 &  $\pz0.096\pm0.031\pm0.024$
 &  $-0.003\pm0.037\pm0.033$
\\[ 1.5mm]
\hline
\\[-2.5mm]
 \multirow{4}{*}{$2-4$}
 &  $\uplambda_{\theta}$
 &  $\pz0.251\pm0.032\pm0.022$
 &  $\pz0.109\pm0.019\pm0.014$
 &  $-0.045\pm0.024\pm0.024$
\\
 &  $\uplambda_{\theta\phi}$
 &  $\pz0.029\pm0.009\pm0.006$
 &  $\pz0.031\pm0.007\pm0.003$
 &  $\pz0.043\pm0.008\pm0.005$
\\
 &  $\uplambda_{\phi}$
 &  $-0.015\pm0.004\pm0.002$
 &  $-0.027\pm0.005\pm0.002$
 &  $-0.017\pm0.005\pm0.002$
\\
 &  $\tilde \uplambda$
 &  $\pz0.202\pm0.035\pm0.023$
 &  $\pz0.027\pm0.024\pm0.017$
 &  $-0.095\pm0.028\pm0.025$
\\[ 1.5mm]
\hline
\\[-2.5mm]
 \multirow{4}{*}{$4-6$}
 &  $\uplambda_{\theta}$
 &  $\pz0.197\pm0.034\pm0.021$
 &  $\pz0.071\pm0.021\pm0.015$
 &  $-0.049\pm0.025\pm0.025$
\\
 &  $\uplambda_{\theta\phi}$
 &  $\pz0.066\pm0.011\pm0.006$
 &  $\pz0.052\pm0.009\pm0.004$
 &  $\pz0.067\pm0.010\pm0.006$
\\
 &  $\uplambda_{\phi}$
 &  $-0.026\pm0.005\pm0.003$
 &  $-0.021\pm0.006\pm0.003$
 &  $-0.028\pm0.006\pm0.003$
\\
 &  $\tilde \uplambda$
 &  $\pz0.116\pm0.039\pm0.026$
 &  $\pz0.008\pm0.028\pm0.018$
 &  $-0.130\pm0.031\pm0.027$
\\[ 1.5mm]
\hline
\\[-2.5mm]
 \multirow{4}{*}{$6-8$}
 &  $\uplambda_{\theta}$
 &  $\pz0.125\pm0.038\pm0.021$
 &  $\pz0.051\pm0.021\pm0.017$
 &  $-0.062\pm0.026\pm0.021$
\\
 &  $\uplambda_{\theta\phi}$
 &  $\pz0.048\pm0.016\pm0.008$
 &  $\pz0.049\pm0.013\pm0.004$
 &  $\pz0.050\pm0.014\pm0.008$
\\
 &  $\uplambda_{\phi}$
 &  $-0.040\pm0.007\pm0.003$
 &  $-0.020\pm0.008\pm0.003$
 &  $-0.007\pm0.008\pm0.004$
\\
 &  $\tilde \uplambda$
 &  $\pz0.006\pm0.044\pm0.026$
 &  $-0.010\pm0.033\pm0.021$
 &  $-0.082\pm0.038\pm0.027$
\\[ 1.5mm]
\hline
\\[-2.5mm]
 \multirow{4}{*}{$8-10$}
 &  $\uplambda_{\theta}$
 &  $\pz0.069\pm0.038\pm0.020$
 &  $\pz0.030\pm0.021\pm0.012$
 &  $-0.100\pm0.027\pm0.015$
\\
 &  $\uplambda_{\theta\phi}$
 &  $\pz0.035\pm0.022\pm0.010$
 &  $\pz0.086\pm0.018\pm0.007$
 &  $\pz0.123\pm0.021\pm0.011$
\\
 &  $\uplambda_{\phi}$
 &  $-0.025\pm0.009\pm0.004$
 &  $-0.028\pm0.011\pm0.005$
 &  $-0.002\pm0.011\pm0.005$
\\
 &  $\tilde \uplambda$
 &  $-0.006\pm0.050\pm0.027$
 &  $-0.053\pm0.039\pm0.018$
 &  $-0.106\pm0.045\pm0.024$
\\[ 1.5mm]
\hline
\\[-2.5mm]
 \multirow{4}{*}{$10-15$}
 &  $\uplambda_{\theta}$
 &  $\pz0.058\pm0.025\pm0.014$
 &  $-0.012\pm0.018\pm0.011$
 &  $-0.017\pm0.024\pm0.014$
\\
 &  $\uplambda_{\theta\phi}$
 &  $\pz0.069\pm0.022\pm0.009$
 &  $\pz0.048\pm0.017\pm0.007$
 &  $\pz0.090\pm0.022\pm0.009$
\\
 &  $\uplambda_{\phi}$
 &  $-0.026\pm0.011\pm0.006$
 &  $-0.001\pm0.013\pm0.007$
 &  $\pz0.005\pm0.014\pm0.008$
\\
 &  $\tilde \uplambda$
 &  $-0.019\pm0.040\pm0.022$
 &  $-0.014\pm0.040\pm0.019$
 &  $-0.001\pm0.049\pm0.028$
\\[ 1.5mm]
\hline
\\[-2.5mm]
 \multirow{4}{*}{$15-20$}
 &  $\uplambda_{\theta}$
 &  $-0.032\pm0.037\pm0.023$
 &  $-0.068\pm0.036\pm0.027$
 &  $-0.176\pm0.046\pm0.035$
\\
 &  $\uplambda_{\theta\phi}$
 &  $\pz0.049\pm0.035\pm0.013$
 &  $\pz0.105\pm0.030\pm0.013$
 &  $\pz0.072\pm0.043\pm0.020$
\\
 &  $\uplambda_{\phi}$
 &  $-0.038\pm0.028\pm0.020$
 &  $\pz0.055\pm0.031\pm0.019$
 &  $\pz0.080\pm0.040\pm0.030$
\\
 &  $\tilde \uplambda$
 &  $-0.140\pm0.075\pm0.043$
 &  $\pz0.102\pm0.098\pm0.046$
 &  $\pz0.068\pm0.127\pm0.071$
\end{tabular*}
\end{footnotesize}
\end{center} 
\end{table}  
%
%
%
%
%
\begin{table}[t] 
\begin{center} 
\caption{\small Values of
 $\uplambda_{\theta}$, $\uplambda_{\theta\phi}$, $\uplambda_{\phi}$
 and $\tilde \uplambda$
 measured in the GJ frame for the \ones produced
 at $\sqrt{s}=7\,\,{\mathrm{TeV}}$.
 The first uncertainty is statistical and the second systematic.
}\label{tab:Y1S_Results_GJ_Slices_7TeV}
\vspace*{2mm}
\begin{footnotesize}
\begin{tabular*}{0.99\textwidth}{@{\hspace{1mm}}c@{\extracolsep{\fill}}cccc@{\hspace{1mm}}}
 $\pty~\left[\!\gevc\right]$ 
 & $\uplambda$ 
 & $2.2<y<3.0$ 
 & $3.0<y<3.5$ 
 & $3.5<y<4.5$ 
\\[ 1.5mm]
\hline
\\[-2.5mm]
 \multirow{4}{*}{$0-2$}
 &  $\uplambda_{\theta}$
 &  $\pz0.218\pm0.057\pm0.042$
 &  $\pz0.106\pm0.034\pm0.022$
 &  $-0.120\pm0.046\pm0.037$
\\
 &  $\uplambda_{\theta\phi}$
 &  $\pz0.064\pm0.019\pm0.014$
 &  $\pz0.007\pm0.012\pm0.005$
 &  $\pz0.050\pm0.014\pm0.007$
\\
 &  $\uplambda_{\phi}$
 &  $\pz0.014\pm0.009\pm0.005$
 &  $-0.008\pm0.009\pm0.004$
 &  $-0.014\pm0.009\pm0.004$
\\
 &  $\tilde \uplambda$
 &  $\pz0.264\pm0.067\pm0.048$
 &  $\pz0.082\pm0.045\pm0.025$
 &  $-0.161\pm0.052\pm0.040$
\\[ 1.5mm]
\hline
\\[-2.5mm]
 \multirow{4}{*}{$2-4$}
 &  $\uplambda_{\theta}$
 &  $\pz0.153\pm0.036\pm0.032$
 &  $\pz0.018\pm0.026\pm0.017$
 &  $-0.125\pm0.034\pm0.027$
\\
 &  $\uplambda_{\theta\phi}$
 &  $\pz0.108\pm0.019\pm0.019$
 &  $\pz0.041\pm0.011\pm0.005$
 &  $\pz0.034\pm0.011\pm0.006$
\\
 &  $\uplambda_{\phi}$
 &  $\pz0.017\pm0.008\pm0.006$
 &  $\pz0.009\pm0.007\pm0.003$
 &  $-0.014\pm0.007\pm0.003$
\\
 &  $\tilde \uplambda$
 &  $\pz0.208\pm0.052\pm0.049$
 &  $\pz0.046\pm0.037\pm0.022$
 &  $-0.166\pm0.040\pm0.030$
\\[ 1.5mm]
\hline
\\[-2.5mm]
 \multirow{4}{*}{$4-6$}
 &  $\uplambda_{\theta}$
 &  $\pz0.065\pm0.033\pm0.025$
 &  $-0.072\pm0.026\pm0.017$
 &  $-0.201\pm0.036\pm0.023$
\\
 &  $\uplambda_{\theta\phi}$
 &  $\pz0.066\pm0.024\pm0.029$
 &  $\pz0.041\pm0.013\pm0.005$
 &  $-0.025\pm0.014\pm0.008$
\\
 &  $\uplambda_{\phi}$
 &  $-0.005\pm0.013\pm0.015$
 &  $-0.003\pm0.009\pm0.004$
 &  $-0.018\pm0.009\pm0.004$
\\
 &  $\tilde \uplambda$
 &  $\pz0.048\pm0.058\pm0.064$
 &  $-0.079\pm0.041\pm0.021$
 &  $-0.250\pm0.045\pm0.027$
\\[ 1.5mm]
\hline
\\[-2.5mm]
 \multirow{4}{*}{$6-8$}
 &  $\uplambda_{\theta}$
 &  $\pz0.008\pm0.036\pm0.022$
 &  $-0.101\pm0.031\pm0.017$
 &  $-0.144\pm0.042\pm0.026$
\\
 &  $\uplambda_{\theta\phi}$
 &  $\pz0.045\pm0.027\pm0.030$
 &  $\pz0.002\pm0.015\pm0.008$
 &  $-0.020\pm0.018\pm0.009$
\\
 &  $\uplambda_{\phi}$
 &  $-0.012\pm0.020\pm0.022$
 &  $-0.010\pm0.013\pm0.007$
 &  $\pz0.017\pm0.013\pm0.006$
\\
 &  $\tilde \uplambda$
 &  $-0.027\pm0.066\pm0.063$
 &  $-0.129\pm0.047\pm0.022$
 &  $-0.093\pm0.057\pm0.031$
\\[ 1.5mm]
\hline
\\[-2.5mm]
 \multirow{4}{*}{$8-10$}
 &  $\uplambda_{\theta}$
 &  $\pz0.019\pm0.048\pm0.040$
 &  $-0.119\pm0.042\pm0.030$
 &  $-0.120\pm0.056\pm0.035$
\\
 &  $\uplambda_{\theta\phi}$
 &  $\pz0.068\pm0.027\pm0.020$
 &  $\pz0.026\pm0.016\pm0.006$
 &  $-0.028\pm0.021\pm0.008$
\\
 &  $\uplambda_{\phi}$
 &  $\pz0.012\pm0.027\pm0.028$
 &  $\pz0.030\pm0.018\pm0.008$
 &  $\pz0.006\pm0.018\pm0.008$
\\
 &  $\tilde \uplambda$
 &  $\pz0.054\pm0.077\pm0.059$
 &  $-0.029\pm0.060\pm0.021$
 &  $-0.103\pm0.070\pm0.029$
\\[ 1.5mm]
\hline
\\[-2.5mm]
 \multirow{4}{*}{$10-15$}
 &  $\uplambda_{\theta}$
 &  $\pz0.088\pm0.052\pm0.051$
 &  $-0.086\pm0.043\pm0.031$
 &  $-0.078\pm0.059\pm0.033$
\\
 &  $\uplambda_{\theta\phi}$
 &  $\pz0.018\pm0.020\pm0.007$
 &  $-0.032\pm0.017\pm0.008$
 &  $-0.097\pm0.023\pm0.008$
\\
 &  $\uplambda_{\phi}$
 &  $-0.034\pm0.026\pm0.024$
 &  $\pz0.014\pm0.017\pm0.009$
 &  $\pz0.022\pm0.019\pm0.007$
\\
 &  $\tilde \uplambda$
 &  $-0.012\pm0.062\pm0.029$
 &  $-0.044\pm0.060\pm0.028$
 &  $-0.011\pm0.077\pm0.031$
\\[ 1.5mm]
\hline
\\[-2.5mm]
 \multirow{4}{*}{$15-20$}
 &  $\uplambda_{\theta}$
 &  $\pz0.161\pm0.096\pm0.058$
 &  $\pz0.150\pm0.103\pm0.062$
 &  $\pz0.336\pm0.164\pm0.083$
\\
 &  $\uplambda_{\theta\phi}$
 &  $-0.043\pm0.045\pm0.017$
 &  $\pz0.007\pm0.044\pm0.015$
 &  $-0.008\pm0.069\pm0.028$
\\
 &  $\uplambda_{\phi}$
 &  $-0.039\pm0.045\pm0.024$
 &  $\pz0.079\pm0.035\pm0.012$
 &  $\pz0.054\pm0.049\pm0.017$
\\
 &  $\tilde \uplambda$
 &  $\pz0.043\pm0.129\pm0.051$
 &  $\pz0.419\pm0.177\pm0.078$
 &  $\pz0.527\pm0.254\pm0.101$
\end{tabular*}
\end{footnotesize}
\end{center} 
\end{table}  
%
%
%
%
%
\begin{table}[t] 
\begin{center} 
\caption{\small Values of
 $\uplambda_{\theta}$, $\uplambda_{\theta\phi}$, $\uplambda_{\phi}$
 and $\tilde \uplambda$
 measured in the GJ frame for the \ones produced
 at $\sqrt{s}=8\,\,{\mathrm{TeV}}$.
 The first uncertainty is statistical and the second systematic.
}\label{tab:Y1S_Results_GJ_Slices_8TeV}
\vspace*{2mm}
\begin{footnotesize}
\begin{tabular*}{0.99\textwidth}{@{\hspace{1mm}}c@{\extracolsep{\fill}}cccc@{\hspace{1mm}}}
 $\pty~\left[\!\gevc\right]$
 & $\uplambda$ 
 & $2.2<y<3.0$ 
 & $3.0<y<3.5$ 
 & $3.5<y<4.5$
\\[ 1.5mm]
\hline
\\[-2.5mm]
 \multirow{4}{*}{$0-2$}
 &  $\uplambda_{\theta}$
 &  $\pz0.154\pm0.038\pm0.029$
 &  $\pz0.052\pm0.023\pm0.023$
 &  $-0.022\pm0.032\pm0.033$
\\
 &  $\uplambda_{\theta\phi}$
 &  $\pz0.080\pm0.013\pm0.007$
 &  $\pz0.054\pm0.008\pm0.004$
 &  $\pz0.023\pm0.010\pm0.005$
\\
 &  $\uplambda_{\phi}$
 &  $\pz0.005\pm0.006\pm0.003$
 &  $\pz0.013\pm0.006\pm0.003$
 &  $\pz0.001\pm0.006\pm0.003$
\\
 &  $\tilde \uplambda$
 &  $\pz0.170\pm0.045\pm0.034$
 &  $\pz0.092\pm0.031\pm0.025$
 &  $-0.018\pm0.037\pm0.034$
\\[ 1.5mm]
\hline
\\[-2.5mm]
 \multirow{4}{*}{$2-4$}
 &  $\uplambda_{\theta}$
 &  $\pz0.144\pm0.024\pm0.024$
 &  $\pz0.049\pm0.017\pm0.016$
 &  $-0.106\pm0.023\pm0.023$
\\
 &  $\uplambda_{\theta\phi}$
 &  $\pz0.111\pm0.013\pm0.014$
 &  $\pz0.072\pm0.007\pm0.005$
 &  $\pz0.045\pm0.008\pm0.006$
\\
 &  $\uplambda_{\phi}$
 &  $\pz0.014\pm0.005\pm0.004$
 &  $-0.009\pm0.005\pm0.003$
 &  $-0.002\pm0.005\pm0.002$
\\
 &  $\tilde \uplambda$
 &  $\pz0.187\pm0.035\pm0.036$
 &  $\pz0.023\pm0.024\pm0.023$
 &  $-0.112\pm0.028\pm0.025$
\\[ 1.5mm]
\hline
\\[-2.5mm]
 \multirow{4}{*}{$4-6$}
 &  $\uplambda_{\theta}$
 &  $\pz0.027\pm0.022\pm0.022$
 &  $-0.031\pm0.018\pm0.012$
 &  $-0.138\pm0.024\pm0.021$
\\
 &  $\uplambda_{\theta\phi}$
 &  $\pz0.124\pm0.015\pm0.016$
 &  $\pz0.066\pm0.009\pm0.005$
 &  $\pz0.030\pm0.010\pm0.007$
\\
 &  $\uplambda_{\phi}$
 &  $\pz0.026\pm0.008\pm0.008$
 &  $\pz0.011\pm0.006\pm0.003$
 &  $-0.002\pm0.006\pm0.003$
\\
 &  $\tilde \uplambda$
 &  $\pz0.109\pm0.039\pm0.042$
 &  $\pz0.002\pm0.028\pm0.018$
 &  $-0.142\pm0.031\pm0.024$
\\[ 1.5mm]
\hline
\\[-2.5mm]
 \multirow{4}{*}{$6-8$}
 &  $\uplambda_{\theta}$
 &  $-0.011\pm0.024\pm0.019$
 &  $-0.056\pm0.021\pm0.017$
 &  $-0.111\pm0.028\pm0.020$
\\
 &  $\uplambda_{\theta\phi}$
 &  $\pz0.078\pm0.018\pm0.019$
 &  $\pz0.046\pm0.010\pm0.006$
 &  $-0.012\pm0.012\pm0.007$
\\
 &  $\uplambda_{\phi}$
 &  $-0.001\pm0.013\pm0.015$
 &  $\pz0.014\pm0.009\pm0.005$
 &  $\pz0.007\pm0.008\pm0.004$
\\
 &  $\tilde \uplambda$
 &  $-0.014\pm0.044\pm0.045$
 &  $-0.013\pm0.033\pm0.020$
 &  $-0.090\pm0.038\pm0.025$
\\[ 1.5mm]
\hline
\\[-2.5mm]
 \multirow{4}{*}{$8-10$}
 &  $\uplambda_{\theta}$
 &  $-0.014\pm0.031\pm0.026$
 &  $-0.125\pm0.028\pm0.019$
 &  $-0.200\pm0.035\pm0.024$
\\
 &  $\uplambda_{\theta\phi}$
 &  $\pz0.039\pm0.018\pm0.015$
 &  $\pz0.030\pm0.011\pm0.005$
 &  $-0.044\pm0.013\pm0.006$
\\
 &  $\uplambda_{\phi}$
 &  $-0.004\pm0.018\pm0.019$
 &  $\pz0.024\pm0.012\pm0.007$
 &  $\pz0.033\pm0.011\pm0.006$
\\
 &  $\tilde \uplambda$
 &  $-0.026\pm0.049\pm0.043$
 &  $-0.054\pm0.039\pm0.017$
 &  $-0.106\pm0.046\pm0.024$
\\[ 1.5mm]
\hline
\\[-2.5mm]
 \multirow{4}{*}{$10-15$}
 &  $\uplambda_{\theta}$
 &  $-0.073\pm0.030\pm0.032$
 &  $-0.053\pm0.029\pm0.020$
 &  $-0.118\pm0.037\pm0.028$
\\
 &  $\uplambda_{\theta\phi}$
 &  $\pz0.025\pm0.012\pm0.006$
 &  $-0.015\pm0.011\pm0.005$
 &  $-0.031\pm0.014\pm0.006$
\\
 &  $\uplambda_{\phi}$
 &  $\pz0.012\pm0.016\pm0.017$
 &  $\pz0.014\pm0.012\pm0.006$
 &  $\pz0.039\pm0.012\pm0.007$
\\
 &  $\tilde \uplambda$
 &  $-0.036\pm0.040\pm0.025$
 &  $-0.010\pm0.040\pm0.021$
 &  $-0.001\pm0.050\pm0.030$
\\[ 1.5mm]
\hline
\\[-2.5mm]
 \multirow{4}{*}{$15-20$}
 &  $\uplambda_{\theta}$
 &  $-0.079\pm0.052\pm0.043$
 &  $-0.055\pm0.060\pm0.031$
 &  $\pz0.047\pm0.091\pm0.055$
\\
 &  $\uplambda_{\theta\phi}$
 &  $-0.018\pm0.027\pm0.013$
 &  $-0.103\pm0.026\pm0.010$
 &  $-0.158\pm0.039\pm0.014$
\\
 &  $\uplambda_{\phi}$
 &  $-0.023\pm0.026\pm0.020$
 &  $\pz0.051\pm0.022\pm0.010$
 &  $\pz0.012\pm0.029\pm0.011$
\\
 &  $\tilde \uplambda$
 &  $-0.144\pm0.075\pm0.039$
 &  $\pz0.102\pm0.098\pm0.044$
 &  $\pz0.084\pm0.128\pm0.060$
\end{tabular*}
\end{footnotesize}
\end{center} 
\end{table}  

\begin{table}[t] 
\begin{center} 
\caption{\small Values of
 $\uplambda_{\theta}$, $\uplambda_{\theta\phi}$ and $\uplambda_{\phi}$
 measured in the HX, CS and GJ frames for the \ones produced
 at $\sqrt{s}=7\,\,{\mathrm{TeV}}$ in the rapidity range
 $2.2<\yy<4.5$.
 The first uncertainty is statistical and the second systematic.
}\label{tab:Y1S_Results_HX_CS_GJ_2.2y4.5_7TeV}
\vspace*{2mm}
\begin{footnotesize}
  \begin{tabular*}{0.99\textwidth}{@{\hspace{1mm}}c@{\extracolsep{\fill}}cccc@{\hspace{1mm}}}
 $\pty~~\left[\!\gevc\right]$ 
 & 
 & $\uplambda_{\theta}$ 
 & $\uplambda_{\theta\phi}$ 
 & $\uplambda_{\phi}$ 
\\[ 1.5mm]
\hline
\\[-2.5mm]
 \multirow{3}{*}{~ 0$\,-$~ 2}
 &  HX
 &  $\pz0.065\pm0.023\pm0.017$
 &  $-0.022\pm0.008\pm0.005$
 &  $-0.004\pm0.005\pm0.002$
\\
 &  CS
 &  $\pz0.073\pm0.024\pm0.018$
 &  $\pz0.002\pm0.008\pm0.004$
 &  $-0.006\pm0.005\pm0.002$
\\
 &  GJ
 &  $\pz0.064\pm0.023\pm0.016$
 &  $\pz0.026\pm0.008\pm0.004$
 &  $-0.003\pm0.005\pm0.002$
\\[ 1.5mm]
\hline
\\[-2.5mm]
 \multirow{3}{*}{~ 2$\,-$~ 4}
 &  HX
 &  $\pz0.036\pm0.016\pm0.012$
 &  $-0.019\pm0.007\pm0.004$
 &  $-0.009\pm0.004\pm0.002$
\\
 &  CS
 &  $\pz0.047\pm0.019\pm0.012$
 &  $\pz0.016\pm0.006\pm0.004$
 &  $-0.013\pm0.004\pm0.002$
\\
 &  GJ
 &  $\pz0.007\pm0.017\pm0.013$
 &  $\pz0.043\pm0.007\pm0.006$
 &  $\pz0.000\pm0.004\pm0.003$
\\[ 1.5mm]
\hline
\\[-2.5mm]
 \multirow{3}{*}{~ 4$\,-$~ 6}
 &  HX
 &  $\pz0.003\pm0.016\pm0.012$
 &  $\pz0.003\pm0.010\pm0.006$
 &  $-0.030\pm0.006\pm0.003$
\\
 &  CS
 &  $-0.014\pm0.020\pm0.011$
 &  $\pz0.023\pm0.008\pm0.005$
 &  $-0.025\pm0.004\pm0.002$
\\
 &  GJ
 &  $-0.060\pm0.017\pm0.010$
 &  $\pz0.020\pm0.009\pm0.006$
 &  $-0.012\pm0.005\pm0.004$
\\[ 1.5mm]
\hline
\\[-2.5mm]
 \multirow{3}{*}{~ 6$\,-$~ 8}
 &  HX
 &  $\pz0.041\pm0.019\pm0.015$
 &  $\pz0.019\pm0.011\pm0.007$
 &  $-0.045\pm0.009\pm0.006$
\\
 &  CS
 &  $-0.030\pm0.021\pm0.011$
 &  $\pz0.042\pm0.011\pm0.007$
 &  $-0.020\pm0.006\pm0.003$
\\
 &  GJ
 &  $-0.078\pm0.020\pm0.016$
 &  $\pz0.006\pm0.010\pm0.006$
 &  $-0.004\pm0.008\pm0.006$
\\[ 1.5mm]
\hline
\\[-2.5mm]
 \multirow{3}{*}{~ 8$\,-$~10}
 &  HX
 &  $\pz0.068\pm0.026\pm0.019$
 &  $-0.014\pm0.012\pm0.005$
 &  $-0.035\pm0.013\pm0.008$
\\
 &  CS
 &  $\pz0.016\pm0.022\pm0.012$
 &  $\pz0.045\pm0.016\pm0.008$
 &  $-0.017\pm0.009\pm0.003$
\\
 &  GJ
 &  $-0.065\pm0.026\pm0.023$
 &  $\pz0.017\pm0.011\pm0.004$
 &  $\pz0.010\pm0.011\pm0.007$
\\[ 1.5mm]
\hline
\\[-2.5mm]
 \multirow{3}{*}{~10$\,-$~15}
 &  HX
 &  $\pz0.073\pm0.025\pm0.024$
 &  $\pz0.015\pm0.012\pm0.005$
 &  $-0.027\pm0.013\pm0.010$
\\
 &  CS
 &  $-0.040\pm0.018\pm0.009$
 &  $\pz0.034\pm0.016\pm0.006$
 &  $\pz0.011\pm0.011\pm0.005$
\\
 &  GJ
 &  $-0.025\pm0.027\pm0.018$
 &  $-0.031\pm0.011\pm0.004$
 &  $\pz0.007\pm0.011\pm0.006$
\\[ 1.5mm]
\hline
\\[-2.5mm]
 \multirow{3}{*}{~15$\,-$~20}
 &  HX
 &  $\pz0.063\pm0.051\pm0.030$
 &  $\pz0.051\pm0.030\pm0.014$
 &  $\pz0.058\pm0.023\pm0.012$
\\
 &  CS
 &  $-0.011\pm0.037\pm0.023$
 &  $-0.032\pm0.030\pm0.010$
 &  $\pz0.082\pm0.028\pm0.017$
\\
 &  GJ
 &  $\pz0.158\pm0.062\pm0.039$
 &  $-0.022\pm0.028\pm0.010$
 &  $\pz0.032\pm0.022\pm0.009$
\\[ 1.5mm]
\hline
\\[-2.5mm]
 \multirow{3}{*}{~20$\,-$~30}
 &  HX
 &  $\pz0.077\pm0.098\pm0.049$
 &  $-0.009\pm0.055\pm0.018$
 &  $-0.004\pm0.036\pm0.011$
\\
 &  CS
 &  $-0.019\pm0.067\pm0.043$
 &  $\pz0.026\pm0.048\pm0.013$
 &  $\pz0.029\pm0.056\pm0.035$
\\
 &  GJ
 &  $\pz0.027\pm0.105\pm0.054$
 &  $-0.031\pm0.048\pm0.017$
 &  $\pz0.014\pm0.036\pm0.013$
\end{tabular*}
\end{footnotesize}
\end{center} 
\end{table}  
\begin{table}[t]
\begin{center}
\caption{\small Values of
 $\uplambda_{\theta}$, $\uplambda_{\theta\phi}$ and $\uplambda_{\phi}$
 measured in the HX, CS and GJ frames for the \ones produced
 at $\sqrt{s}=8\,\,{\mathrm{TeV}}$ in the rapidity range
 $2.2<\yy<4.5$.
 The first uncertainty is statistical and the second systematic.
}\label{tab:Y1S_Results_HX_CS_GJ_2.2y4.5_8TeV}
\vspace*{2mm}
\begin{footnotesize}
\begin{tabular*}{0.99\textwidth}{@{\hspace{1mm}}c@{\extracolsep{\fill}}cccc@{\hspace{1mm}}}
 $\pty~~\left[\!\gevc\right]$ 
 & 
 & $\uplambda_{\theta}$ 
 & $\uplambda_{\theta\phi}$ 
 & $\uplambda_{\phi}$ 
\\[ 1.5mm]
\hline
\\[-2.5mm]
 \multirow{3}{*}{~ 0$\,-$~ 2}
 &  HX
 &  $\pz0.072\pm0.016\pm0.014$
 &  $\pz0.003\pm0.006\pm0.003$
 &  $-0.002\pm0.004\pm0.001$
\\
 &  CS
 &  $\pz0.064\pm0.016\pm0.014$
 &  $\pz0.026\pm0.005\pm0.003$
 &  $-0.001\pm0.003\pm0.002$
\\
 &  GJ
 &  $\pz0.040\pm0.016\pm0.012$
 &  $\pz0.046\pm0.005\pm0.004$
 &  $\pz0.005\pm0.004\pm0.001$
\\[ 1.5mm]
\hline
\\[-2.5mm]
 \multirow{3}{*}{~ 2$\,-$~ 4}
 &  HX
 &  $\pz0.077\pm0.011\pm0.011$
 &  $-0.028\pm0.005\pm0.004$
 &  $-0.016\pm0.003\pm0.002$
\\
 &  CS
 &  $\pz0.086\pm0.013\pm0.010$
 &  $\pz0.024\pm0.004\pm0.005$
 &  $-0.020\pm0.003\pm0.001$
\\
 &  GJ
 &  $\pz0.027\pm0.011\pm0.007$
 &  $\pz0.062\pm0.005\pm0.005$
 &  $-0.003\pm0.003\pm0.002$
\\[ 1.5mm]
\hline
\\[-2.5mm]
 \multirow{3}{*}{~ 4$\,-$~ 6}
 &  HX
 &  $\pz0.078\pm0.011\pm0.012$
 &  $-0.008\pm0.006\pm0.005$
 &  $-0.035\pm0.004\pm0.003$
\\
 &  CS
 &  $\pz0.049\pm0.014\pm0.009$
 &  $\pz0.049\pm0.005\pm0.006$
 &  $-0.026\pm0.003\pm0.002$
\\
 &  GJ
 &  $-0.047\pm0.012\pm0.010$
 &  $\pz0.058\pm0.006\pm0.006$
 &  $\pz0.004\pm0.004\pm0.004$
\\[ 1.5mm]
\hline
\\[-2.5mm]
 \multirow{3}{*}{~ 6$\,-$~ 8}
 &  HX
 &  $\pz0.063\pm0.013\pm0.015$
 &  $-0.010\pm0.007\pm0.005$
 &  $-0.039\pm0.006\pm0.005$
\\
 &  CS
 &  $\pz0.023\pm0.014\pm0.010$
 &  $\pz0.042\pm0.008\pm0.007$
 &  $-0.026\pm0.004\pm0.002$
\\
 &  GJ
 &  $-0.058\pm0.013\pm0.013$
 &  $\pz0.033\pm0.007\pm0.005$
 &  $\pz0.001\pm0.005\pm0.005$
\\[ 1.5mm]
\hline
\\[-2.5mm]
 \multirow{3}{*}{~ 8$\,-$~10}
 &  HX
 &  $\pz0.114\pm0.017\pm0.019$
 &  $-0.007\pm0.008\pm0.004$
 &  $-0.056\pm0.009\pm0.008$
\\
 &  CS
 &  $\pz0.002\pm0.014\pm0.008$
 &  $\pz0.072\pm0.011\pm0.007$
 &  $-0.018\pm0.006\pm0.003$
\\
 &  GJ
 &  $-0.101\pm0.017\pm0.016$
 &  $\pz0.011\pm0.007\pm0.003$
 &  $\pz0.017\pm0.007\pm0.006$
\\[ 1.5mm]
\hline
\\[-2.5mm]
 \multirow{3}{*}{~10$\,-$~15}
 &  HX
 &  $\pz0.099\pm0.017\pm0.017$
 &  $-0.025\pm0.008\pm0.004$
 &  $-0.042\pm0.009\pm0.006$
\\
 &  CS
 &  $\pz0.005\pm0.012\pm0.008$
 &  $\pz0.061\pm0.011\pm0.007$
 &  $-0.010\pm0.007\pm0.004$
\\
 &  GJ
 &  $-0.077\pm0.017\pm0.016$
 &  $-0.005\pm0.007\pm0.003$
 &  $\pz0.018\pm0.007\pm0.006$
\\[ 1.5mm]
\hline
\\[-2.5mm]
 \multirow{3}{*}{~15$\,-$~20}
 &  HX
 &  $\pz0.160\pm0.035\pm0.021$
 &  $-0.012\pm0.020\pm0.009$
 &  $-0.060\pm0.016\pm0.008$
\\
 &  CS
 &  $-0.073\pm0.022\pm0.017$
 &  $\pz0.084\pm0.019\pm0.007$
 &  $\pz0.019\pm0.018\pm0.013$
\\
 &  GJ
 &  $-0.060\pm0.035\pm0.020$
 &  $-0.080\pm0.016\pm0.007$
 &  $\pz0.016\pm0.014\pm0.006$
\\[ 1.5mm]
\hline
\\[-2.5mm]
 \multirow{3}{*}{~20$\,-$~30}
 &  HX
 &  $\pz0.183\pm0.064\pm0.033$
 &  $\pz0.084\pm0.035\pm0.013$
 &  $-0.023\pm0.023\pm0.009$
\\
 &  CS
 &  $-0.155\pm0.037\pm0.029$
 &  $\pz0.004\pm0.028\pm0.011$
 &  $\pz0.086\pm0.032\pm0.026$
\\
 &  GJ
 &  $\pz0.176\pm0.069\pm0.043$
 &  $-0.093\pm0.031\pm0.014$
 &  $-0.019\pm0.023\pm0.009$
\end{tabular*}
\end{footnotesize} 
\end{center} 
\end{table}  
\begin{table}[t]
\begin{center}
\caption{\small Values of
 $\tilde \uplambda$
 measured in the HX, CS and GJ frames for the \ones produced
 at $\sqrt{s}=7$ and $8\,\,{\mathrm{TeV}}$ in the rapidity range
 $2.2<\yy<4.5$.
 The first uncertainty is statistical and the second systematic.
}\label{tab:Y1S_lamT_Results_HX_CS_GJ_2.2y4.5}
\vspace*{7mm}
\begin{footnotesize} 
\begin{tabular*}{0.90\textwidth}{@{\hspace{1mm}}c@{\extracolsep{\fill}}ccc@{\hspace{1mm}}}
 $\pty~~~\left[\!\gevc\right]$ & $\tilde \uplambda$ & $\sqrt{s}=7\tev$ & $\sqrt{s}=8\tev$ 
\\[ 1.5mm]
\hline
\\[-2.5mm]
 \multirow{3}{*}{~ 0$\,-$~ 2}
 &  HX
 &  $\pz0.054\pm0.028\pm0.019$
 &  $\pz0.064\pm0.020\pm0.014$
\\
 &  CS
 &  $\pz0.055\pm0.029\pm0.020$
 &  $\pz0.060\pm0.020\pm0.016$
\\
 &  GJ
 &  $\pz0.055\pm0.029\pm0.018$
 &  $\pz0.056\pm0.020\pm0.014$
\\[ 1.5mm]
\hline
\\[-2.5mm]
 \multirow{3}{*}{~ 2$\,-$~ 4}
 &  HX
 &  $\pz0.009\pm0.022\pm0.014$
 &  $\pz0.029\pm0.015\pm0.011$
\\
 &  CS
 &  $\pz0.009\pm0.022\pm0.014$
 &  $\pz0.027\pm0.015\pm0.012$
\\
 &  GJ
 &  $\pz0.006\pm0.022\pm0.016$
 &  $\pz0.019\pm0.015\pm0.010$
\\[ 1.5mm]
\hline
\\[-2.5mm]
 \multirow{3}{*}{~ 4$\,-$~ 6}
 &  HX
 &  $-0.085\pm0.024\pm0.013$
 &  $-0.026\pm0.017\pm0.012$
\\
 &  CS
 &  $-0.088\pm0.025\pm0.013$
 &  $-0.029\pm0.017\pm0.012$
\\
 &  GJ
 &  $-0.094\pm0.025\pm0.013$
 &  $-0.036\pm0.017\pm0.013$
\\[ 1.5mm]
\hline
\\[-2.5mm]
 \multirow{3}{*}{~ 6$\,-$~ 8}
 &  HX
 &  $-0.089\pm0.029\pm0.016$
 &  $-0.053\pm0.019\pm0.014$
\\
 &  CS
 &  $-0.088\pm0.029\pm0.014$
 &  $-0.052\pm0.020\pm0.013$
\\
 &  GJ
 &  $-0.089\pm0.029\pm0.017$
 &  $-0.054\pm0.020\pm0.013$
\\[ 1.5mm]
\hline
\\[-2.5mm]
 \multirow{3}{*}{~ 8$\,-$~10}
 &  HX
 &  $-0.036\pm0.035\pm0.017$
 &  $-0.051\pm0.023\pm0.012$
\\
 &  CS
 &  $-0.035\pm0.035\pm0.016$
 &  $-0.050\pm0.023\pm0.012$
\\
 &  GJ
 &  $-0.035\pm0.036\pm0.017$
 &  $-0.050\pm0.023\pm0.012$
\\[ 1.5mm]
\hline
\\[-2.5mm]
 \multirow{3}{*}{~10$\,-$~15}
 &  HX
 &  $-0.009\pm0.036\pm0.014$
 &  $-0.025\pm0.023\pm0.010$
\\
 &  CS
 &  $-0.007\pm0.036\pm0.016$
 &  $-0.024\pm0.023\pm0.010$
\\
 &  GJ
 &  $-0.004\pm0.036\pm0.013$
 &  $-0.022\pm0.023\pm0.011$
\\[ 1.5mm]
\hline
\\[-2.5mm]
 \multirow{3}{*}{~15$\,-$~20}
 &  HX
 &  $\pz0.250\pm0.094\pm0.041$
 &  $-0.019\pm0.052\pm0.025$
\\
 &  CS
 &  $\pz0.256\pm0.094\pm0.042$
 &  $-0.016\pm0.052\pm0.028$
\\
 &  GJ
 &  $\pz0.261\pm0.094\pm0.036$
 &  $-0.012\pm0.052\pm0.022$
\\[ 1.5mm]
\hline
\\[-2.5mm]
 \multirow{3}{*}{~20$\,-$~30}
 &  HX
 &  $\pz0.066\pm0.156\pm0.063$
 &  $\pz0.110\pm0.098\pm0.048$
\\
 &  CS
 &  $\pz0.068\pm0.156\pm0.078$
 &  $\pz0.114\pm0.098\pm0.061$
\\
 &  GJ
 &  $\pz0.071\pm0.157\pm0.061$
 &  $\pz0.115\pm0.098\pm0.051$
\end{tabular*}
\end{footnotesize} 
\end{center} 
\end{table}  

\clearpage 
{\boldmath{\section{Polarization results for the \twos state}\label{sec:upstwospol}}}

Values of the polarization parameters $\pmb{\uplambda}$ and $\tilde\uplambda$
for the \twos~meson
are presented 
in Tables~\ref{tab:Y2S_Results_HX_Slices_7TeV} and~\ref{tab:Y2S_Results_HX_Slices_8TeV}
for the~HX~frame,
in Tables~\ref{tab:Y2S_Results_CS_Slices_7TeV} and~\ref{tab:Y2S_Results_CS_Slices_8TeV}
for the~CS~frame and
in Tables~\ref{tab:Y2S_Results_GJ_Slices_7TeV} and~\ref{tab:Y2S_Results_GJ_Slices_8TeV}
for the~GJ~frame for $\sqs=7$~and~8\tev, respectively.
The~polarization parameters $\pmb{\uplambda}$ 
measured in the~wide rapidity bin $2.2<\yy<4.5$ are presented 
in Tables~\ref{tab:Y2S_Results_HX_CS_GJ_2.2y4.5_7TeV} 
      and~\ref{tab:Y2S_Results_HX_CS_GJ_2.2y4.5_8TeV}, while 
the~parameters $\tilde\uplambda$ are listed 
in Table~\ref{tab:Y2S_lamT_Results_HX_CS_GJ_2.2y4.5}.
 
\begin{table}[htb] 
\begin{center} 
\caption{\small Values of
  $\uplambda_{\theta}$, $\uplambda_{\theta\phi}$, $\uplambda_{\phi}$
  and $\tilde \uplambda$
  measured in the HX frame for the \twos produced
  at $\sqrt{s}=7\,\,{\mathrm{TeV}}$.
  The first uncertainty is statistical and the second systematic.
}\label{tab:Y2S_Results_HX_Slices_7TeV}
\vspace*{2mm}
\begin{footnotesize}
\begin{tabular*}{0.99\textwidth}{@{\hspace{1mm}}c@{\extracolsep{\fill}}cccc@{\hspace{1mm}}}
 $\pty~\left[\!\gevc\right]$ 
 & $\uplambda$ 
 & $2.2<y<3.0$
 & $3.0<y<3.5$ 
 & $3.5<y<4.5$ 
\\[ 1.5mm]
\hline
\\[-2.5mm]
 \multirow{4}{*}{$0-2$}
 &  $\uplambda_{\theta}$
 &  $\pz0.167\pm0.148\pm0.068$
 &  $-0.244\pm0.072\pm0.046$
 &  $-0.190\pm0.097\pm0.071$
\\
 &  $\uplambda_{\theta\phi}$
 &  $\pz0.021\pm0.037\pm0.010$
 &  $\pz0.013\pm0.026\pm0.009$
 &  $-0.011\pm0.036\pm0.017$
\\
 &  $\uplambda_{\phi}$
 &  $-0.014\pm0.020\pm0.006$
 &  $\pz0.007\pm0.021\pm0.007$
 &  $\pz0.012\pm0.022\pm0.007$
\\
 &  $\tilde \uplambda$
 &  $\pz0.123\pm0.158\pm0.071$
 &  $-0.224\pm0.095\pm0.052$
 &  $-0.155\pm0.119\pm0.074$
\\[ 1.5mm]
\hline
\\[-2.5mm]
 \multirow{4}{*}{$2-4$}
 &  $\uplambda_{\theta}$
 &  $\pz0.247\pm0.104\pm0.055$
 &  $-0.032\pm0.056\pm0.033$
 &  $-0.215\pm0.061\pm0.036$
\\
 &  $\uplambda_{\theta\phi}$
 &  $-0.055\pm0.031\pm0.012$
 &  $\pz0.019\pm0.025\pm0.010$
 &  $\pz0.096\pm0.035\pm0.018$
\\
 &  $\uplambda_{\phi}$
 &  $\pz0.009\pm0.015\pm0.004$
 &  $\pz0.015\pm0.017\pm0.005$
 &  $-0.003\pm0.019\pm0.008$
\\
 &  $\tilde \uplambda$
 &  $\pz0.278\pm0.120\pm0.062$
 &  $\pz0.014\pm0.082\pm0.042$
 &  $-0.223\pm0.088\pm0.051$
\\[ 1.5mm]
\hline
\\[-2.5mm]
 \multirow{4}{*}{$4-6$}
 &  $\uplambda_{\theta}$
 &  $\pz0.023\pm0.097\pm0.056$
 &  $-0.098\pm0.053\pm0.024$
 &  $-0.100\pm0.055\pm0.034$
\\
 &  $\uplambda_{\theta\phi}$
 &  $\pz0.012\pm0.040\pm0.019$
 &  $\pz0.073\pm0.031\pm0.012$
 &  $\pz0.124\pm0.044\pm0.030$
\\
 &  $\uplambda_{\phi}$
 &  $\pz0.042\pm0.019\pm0.007$
 &  $-0.015\pm0.021\pm0.008$
 &  $-0.027\pm0.027\pm0.015$
\\
 &  $\tilde \uplambda$
 &  $\pz0.157\pm0.129\pm0.070$
 &  $-0.142\pm0.085\pm0.032$
 &  $-0.176\pm0.099\pm0.063$
\\[ 1.5mm]
\hline
\\[-2.5mm]
 \multirow{4}{*}{$6-8$}
 &  $\uplambda_{\theta}$
 &  $-0.103\pm0.097\pm0.054$
 &  $-0.053\pm0.063\pm0.029$
 &  $\pz0.002\pm0.062\pm0.033$
\\
 &  $\uplambda_{\theta\phi}$
 &  $\pz0.020\pm0.049\pm0.030$
 &  $\pz0.244\pm0.036\pm0.016$
 &  $\pz0.083\pm0.049\pm0.030$
\\
 &  $\uplambda_{\phi}$
 &  $\pz0.037\pm0.028\pm0.016$
 &  $-0.070\pm0.032\pm0.015$
 &  $-0.015\pm0.040\pm0.025$
\\
 &  $\tilde \uplambda$
 &  $\pz0.009\pm0.140\pm0.079$
 &  $-0.245\pm0.095\pm0.032$
 &  $-0.042\pm0.121\pm0.069$
\\[ 1.5mm]
\hline
\\[-2.5mm]
 \multirow{4}{*}{$8-10$}
 &  $\uplambda_{\theta}$
 &  $-0.023\pm0.114\pm0.063$
 &  $\pz0.209\pm0.090\pm0.065$
 &  $\pz0.088\pm0.092\pm0.068$
\\
 &  $\uplambda_{\theta\phi}$
 &  $\pz0.025\pm0.055\pm0.027$
 &  $\pz0.147\pm0.037\pm0.013$
 &  $\pz0.174\pm0.051\pm0.029$
\\
 &  $\uplambda_{\phi}$
 &  $-0.013\pm0.041\pm0.026$
 &  $-0.055\pm0.044\pm0.024$
 &  $-0.067\pm0.060\pm0.041$
\\
 &  $\tilde \uplambda$
 &  $-0.063\pm0.148\pm0.072$
 &  $\pz0.042\pm0.119\pm0.037$
 &  $-0.105\pm0.138\pm0.061$
\\[ 1.5mm]
\hline
\\[-2.5mm]
 \multirow{4}{*}{$10-15$}
 &  $\uplambda_{\theta}$
 &  $\pz0.172\pm0.103\pm0.087$
 &  $\pz0.114\pm0.077\pm0.053$
 &  $\pz0.237\pm0.102\pm0.095$
\\
 &  $\uplambda_{\theta\phi}$
 &  $\pz0.007\pm0.043\pm0.022$
 &  $\pz0.004\pm0.034\pm0.013$
 &  $\pz0.123\pm0.044\pm0.015$
\\
 &  $\uplambda_{\phi}$
 &  $-0.033\pm0.038\pm0.028$
 &  $\pz0.020\pm0.039\pm0.024$
 &  $-0.116\pm0.062\pm0.057$
\\
 &  $\tilde \uplambda$
 &  $\pz0.071\pm0.119\pm0.057$
 &  $\pz0.178\pm0.123\pm0.055$
 &  $-0.099\pm0.131\pm0.080$
\\[ 1.5mm]
\hline
\\[-2.5mm]
 \multirow{4}{*}{$15-20$}
 &  $\uplambda_{\theta}$
 &  $\pz0.095\pm0.149\pm0.150$
 &  $\pz0.112\pm0.136\pm0.132$
 &  $\pz0.376\pm0.229\pm0.216$
\\
 &  $\uplambda_{\theta\phi}$
 &  $\pz0.063\pm0.077\pm0.051$
 &  $-0.008\pm0.083\pm0.057$
 &  $\pz0.076\pm0.135\pm0.098$
\\
 &  $\uplambda_{\phi}$
 &  $\pz0.018\pm0.054\pm0.050$
 &  $-0.057\pm0.073\pm0.056$
 &  $-0.190\pm0.127\pm0.113$
\\
 &  $\tilde \uplambda$
 &  $\pz0.152\pm0.207\pm0.162$
 &  $-0.055\pm0.229\pm0.164$
 &  $-0.163\pm0.312\pm0.191$
\end{tabular*}
\end{footnotesize} 
\end{center} 
\end{table}  
%
%
%
%
%
\begin{table}[t] 
\begin{center} 
\caption{\small Values of
 $\uplambda_{\theta}$, $\uplambda_{\theta\phi}$, $\uplambda_{\phi}$
 and $\tilde \uplambda$
 measured in the HX frame for the \twos produced
 at $\sqrt{s}=8\,\,{\mathrm{TeV}}$.
 The first uncertainty is statistical and the second systematic.
}\label{tab:Y2S_Results_HX_Slices_8TeV}
\vspace*{2mm}
\begin{footnotesize}
\begin{tabular*}{0.99\textwidth}{@{\hspace{1mm}}c@{\extracolsep{\fill}}cccc@{\hspace{1mm}}}
 $\pty~\left[\!\gevc\right]$ 
 & $\uplambda$ 
 & $2.2<y<3.0$ 
 & $3.0<y<3.5$ 
 & $3.5<y<4.5$ 
\\[ 1.5mm]
\hline
\\[-2.5mm]
 \multirow{4}{*}{$0-2$}
 &  $\uplambda_{\theta}$
 &  $\pz0.252\pm0.101\pm0.054$
 &  $\pz0.131\pm0.057\pm0.035$
 &  $-0.156\pm0.067\pm0.054$
\\
 &  $\uplambda_{\theta\phi}$
 &  $-0.033\pm0.026\pm0.009$
 &  $-0.002\pm0.020\pm0.007$
 &  $-0.041\pm0.025\pm0.014$
\\
 &  $\uplambda_{\phi}$
 &  $\pz0.002\pm0.014\pm0.004$
 &  $-0.009\pm0.015\pm0.005$
 &  $\pz0.017\pm0.015\pm0.005$
\\
 &  $\tilde \uplambda$
 &  $\pz0.259\pm0.111\pm0.061$
 &  $\pz0.104\pm0.075\pm0.039$
 &  $-0.107\pm0.083\pm0.059$
\\[ 1.5mm]
\hline
\\[-2.5mm]
 \multirow{4}{*}{$2-4$}
 &  $\uplambda_{\theta}$
 &  $\pz0.210\pm0.070\pm0.037$
 &  $\pz0.131\pm0.040\pm0.029$
 &  $-0.081\pm0.042\pm0.033$
\\
 &  $\uplambda_{\theta\phi}$
 &  $-0.035\pm0.021\pm0.009$
 &  $-0.052\pm0.018\pm0.008$
 &  $\pz0.030\pm0.024\pm0.014$
\\
 &  $\uplambda_{\phi}$
 &  $-0.017\pm0.010\pm0.004$
 &  $\pz0.018\pm0.012\pm0.004$
 &  $-0.002\pm0.013\pm0.005$
\\
 &  $\tilde \uplambda$
 &  $\pz0.156\pm0.078\pm0.040$
 &  $\pz0.190\pm0.060\pm0.037$
 &  $-0.088\pm0.061\pm0.042$
\\[ 1.5mm]
\hline
\\[-2.5mm]
 \multirow{4}{*}{$4-6$}
 &  $\uplambda_{\theta}$
 &  $\pz0.121\pm0.066\pm0.037$
 &  $\pz0.048\pm0.039\pm0.023$
 &  $-0.086\pm0.039\pm0.022$
\\
 &  $\uplambda_{\theta\phi}$
 &  $-0.046\pm0.027\pm0.016$
 &  $\pz0.092\pm0.022\pm0.010$
 &  $\pz0.180\pm0.030\pm0.024$
\\
 &  $\uplambda_{\phi}$
 &  $\pz0.015\pm0.013\pm0.005$
 &  $-0.030\pm0.015\pm0.006$
 &  $-0.035\pm0.019\pm0.013$
\\
 &  $\tilde \uplambda$
 &  $\pz0.168\pm0.085\pm0.049$
 &  $-0.040\pm0.061\pm0.034$
 &  $-0.185\pm0.067\pm0.048$
\\[ 1.5mm]
\hline
\\[-2.5mm]
 \multirow{4}{*}{$6-8$}
 &  $\uplambda_{\theta}$
 &  $-0.018\pm0.066\pm0.043$
 &  $\pz0.109\pm0.045\pm0.027$
 &  $\pz0.049\pm0.044\pm0.026$
\\
 &  $\uplambda_{\theta\phi}$
 &  $\pz0.003\pm0.033\pm0.023$
 &  $\pz0.168\pm0.025\pm0.013$
 &  $\pz0.216\pm0.036\pm0.024$
\\
 &  $\uplambda_{\phi}$
 &  $\pz0.024\pm0.019\pm0.011$
 &  $-0.090\pm0.023\pm0.011$
 &  $-0.050\pm0.029\pm0.019$
\\
 &  $\tilde \uplambda$
 &  $\pz0.056\pm0.095\pm0.062$
 &  $-0.146\pm0.066\pm0.029$
 &  $-0.096\pm0.082\pm0.044$
\\[ 1.5mm]
\hline
\\[-2.5mm]
 \multirow{4}{*}{$8-10$}
 &  $\uplambda_{\theta}$
 &  $\pz0.083\pm0.077\pm0.050$
 &  $\pz0.078\pm0.055\pm0.041$
 &  $\pz0.018\pm0.055\pm0.051$
\\
 &  $\uplambda_{\theta\phi}$
 &  $-0.022\pm0.037\pm0.019$
 &  $\pz0.099\pm0.024\pm0.010$
 &  $\pz0.125\pm0.032\pm0.021$
\\
 &  $\uplambda_{\phi}$
 &  $\pz0.023\pm0.027\pm0.017$
 &  $-0.029\pm0.029\pm0.019$
 &  $\pz0.013\pm0.036\pm0.031$
\\
 &  $\tilde \uplambda$
 &  $\pz0.155\pm0.107\pm0.049$
 &  $-0.010\pm0.078\pm0.026$
 &  $\pz0.057\pm0.096\pm0.058$
\\[ 1.5mm]
\hline
\\[-2.5mm]
 \multirow{4}{*}{$10-15$}
 &  $\uplambda_{\theta}$
 &  $\pz0.388\pm0.076\pm0.062$
 &  $\pz0.209\pm0.054\pm0.047$
 &  $\pz0.200\pm0.065\pm0.077$
\\
 &  $\uplambda_{\theta\phi}$
 &  $-0.045\pm0.030\pm0.015$
 &  $\pz0.069\pm0.022\pm0.009$
 &  $\pz0.077\pm0.028\pm0.012$
\\
 &  $\uplambda_{\phi}$
 &  $-0.055\pm0.026\pm0.019$
 &  $-0.042\pm0.027\pm0.019$
 &  $-0.022\pm0.038\pm0.039$
\\
 &  $\tilde \uplambda$
 &  $\pz0.212\pm0.082\pm0.033$
 &  $\pz0.079\pm0.075\pm0.028$
 &  $\pz0.130\pm0.095\pm0.058$
\\[ 1.5mm]
\hline
\\[-2.5mm]
 \multirow{4}{*}{$15-20$}
 &  $\uplambda_{\theta}$
 &  $\pz0.019\pm0.096\pm0.103$
 &  $\pz0.246\pm0.095\pm0.113$
 &  $\pz0.065\pm0.114\pm0.118$
\\
 &  $\uplambda_{\theta\phi}$
 &  $\pz0.043\pm0.051\pm0.037$
 &  $-0.062\pm0.057\pm0.046$
 &  $\pz0.201\pm0.072\pm0.057$
\\
 &  $\uplambda_{\phi}$
 &  $-0.015\pm0.037\pm0.028$
 &  $-0.073\pm0.049\pm0.045$
 &  $-0.016\pm0.068\pm0.060$
\\
 &  $\tilde \uplambda$
 &  $-0.025\pm0.129\pm0.083$
 &  $\pz0.025\pm0.155\pm0.118$
 &  $\pz0.018\pm0.208\pm0.167$
\end{tabular*}
\end{footnotesize}
\end{center} 
\end{table}  
%
%
%
%
%
\begin{table}[t] 
\begin{center} 
\caption{\small Values of
 $\uplambda_{\theta}$, $\uplambda_{\theta\phi}$, $\uplambda_{\phi}$
 and $\tilde \uplambda$
 measured in the CS frame for the \twos produced
 at $\sqrt{s}=7\,\,{\mathrm{TeV}}$.
 The first uncertainty is statistical and the second systematic.
}\label{tab:Y2S_Results_CS_Slices_7TeV}
\vspace*{2mm}
\begin{footnotesize}
\begin{tabular*}{0.99\textwidth}{@{\hspace{1mm}}c@{\extracolsep{\fill}}cccc@{\hspace{1mm}}}
 $\pty~\left[\!\gevc\right]$ 
 & $\uplambda$ 
 & 2.2~$<y<$~3.0 
 & 3.0~$<y<$~3.5 
 & 3.5~$<y<$~4.5 
\\[ 1.5mm]
\hline
\\[-2.5mm]
 \multirow{4}{*}{$0-2$}
 &  $\uplambda_{\theta}$
 &  $\pz0.175\pm0.146\pm0.066$
 &  $-0.249\pm0.073\pm0.045$
 &  $-0.198\pm0.103\pm0.059$
\\
 &  $\uplambda_{\theta\phi}$
 &  $\pz0.088\pm0.038\pm0.013$
 &  $-0.003\pm0.025\pm0.008$
 &  $-0.012\pm0.033\pm0.012$
\\
 &  $\uplambda_{\phi}$
 &  $-0.008\pm0.020\pm0.006$
 &  $\pz0.007\pm0.021\pm0.006$
 &  $\pz0.008\pm0.021\pm0.007$
\\
 &  $\tilde \uplambda$
 &  $\pz0.152\pm0.159\pm0.070$
 &  $-0.230\pm0.095\pm0.050$
 &  $-0.177\pm0.120\pm0.059$
\\[ 1.5mm]
\hline
\\[-2.5mm]
 \multirow{4}{*}{$2-4$}
 &  $\uplambda_{\theta}$
 &  $\pz0.237\pm0.107\pm0.048$
 &  $-0.036\pm0.062\pm0.032$
 &  $-0.212\pm0.075\pm0.036$
\\
 &  $\uplambda_{\theta\phi}$
 &  $\pz0.089\pm0.030\pm0.009$
 &  $\pz0.027\pm0.021\pm0.006$
 &  $\pz0.046\pm0.026\pm0.008$
\\
 &  $\uplambda_{\phi}$
 &  $\pz0.011\pm0.014\pm0.004$
 &  $\pz0.018\pm0.016\pm0.004$
 &  $\pz0.009\pm0.016\pm0.005$
\\
 &  $\tilde \uplambda$
 &  $\pz0.274\pm0.122\pm0.052$
 &  $\pz0.017\pm0.083\pm0.040$
 &  $-0.186\pm0.091\pm0.043$
\\[ 1.5mm]
\hline
\\[-2.5mm]
 \multirow{4}{*}{$4-6$}
 &  $\uplambda_{\theta}$
 &  $\pz0.054\pm0.106\pm0.039$
 &  $-0.148\pm0.062\pm0.026$
 &  $-0.163\pm0.079\pm0.039$
\\
 &  $\uplambda_{\theta\phi}$
 &  $\pz0.069\pm0.035\pm0.011$
 &  $\pz0.032\pm0.026\pm0.007$
 &  $\pz0.065\pm0.031\pm0.012$
\\
 &  $\uplambda_{\phi}$
 &  $\pz0.052\pm0.015\pm0.004$
 &  $\pz0.004\pm0.017\pm0.004$
 &  $\pz0.009\pm0.018\pm0.006$
\\
 &  $\tilde \uplambda$
 &  $\pz0.220\pm0.133\pm0.049$
 &  $-0.137\pm0.086\pm0.031$
 &  $-0.138\pm0.101\pm0.048$
\\[ 1.5mm]
\hline
\\[-2.5mm]
 \multirow{4}{*}{$6-8$}
 &  $\uplambda_{\theta}$
 &  $-0.027\pm0.113\pm0.042$
 &  $-0.354\pm0.057\pm0.025$
 &  $-0.074\pm0.085\pm0.039$
\\
 &  $\uplambda_{\theta\phi}$
 &  $-0.017\pm0.046\pm0.013$
 &  $\pz0.088\pm0.034\pm0.009$
 &  $\pz0.046\pm0.043\pm0.014$
\\
 &  $\uplambda_{\phi}$
 &  $\pz0.027\pm0.019\pm0.005$
 &  $\pz0.041\pm0.021\pm0.008$
 &  $\pz0.021\pm0.024\pm0.008$
\\
 &  $\tilde \uplambda$
 &  $\pz0.056\pm0.143\pm0.052$
 &  $-0.241\pm0.095\pm0.043$
 &  $-0.011\pm0.123\pm0.052$
\\[ 1.5mm]
\hline
\\[-2.5mm]
 \multirow{4}{*}{$8-10$}
 &  $\uplambda_{\theta}$
 &  $-0.036\pm0.115\pm0.047$
 &  $-0.161\pm0.057\pm0.026$
 &  $-0.217\pm0.077\pm0.034$
\\
 &  $\uplambda_{\theta\phi}$
 &  $\pz0.038\pm0.062\pm0.019$
 &  $\pz0.135\pm0.048\pm0.015$
 &  $\pz0.077\pm0.059\pm0.026$
\\
 &  $\uplambda_{\phi}$
 &  $-0.001\pm0.025\pm0.010$
 &  $\pz0.066\pm0.029\pm0.010$
 &  $\pz0.046\pm0.033\pm0.013$
\\
 &  $\tilde \uplambda$
 &  $-0.040\pm0.150\pm0.059$
 &  $\pz0.040\pm0.119\pm0.046$
 &  $-0.081\pm0.139\pm0.059$
\\[ 1.5mm]
\hline
\\[-2.5mm]
 \multirow{4}{*}{$10-15$}
 &  $\uplambda_{\theta}$
 &  $-0.019\pm0.072\pm0.036$
 &  $\pz0.009\pm0.051\pm0.028$
 &  $-0.224\pm0.062\pm0.035$
\\
 &  $\uplambda_{\theta\phi}$
 &  $\pz0.046\pm0.058\pm0.024$
 &  $\pz0.055\pm0.048\pm0.018$
 &  $\pz0.137\pm0.059\pm0.028$
\\
 &  $\uplambda_{\phi}$
 &  $\pz0.023\pm0.028\pm0.014$
 &  $\pz0.053\pm0.033\pm0.016$
 &  $\pz0.040\pm0.038\pm0.021$
\\
 &  $\tilde \uplambda$
 &  $\pz0.052\pm0.118\pm0.055$
 &  $\pz0.177\pm0.123\pm0.053$
 &  $-0.108\pm0.133\pm0.063$
\\[ 1.5mm]
\hline
\\[-2.5mm]
 \multirow{4}{*}{$15-20$}
 &  $\uplambda_{\theta}$
 &  $-0.068\pm0.091\pm0.083$
 &  $-0.072\pm0.093\pm0.100$
 &  $-0.291\pm0.118\pm0.139$
\\
 &  $\uplambda_{\theta\phi}$
 &  $\pz0.013\pm0.089\pm0.053$
 &  $\pz0.071\pm0.080\pm0.048$
 &  $\pz0.182\pm0.115\pm0.074$
\\
 &  $\uplambda_{\phi}$
 &  $\pz0.071\pm0.060\pm0.049$
 &  $\pz0.011\pm0.080\pm0.068$
 &  $\pz0.047\pm0.109\pm0.121$
\\
 &  $\tilde \uplambda$
 &  $\pz0.157\pm0.208\pm0.167$
 &  $-0.040\pm0.230\pm0.154$
 &  $-0.159\pm0.315\pm0.290$
\end{tabular*}
\end{footnotesize}
\end{center} 
\end{table}  
%
%
%
%
%
\begin{table}[t] 
\begin{center} 
\caption{\small Values of
 $\uplambda_{\theta}$, $\uplambda_{\theta\phi}$, $\uplambda_{\phi}$
 and $\tilde \uplambda$
 measured in the CS frame for the \twos produced
 at $\sqrt{s}=8\,\,{\mathrm{TeV}}$.
 The first uncertainty is statistical and the second systematic.
}\label{tab:Y2S_Results_CS_Slices_8TeV}
\vspace*{2mm}
\begin{footnotesize}
\begin{tabular*}{0.99\textwidth}{@{\hspace{1mm}}c@{\extracolsep{\fill}}cccc@{\hspace{1mm}}}
 $\pty~\left[\!\gevc\right]$ 
 & $\uplambda$ 
 & $2.2<y<3.0$ 
 & $3.0<y<3.5$
 & $3.5<y<4.5$ 
\\[ 1.5mm]
\hline
\\[-2.5mm]
 \multirow{4}{*}{$0-2$}
 &  $\uplambda_{\theta}$
 &  $\pz0.254\pm0.101\pm0.048$
 &  $\pz0.119\pm0.058\pm0.038$
 &  $-0.139\pm0.072\pm0.051$
\\
 &  $\uplambda_{\theta\phi}$
 &  $\pz0.037\pm0.026\pm0.010$
 &  $\pz0.026\pm0.019\pm0.007$
 &  $-0.035\pm0.023\pm0.008$
\\
 &  $\uplambda_{\phi}$
 &  $\pz0.002\pm0.014\pm0.004$
 &  $-0.007\pm0.015\pm0.005$
 &  $\pz0.008\pm0.015\pm0.005$
\\
 &  $\tilde \uplambda$
 &  $\pz0.260\pm0.111\pm0.047$
 &  $\pz0.097\pm0.075\pm0.042$
 &  $-0.116\pm0.084\pm0.058$
\\[ 1.5mm]
\hline
\\[-2.5mm]
 \multirow{4}{*}{$2-4$}
 &  $\uplambda_{\theta}$
 &  $\pz0.193\pm0.072\pm0.046$
 &  $\pz0.184\pm0.045\pm0.030$
 &  $-0.054\pm0.053\pm0.038$
\\
 &  $\uplambda_{\theta\phi}$
 &  $\pz0.096\pm0.020\pm0.009$
 &  $\pz0.008\pm0.015\pm0.005$
 &  $\pz0.037\pm0.018\pm0.006$
\\
 &  $\uplambda_{\phi}$
 &  $-0.011\pm0.010\pm0.003$
 &  $\pz0.008\pm0.011\pm0.004$
 &  $-0.004\pm0.011\pm0.003$
\\
 &  $\tilde \uplambda$
 &  $\pz0.160\pm0.079\pm0.051$
 &  $\pz0.209\pm0.061\pm0.038$
 &  $-0.065\pm0.063\pm0.040$
\\[ 1.5mm]
\hline
\\[-2.5mm]
 \multirow{4}{*}{$4-6$}
 &  $\uplambda_{\theta}$
 &  $\pz0.148\pm0.074\pm0.029$
 &  $-0.075\pm0.043\pm0.024$
 &  $-0.207\pm0.053\pm0.031$
\\
 &  $\uplambda_{\theta\phi}$
 &  $\pz0.077\pm0.023\pm0.008$
 &  $\pz0.100\pm0.018\pm0.006$
 &  $\pz0.106\pm0.021\pm0.008$
\\
 &  $\uplambda_{\phi}$
 &  $\pz0.013\pm0.010\pm0.003$
 &  $\pz0.012\pm0.012\pm0.004$
 &  $\pz0.023\pm0.012\pm0.004$
\\
 &  $\tilde \uplambda$
 &  $\pz0.191\pm0.088\pm0.034$
 &  $-0.038\pm0.062\pm0.033$
 &  $-0.141\pm0.069\pm0.038$
\\[ 1.5mm]
\hline
\\[-2.5mm]
 \multirow{4}{*}{$6-8$}
 &  $\uplambda_{\theta}$
 &  $\pz0.023\pm0.077\pm0.031$
 &  $-0.190\pm0.041\pm0.018$
 &  $-0.228\pm0.055\pm0.028$
\\
 &  $\uplambda_{\theta\phi}$
 &  $\pz0.034\pm0.030\pm0.009$
 &  $\pz0.143\pm0.024\pm0.007$
 &  $\pz0.121\pm0.028\pm0.011$
\\
 &  $\uplambda_{\phi}$
 &  $\pz0.023\pm0.013\pm0.005$
 &  $\pz0.017\pm0.015\pm0.005$
 &  $\pz0.056\pm0.015\pm0.005$
\\
 &  $\tilde \uplambda$
 &  $\pz0.093\pm0.097\pm0.040$
 &  $-0.140\pm0.066\pm0.027$
 &  $-0.065\pm0.083\pm0.038$
\\[ 1.5mm]
\hline
\\[-2.5mm]
 \multirow{4}{*}{$8-10$}
 &  $\uplambda_{\theta}$
 &  $\pz0.096\pm0.080\pm0.032$
 &  $-0.124\pm0.039\pm0.017$
 &  $-0.143\pm0.053\pm0.025$
\\
 &  $\uplambda_{\theta\phi}$
 &  $\pz0.057\pm0.042\pm0.015$
 &  $\pz0.054\pm0.033\pm0.010$
 &  $\pz0.020\pm0.039\pm0.016$
\\
 &  $\uplambda_{\phi}$
 &  $\pz0.026\pm0.017\pm0.007$
 &  $\pz0.040\pm0.020\pm0.006$
 &  $\pz0.068\pm0.021\pm0.008$
\\
 &  $\tilde \uplambda$
 &  $\pz0.180\pm0.108\pm0.041$
 &  $-0.005\pm0.079\pm0.024$
 &  $\pz0.066\pm0.097\pm0.042$
\\[ 1.5mm]
\hline
\\[-2.5mm]
 \multirow{4}{*}{$10-15$}
 &  $\uplambda_{\theta}$
 &  $\pz0.041\pm0.048\pm0.027$
 &  $-0.104\pm0.030\pm0.020$
 &  $-0.104\pm0.043\pm0.027$
\\
 &  $\uplambda_{\theta\phi}$
 &  $\pz0.155\pm0.039\pm0.018$
 &  $\pz0.106\pm0.031\pm0.010$
 &  $\pz0.093\pm0.040\pm0.019$
\\
 &  $\uplambda_{\phi}$
 &  $\pz0.043\pm0.018\pm0.009$
 &  $\pz0.058\pm0.021\pm0.012$
 &  $\pz0.072\pm0.025\pm0.011$
\\
 &  $\tilde \uplambda$
 &  $\pz0.177\pm0.081\pm0.036$
 &  $\pz0.073\pm0.075\pm0.039$
 &  $\pz0.122\pm0.096\pm0.044$
\\[ 1.5mm]
\hline
\\[-2.5mm]
 \multirow{4}{*}{$15-20$}
 &  $\uplambda_{\theta}$
 &  $-0.063\pm0.063\pm0.058$
 &  $-0.030\pm0.062\pm0.070$
 &  $-0.261\pm0.074\pm0.080$
\\
 &  $\uplambda_{\theta\phi}$
 &  $-0.008\pm0.059\pm0.033$
 &  $\pz0.159\pm0.054\pm0.036$
 &  $-0.051\pm0.069\pm0.042$
\\
 &  $\uplambda_{\phi}$
 &  $\pz0.010\pm0.042\pm0.042$
 &  $\pz0.020\pm0.052\pm0.039$
 &  $\pz0.093\pm0.064\pm0.061$
\\
 &  $\tilde \uplambda$
 &  $-0.032\pm0.129\pm0.096$
 &  $\pz0.029\pm0.155\pm0.081$
 &  $\pz0.022\pm0.209\pm0.171$
\end{tabular*}
\end{footnotesize}
\end{center} 
\end{table}  
%
%
%
%
%
\begin{table}[t] 
\begin{center} 
\caption{\small Values of
 $\uplambda_{\theta}$, $\uplambda_{\theta\phi}$, $\uplambda_{\phi}$
 and $\tilde \uplambda$
 measured in the GJ frame for the \twos produced
 at $\sqrt{s}=7\,\,{\mathrm{TeV}}$.
 The first uncertainty is statistical and the second systematic.
}\label{tab:Y2S_Results_GJ_Slices_7TeV}
\vspace*{2mm}
\begin{footnotesize}
\begin{tabular*}{0.99\textwidth}{@{\hspace{1mm}}c@{\extracolsep{\fill}}cccc@{\hspace{1mm}}}
 $\pty~\left[\!\gevc\right]$ 
 & $\uplambda$ 
 & $2.2<y<3.0$ 
 & $3.0<y<3.5$ 
 & $3.5<y<4.5$ 
\\[ 1.5mm]
\hline
\\[-2.5mm]
 \multirow{4}{*}{$0-2$}
 &  $\uplambda_{\theta}$
 &  $\pz0.129\pm0.133\pm0.071$
 &  $-0.248\pm0.072\pm0.043$
 &  $-0.209\pm0.103\pm0.068$
\\
 &  $\uplambda_{\theta\phi}$
 &  $\pz0.146\pm0.044\pm0.021$
 &  $-0.018\pm0.027\pm0.009$
 &  $-0.015\pm0.033\pm0.014$
\\
 &  $\uplambda_{\phi}$
 &  $\pz0.012\pm0.020\pm0.007$
 &  $\pz0.007\pm0.021\pm0.007$
 &  $\pz0.006\pm0.021\pm0.007$
\\
 &  $\tilde \uplambda$
 &  $\pz0.168\pm0.156\pm0.085$
 &  $-0.229\pm0.095\pm0.052$
 &  $-0.192\pm0.120\pm0.071$
\\[ 1.5mm]
\hline
\\[-2.5mm]
 \multirow{4}{*}{$2-4$}
 &  $\uplambda_{\theta}$
 &  $\pz0.028\pm0.080\pm0.048$
 &  $-0.069\pm0.057\pm0.033$
 &  $-0.195\pm0.075\pm0.040$
\\
 &  $\uplambda_{\theta\phi}$
 &  $\pz0.156\pm0.041\pm0.025$
 &  $\pz0.028\pm0.024\pm0.010$
 &  $\pz0.016\pm0.025\pm0.011$
\\
 &  $\uplambda_{\phi}$
 &  $\pz0.062\pm0.017\pm0.008$
 &  $\pz0.029\pm0.016\pm0.005$
 &  $\pz0.017\pm0.016\pm0.005$
\\
 &  $\tilde \uplambda$
 &  $\pz0.228\pm0.120\pm0.071$
 &  $\pz0.018\pm0.083\pm0.042$
 &  $-0.147\pm0.091\pm0.048$
\\[ 1.5mm]
\hline
\\[-2.5mm]
 \multirow{4}{*}{$4-6$}
 &  $\uplambda_{\theta}$
 &  $-0.091\pm0.068\pm0.036$
 &  $-0.162\pm0.056\pm0.023$
 &  $-0.184\pm0.078\pm0.035$
\\
 &  $\uplambda_{\theta\phi}$
 &  $\pz0.075\pm0.048\pm0.038$
 &  $-0.024\pm0.028\pm0.010$
 &  $\pz0.006\pm0.030\pm0.016$
\\
 &  $\uplambda_{\phi}$
 &  $\pz0.103\pm0.024\pm0.017$
 &  $\pz0.010\pm0.019\pm0.006$
 &  $\pz0.026\pm0.018\pm0.008$
\\
 &  $\tilde \uplambda$
 &  $\pz0.243\pm0.133\pm0.093$
 &  $-0.133\pm0.086\pm0.034$
 &  $-0.109\pm0.102\pm0.049$
\\[ 1.5mm]
\hline
\\[-2.5mm]
 \multirow{4}{*}{$6-8$}
 &  $\uplambda_{\theta}$
 &  $-0.025\pm0.074\pm0.041$
 &  $-0.296\pm0.059\pm0.025$
 &  $-0.097\pm0.089\pm0.044$
\\
 &  $\uplambda_{\theta\phi}$
 &  $-0.015\pm0.058\pm0.047$
 &  $-0.150\pm0.032\pm0.011$
 &  $-0.003\pm0.038\pm0.015$
\\
 &  $\uplambda_{\phi}$
 &  $\pz0.032\pm0.038\pm0.028$
 &  $\pz0.022\pm0.026\pm0.009$
 &  $\pz0.036\pm0.025\pm0.010$
\\
 &  $\tilde \uplambda$
 &  $\pz0.074\pm0.144\pm0.109$
 &  $-0.235\pm0.096\pm0.033$
 &  $\pz0.010\pm0.125\pm0.058$
\\[ 1.5mm]
\hline
\\[-2.5mm]
 \multirow{4}{*}{$8-10$}
 &  $\uplambda_{\theta}$
 &  $-0.117\pm0.085\pm0.057$
 &  $-0.213\pm0.075\pm0.045$
 &  $-0.148\pm0.109\pm0.063$
\\
 &  $\uplambda_{\theta\phi}$
 &  $-0.009\pm0.057\pm0.043$
 &  $-0.092\pm0.032\pm0.010$
 &  $-0.112\pm0.041\pm0.012$
\\
 &  $\uplambda_{\phi}$
 &  $\pz0.028\pm0.050\pm0.043$
 &  $\pz0.083\pm0.032\pm0.012$
 &  $\pz0.029\pm0.034\pm0.014$
\\
 &  $\tilde \uplambda$
 &  $-0.035\pm0.150\pm0.102$
 &  $\pz0.039\pm0.120\pm0.044$
 &  $-0.064\pm0.141\pm0.058$
\\[ 1.5mm]
\hline
\\[-2.5mm]
 \multirow{4}{*}{$10-15$}
 &  $\uplambda_{\theta}$
 &  $\pz0.020\pm0.087\pm0.099$
 &  $-0.044\pm0.080\pm0.048$
 &  $-0.199\pm0.102\pm0.064$
\\
 &  $\uplambda_{\theta\phi}$
 &  $-0.019\pm0.037\pm0.019$
 &  $-0.042\pm0.031\pm0.010$
 &  $-0.157\pm0.039\pm0.014$
\\
 &  $\uplambda_{\phi}$
 &  $\pz0.006\pm0.045\pm0.055$
 &  $\pz0.069\pm0.032\pm0.012$
 &  $\pz0.030\pm0.034\pm0.016$
\\
 &  $\tilde \uplambda$
 &  $\pz0.038\pm0.116\pm0.081$
 &  $\pz0.175\pm0.123\pm0.053$
 &  $-0.112\pm0.134\pm0.058$
\\[ 1.5mm]
\hline
\\[-2.5mm]
 \multirow{4}{*}{$15-20$}
 &  $\uplambda_{\theta}$
 &  $\pz0.059\pm0.144\pm0.142$
 &  $-0.059\pm0.151\pm0.132$
 &  $-0.174\pm0.220\pm0.173$
\\
 &  $\uplambda_{\theta\phi}$
 &  $-0.074\pm0.065\pm0.042$
 &  $-0.078\pm0.068\pm0.042$
 &  $-0.256\pm0.099\pm0.047$
\\
 &  $\uplambda_{\phi}$
 &  $\pz0.032\pm0.068\pm0.067$
 &  $\pz0.012\pm0.057\pm0.033$
 &  $\pz0.006\pm0.078\pm0.036$
\\
 &  $\tilde \uplambda$
 &  $\pz0.161\pm0.207\pm0.142$
 &  $-0.024\pm0.231\pm0.157$
 &  $-0.157\pm0.315\pm0.194$
\end{tabular*}
\end{footnotesize}
\end{center} 
\end{table}  
%
%
%
%
%
\begin{table}[t] 
\begin{center} 
\caption{\small Values of
 $\uplambda_{\theta}$, $\uplambda_{\theta\phi}$, $\uplambda_{\phi}$
 and $\tilde \uplambda$
 measured in the GJ frame for the \twos produced
 at $\sqrt{s}=8\,\,{\mathrm{TeV}}$.
 The first uncertainty is statistical and the second systematic.
}\label{tab:Y2S_Results_GJ_Slices_8TeV}
\vspace*{2mm}
\begin{footnotesize}
\begin{tabular*}{0.99\textwidth}{@{\hspace{1mm}}c@{\extracolsep{\fill}}cccc@{\hspace{1mm}}}
 $\pty~\left[\!\gevc\right]$ 
 & $\uplambda$ 
 & $2.2<y<3.0$ 
 & $3.0<y<3.5$ 
 & $3.5<y<4.5$
\\[ 1.5mm]
\hline
\\[-2.5mm]
 \multirow{4}{*}{$0-2$}
 &  $\uplambda_{\theta}$
 &  $\pz0.190\pm0.092\pm0.058$
 &  $\pz0.087\pm0.056\pm0.037$
 &  $-0.130\pm0.073\pm0.050$
\\
 &  $\uplambda_{\theta\phi}$
 &  $\pz0.095\pm0.030\pm0.017$
 &  $\pz0.050\pm0.020\pm0.008$
 &  $-0.026\pm0.022\pm0.009$
\\
 &  $\uplambda_{\phi}$
 &  $\pz0.015\pm0.014\pm0.006$
 &  $-0.001\pm0.015\pm0.006$
 &  $\pz0.004\pm0.015\pm0.005$
\\
 &  $\tilde \uplambda$
 &  $\pz0.238\pm0.108\pm0.068$
 &  $\pz0.085\pm0.075\pm0.043$
 &  $-0.119\pm0.085\pm0.053$
\\[ 1.5mm]
\hline
\\[-2.5mm]
 \multirow{4}{*}{$2-4$}
 &  $\uplambda_{\theta}$
 &  $-0.008\pm0.054\pm0.032$
 &  $\pz0.145\pm0.042\pm0.028$
 &  $-0.074\pm0.052\pm0.039$
\\
 &  $\uplambda_{\theta\phi}$
 &  $\pz0.153\pm0.027\pm0.018$
 &  $\pz0.075\pm0.017\pm0.010$
 &  $\pz0.057\pm0.017\pm0.010$
\\
 &  $\uplambda_{\phi}$
 &  $\pz0.040\pm0.012\pm0.006$
 &  $\pz0.022\pm0.011\pm0.005$
 &  $\pz0.010\pm0.011\pm0.004$
\\
 &  $\tilde \uplambda$
 &  $\pz0.117\pm0.077\pm0.049$
 &  $\pz0.216\pm0.061\pm0.040$
 &  $-0.046\pm0.063\pm0.047$
\\[ 1.5mm]
\hline
\\[-2.5mm]
 \multirow{4}{*}{$4-6$}
 &  $\uplambda_{\theta}$
 &  $-0.072\pm0.046\pm0.030$
 &  $-0.179\pm0.037\pm0.018$
 &  $-0.248\pm0.051\pm0.030$
\\
 &  $\uplambda_{\theta\phi}$
 &  $\pz0.107\pm0.032\pm0.027$
 &  $\pz0.037\pm0.019\pm0.010$
 &  $\pz0.011\pm0.021\pm0.013$
\\
 &  $\uplambda_{\phi}$
 &  $\pz0.074\pm0.016\pm0.012$
 &  $\pz0.047\pm0.013\pm0.006$
 &  $\pz0.050\pm0.012\pm0.006$
\\
 &  $\tilde \uplambda$
 &  $\pz0.162\pm0.087\pm0.067$
 &  $-0.038\pm0.062\pm0.031$
 &  $-0.102\pm0.070\pm0.043$
\\[ 1.5mm]
\hline
\\[-2.5mm]
 \multirow{4}{*}{$6-8$}
 &  $\uplambda_{\theta}$
 &  $-0.088\pm0.047\pm0.025$
 &  $-0.282\pm0.039\pm0.017$
 &  $-0.261\pm0.056\pm0.028$
\\
 &  $\uplambda_{\theta\phi}$
 &  $\pz0.021\pm0.038\pm0.036$
 &  $-0.036\pm0.020\pm0.007$
 &  $-0.063\pm0.025\pm0.012$
\\
 &  $\uplambda_{\phi}$
 &  $\pz0.059\pm0.024\pm0.024$
 &  $\pz0.052\pm0.017\pm0.006$
 &  $\pz0.074\pm0.017\pm0.008$
\\
 &  $\tilde \uplambda$
 &  $\pz0.093\pm0.097\pm0.087$
 &  $-0.132\pm0.067\pm0.024$
 &  $-0.041\pm0.084\pm0.041$
\\[ 1.5mm]
\hline
\\[-2.5mm]
 \multirow{4}{*}{$8-10$}
 &  $\uplambda_{\theta}$
 &  $-0.069\pm0.057\pm0.048$
 &  $-0.087\pm0.055\pm0.036$
 &  $-0.046\pm0.074\pm0.048$
\\
 &  $\uplambda_{\theta\phi}$
 &  $\pz0.043\pm0.037\pm0.029$
 &  $-0.072\pm0.023\pm0.008$
 &  $-0.096\pm0.028\pm0.010$
\\
 &  $\uplambda_{\phi}$
 &  $\pz0.079\pm0.032\pm0.030$
 &  $\pz0.029\pm0.023\pm0.010$
 &  $\pz0.037\pm0.023\pm0.010$
\\
 &  $\tilde \uplambda$
 &  $\pz0.185\pm0.109\pm0.077$
 &  $-0.001\pm0.079\pm0.028$
 &  $\pz0.069\pm0.097\pm0.049$
\\[ 1.5mm]
\hline
\\[-2.5mm]
 \multirow{4}{*}{$10-15$}
 &  $\uplambda_{\theta}$
 &  $-0.101\pm0.054\pm0.059$
 &  $-0.122\pm0.050\pm0.041$
 &  $-0.097\pm0.069\pm0.054$
\\
 &  $\uplambda_{\theta\phi}$
 &  $-0.025\pm0.023\pm0.014$
 &  $-0.098\pm0.019\pm0.008$
 &  $-0.107\pm0.026\pm0.012$
\\
 &  $\uplambda_{\phi}$
 &  $\pz0.078\pm0.027\pm0.031$
 &  $\pz0.063\pm0.020\pm0.009$
 &  $\pz0.068\pm0.022\pm0.012$
\\
 &  $\tilde \uplambda$
 &  $\pz0.144\pm0.080\pm0.055$
 &  $\pz0.070\pm0.075\pm0.035$
 &  $\pz0.115\pm0.097\pm0.046$
\\[ 1.5mm]
\hline
\\[-2.5mm]
 \multirow{4}{*}{$15-20$}
 &  $\uplambda_{\theta}$
 &  $\pz0.028\pm0.095\pm0.141$
 &  $-0.174\pm0.093\pm0.091$
 &  $\pz0.226\pm0.168\pm0.172$
\\
 &  $\uplambda_{\theta\phi}$
 &  $-0.021\pm0.044\pm0.035$
 &  $-0.098\pm0.041\pm0.026$
 &  $-0.159\pm0.068\pm0.034$
\\
 &  $\uplambda_{\phi}$
 &  $-0.022\pm0.047\pm0.067$
 &  $\pz0.069\pm0.036\pm0.024$
 &  $-0.068\pm0.052\pm0.029$
\\
 &  $\tilde \uplambda$
 &  $-0.039\pm0.128\pm0.109$
 &  $\pz0.035\pm0.155\pm0.118$
 &  $\pz0.020\pm0.209\pm0.174$
\end{tabular*}
\end{footnotesize}
\end{center} 
\end{table}  

\begin{table}[t] 
\begin{center} 
\caption{\small Values of
 $\uplambda_{\theta}$, $\uplambda_{\theta\phi}$ and $\uplambda_{\phi}$
 measured in the HX, CS and GJ frames for the \twos produced
 at $\sqrt{s}=7\,\,{\mathrm{TeV}}$ in the rapidity range
 $2.2<\yy<4.5$.
 The first uncertainty is statistical and the second systematic.
}\label{tab:Y2S_Results_HX_CS_GJ_2.2y4.5_7TeV}
\vspace*{2mm}
\begin{footnotesize}
  \begin{tabular*}{0.99\textwidth}{@{\hspace{1mm}}c@{\extracolsep{\fill}}cccc@{\hspace{1mm}}}
 $\pty~~\left[\!\gevc\right]$ 
 & 
 & $\uplambda_{\theta}$ 
 & $\uplambda_{\theta\phi}$ 
 & $\uplambda_{\phi}$ 
\\[ 1.5mm]
\hline
\\[-2.5mm]
 \multirow{3}{*}{~ 0$\,-$~ 2}
 &  HX
 &  $-0.160\pm0.052\pm0.039$
 &  $\pz0.007\pm0.018\pm0.007$
 &  $\pz0.001\pm0.012\pm0.004$
\\
 &  CS
 &  $-0.165\pm0.053\pm0.039$
 &  $\pz0.011\pm0.017\pm0.005$
 &  $\pz0.000\pm0.012\pm0.004$
\\
 &  GJ
 &  $-0.176\pm0.051\pm0.037$
 &  $\pz0.013\pm0.018\pm0.007$
 &  $\pz0.004\pm0.012\pm0.004$
\\[ 1.5mm]
\hline
\\[-2.5mm]
 \multirow{3}{*}{~ 2$\,-$~ 4}
 &  HX
 &  $-0.045\pm0.037\pm0.025$
 &  $\pz0.016\pm0.016\pm0.009$
 &  $\pz0.009\pm0.009\pm0.004$
\\
 &  CS
 &  $-0.051\pm0.041\pm0.027$
 &  $\pz0.035\pm0.014\pm0.005$
 &  $\pz0.011\pm0.009\pm0.003$
\\
 &  GJ
 &  $-0.104\pm0.037\pm0.022$
 &  $\pz0.040\pm0.015\pm0.009$
 &  $\pz0.027\pm0.009\pm0.004$
\\[ 1.5mm]
\hline
\\[-2.5mm]
 \multirow{3}{*}{~ 4$\,-$~ 6}
 &  HX
 &  $-0.069\pm0.034\pm0.019$
 &  $\pz0.064\pm0.021\pm0.009$
 &  $\pz0.008\pm0.012\pm0.006$
\\
 &  CS
 &  $-0.106\pm0.042\pm0.020$
 &  $\pz0.040\pm0.016\pm0.006$
 &  $\pz0.024\pm0.009\pm0.004$
\\
 &  GJ
 &  $-0.147\pm0.037\pm0.019$
 &  $-0.004\pm0.018\pm0.010$
 &  $\pz0.039\pm0.011\pm0.007$
\\[ 1.5mm]
\hline
\\[-2.5mm]
 \multirow{3}{*}{~ 6$\,-$~ 8}
 &  HX
 &  $-0.050\pm0.037\pm0.021$
 &  $\pz0.131\pm0.024\pm0.013$
 &  $-0.022\pm0.019\pm0.010$
\\
 &  CS
 &  $-0.193\pm0.042\pm0.016$
 &  $\pz0.047\pm0.022\pm0.007$
 &  $\pz0.032\pm0.012\pm0.004$
\\
 &  GJ
 &  $-0.170\pm0.040\pm0.024$
 &  $-0.077\pm0.021\pm0.009$
 &  $\pz0.026\pm0.015\pm0.008$
\\[ 1.5mm]
\hline
\\[-2.5mm]
 \multirow{3}{*}{~ 8$\,-$~10}
 &  HX
 &  $\pz0.089\pm0.050\pm0.028$
 &  $\pz0.114\pm0.025\pm0.010$
 &  $-0.051\pm0.026\pm0.015$
\\
 &  CS
 &  $-0.153\pm0.041\pm0.016$
 &  $\pz0.085\pm0.030\pm0.010$
 &  $\pz0.032\pm0.016\pm0.007$
\\
 &  GJ
 &  $-0.170\pm0.048\pm0.032$
 &  $-0.077\pm0.022\pm0.008$
 &  $\pz0.038\pm0.020\pm0.010$
\\[ 1.5mm]
\hline
\\[-2.5mm]
 \multirow{3}{*}{~10$\,-$~15}
 &  HX
 &  $\pz0.139\pm0.049\pm0.040$
 &  $\pz0.036\pm0.021\pm0.009$
 &  $-0.025\pm0.024\pm0.016$
\\
 &  CS
 &  $-0.057\pm0.033\pm0.018$
 &  $\pz0.063\pm0.029\pm0.012$
 &  $\pz0.039\pm0.018\pm0.009$
\\
 &  GJ
 &  $-0.051\pm0.048\pm0.037$
 &  $-0.060\pm0.020\pm0.007$
 &  $\pz0.037\pm0.019\pm0.012$
\\[ 1.5mm]
\hline
\\[-2.5mm]
 \multirow{3}{*}{~15$\,-$~20}
 &  HX
 &  $\pz0.155\pm0.088\pm0.079$
 &  $\pz0.043\pm0.050\pm0.029$
 &  $-0.036\pm0.042\pm0.034$
\\
 &  CS
 &  $-0.110\pm0.056\pm0.055$
 &  $\pz0.062\pm0.050\pm0.027$
 &  $\pz0.053\pm0.044\pm0.043$
\\
 &  GJ
 &  $-0.013\pm0.091\pm0.070$
 &  $-0.111\pm0.042\pm0.028$
 &  $\pz0.023\pm0.036\pm0.021$
\\[ 1.5mm]
\hline
\\[-2.5mm]
 \multirow{3}{*}{~20$\,-$~30}
 &  HX
 &  $\pz0.267\pm0.164\pm0.149$
 &  $\pz0.002\pm0.091\pm0.081$
 &  $\pz0.060\pm0.059\pm0.048$
\\
 &  CS
 &  $-0.001\pm0.107\pm0.155$
 &  $\pz0.094\pm0.078\pm0.058$
 &  $\pz0.139\pm0.079\pm0.107$
\\
 &  GJ
 &  $\pz0.051\pm0.156\pm0.214$
 &  $-0.138\pm0.077\pm0.064$
 &  $\pz0.127\pm0.056\pm0.046$
\end{tabular*}
\end{footnotesize}
\end{center} 
\end{table}  
\begin{table}[t]
\begin{center}
\caption{\small Values of
 $\uplambda_{\theta}$, $\uplambda_{\theta\phi}$ and $\uplambda_{\phi}$
 measured in the HX, CS and GJ frames for the \twos produced
 at $\sqrt{s}=8\,\,{\mathrm{TeV}}$ in the rapidity range
 $2.2<\yy<4.5$.
 The first uncertainty is statistical and the second systematic.
}\label{tab:Y2S_Results_HX_CS_GJ_2.2y4.5_8TeV}
\vspace*{2mm}
\begin{footnotesize}
\begin{tabular*}{0.99\textwidth}{@{\hspace{1mm}}c@{\extracolsep{\fill}}cccc@{\hspace{1mm}}}
 $\pty~~\left[\!\gevc\right]$ 
 & 
 & $\uplambda_{\theta}$ 
 & $\uplambda_{\theta\phi}$ 
 & $\uplambda_{\phi}$ 
\\[ 1.5mm]
\hline
\\[-2.5mm]
 \multirow{3}{*}{~ 0$\,-$~ 2}
 &  HX
 &  $\pz0.049\pm0.038\pm0.031$
 &  $-0.023\pm0.013\pm0.005$
 &  $\pz0.002\pm0.008\pm0.003$
\\
 &  CS
 &  $\pz0.051\pm0.039\pm0.030$
 &  $\pz0.005\pm0.012\pm0.004$
 &  $\pz0.000\pm0.008\pm0.003$
\\
 &  GJ
 &  $\pz0.031\pm0.038\pm0.030$
 &  $\pz0.031\pm0.013\pm0.007$
 &  $\pz0.004\pm0.008\pm0.003$
\\[ 1.5mm]
\hline
\\[-2.5mm]
 \multirow{3}{*}{~ 2$\,-$~ 4}
 &  HX
 &  $\pz0.058\pm0.025\pm0.024$
 &  $-0.023\pm0.011\pm0.007$
 &  $-0.001\pm0.006\pm0.003$
\\
 &  CS
 &  $\pz0.080\pm0.029\pm0.024$
 &  $\pz0.033\pm0.009\pm0.005$
 &  $-0.006\pm0.006\pm0.002$
\\
 &  GJ
 &  $\pz0.011\pm0.026\pm0.022$
 &  $\pz0.079\pm0.011\pm0.009$
 &  $\pz0.016\pm0.006\pm0.004$
\\[ 1.5mm]
\hline
\\[-2.5mm]
 \multirow{3}{*}{~ 4$\,-$~ 6}
 &  HX
 &  $\pz0.010\pm0.024\pm0.018$
 &  $\pz0.070\pm0.014\pm0.008$
 &  $-0.012\pm0.008\pm0.005$
\\
 &  CS
 &  $-0.068\pm0.029\pm0.021$
 &  $\pz0.080\pm0.011\pm0.007$
 &  $\pz0.016\pm0.006\pm0.003$
\\
 &  GJ
 &  $-0.168\pm0.024\pm0.015$
 &  $\pz0.034\pm0.012\pm0.009$
 &  $\pz0.049\pm0.007\pm0.006$
\\[ 1.5mm]
\hline
\\[-2.5mm]
 \multirow{3}{*}{~ 6$\,-$~ 8}
 &  HX
 &  $\pz0.044\pm0.026\pm0.019$
 &  $\pz0.125\pm0.017\pm0.008$
 &  $-0.034\pm0.013\pm0.007$
\\
 &  CS
 &  $-0.145\pm0.029\pm0.014$
 &  $\pz0.095\pm0.014\pm0.008$
 &  $\pz0.033\pm0.008\pm0.004$
\\
 &  GJ
 &  $-0.210\pm0.026\pm0.017$
 &  $-0.037\pm0.014\pm0.007$
 &  $\pz0.056\pm0.010\pm0.008$
\\[ 1.5mm]
\hline
\\[-2.5mm]
 \multirow{3}{*}{~ 8$\,-$~10}
 &  HX
 &  $\pz0.039\pm0.032\pm0.024$
 &  $\pz0.080\pm0.016\pm0.008$
 &  $-0.008\pm0.017\pm0.010$
\\
 &  CS
 &  $-0.096\pm0.028\pm0.013$
 &  $\pz0.037\pm0.020\pm0.009$
 &  $\pz0.039\pm0.011\pm0.006$
\\
 &  GJ
 &  $-0.072\pm0.034\pm0.024$
 &  $-0.058\pm0.015\pm0.006$
 &  $\pz0.031\pm0.013\pm0.009$
\\[ 1.5mm]
\hline
\\[-2.5mm]
 \multirow{3}{*}{~10$\,-$~15}
 &  HX
 &  $\pz0.224\pm0.034\pm0.035$
 &  $\pz0.048\pm0.014\pm0.008$
 &  $-0.044\pm0.017\pm0.013$
\\
 &  CS
 &  $-0.076\pm0.021\pm0.013$
 &  $\pz0.105\pm0.019\pm0.010$
 &  $\pz0.052\pm0.012\pm0.008$
\\
 &  GJ
 &  $-0.101\pm0.031\pm0.028$
 &  $-0.079\pm0.013\pm0.008$
 &  $\pz0.060\pm0.012\pm0.010$
\\[ 1.5mm]
\hline
\\[-2.5mm]
 \multirow{3}{*}{~15$\,-$~20}
 &  HX
 &  $\pz0.119\pm0.056\pm0.063$
 &  $\pz0.034\pm0.032\pm0.024$
 &  $-0.036\pm0.027\pm0.027$
\\
 &  CS
 &  $-0.090\pm0.037\pm0.039$
 &  $\pz0.050\pm0.033\pm0.017$
 &  $\pz0.032\pm0.029\pm0.029$
\\
 &  GJ
 &  $-0.025\pm0.059\pm0.056$
 &  $-0.079\pm0.027\pm0.018$
 &  $\pz0.010\pm0.023\pm0.017$
\\[ 1.5mm]
\hline
\\[-2.5mm]
 \multirow{3}{*}{~20$\,-$~30}
 &  HX
 &  $\pz0.043\pm0.092\pm0.118$
 &  $\pz0.060\pm0.054\pm0.047$
 &  $\pz0.013\pm0.037\pm0.027$
\\
 &  CS
 &  $-0.057\pm0.064\pm0.111$
 &  $-0.012\pm0.049\pm0.042$
 &  $\pz0.045\pm0.052\pm0.091$
\\
 &  GJ
 &  $\pz0.068\pm0.104\pm0.150$
 &  $-0.042\pm0.048\pm0.044$
 &  $\pz0.004\pm0.037\pm0.034$
\end{tabular*}
\end{footnotesize} 
\end{center} 
\end{table}  
\begin{table}[t]
\begin{center}
\caption{\small Values of
 $\tilde \uplambda$
 measured in the HX, CS and GJ frames for the \twos produced
 at $\sqrt{s}=7$ and $8\,\,{\mathrm{TeV}}$ in the rapidity range
 $2.2<\yy<4.5$.
 The first uncertainty is statistical and the second systematic.
}\label{tab:Y2S_lamT_Results_HX_CS_GJ_2.2y4.5}
\vspace*{7mm}
\begin{footnotesize}
\begin{tabular*}{0.90\textwidth}{@{\hspace{1mm}}c@{\extracolsep{\fill}}ccc@{\hspace{1mm}}}
 $\pty~~~\left[\!\gevc\right]$ & $\tilde \uplambda$ & $\sqrt{s}=7\tev$ & $\sqrt{s}=8\tev$ 
\\[ 1.5mm]
\hline
\\[-2.5mm]
 \multirow{3}{*}{~ 0$\,-$~ 2}
 &  HX
 &  $-0.158\pm0.063\pm0.042$
 &  $\pz0.054\pm0.046\pm0.034$
\\
 &  CS
 &  $-0.164\pm0.063\pm0.042$
 &  $\pz0.050\pm0.047\pm0.033$
\\
 &  GJ
 &  $-0.164\pm0.063\pm0.038$
 &  $\pz0.042\pm0.046\pm0.033$
\\[ 1.5mm]
\hline
\\[-2.5mm]
 \multirow{3}{*}{~ 2$\,-$~ 4}
 &  HX
 &  $-0.018\pm0.050\pm0.033$
 &  $\pz0.055\pm0.034\pm0.030$
\\
 &  CS
 &  $-0.018\pm0.050\pm0.031$
 &  $\pz0.063\pm0.035\pm0.029$
\\
 &  GJ
 &  $-0.022\pm0.050\pm0.031$
 &  $\pz0.060\pm0.034\pm0.030$
\\[ 1.5mm]
\hline
\\[-2.5mm]
 \multirow{3}{*}{~ 4$\,-$~ 6}
 &  HX
 &  $-0.046\pm0.054\pm0.031$
 &  $-0.026\pm0.037\pm0.026$
\\
 &  CS
 &  $-0.036\pm0.055\pm0.028$
 &  $-0.020\pm0.037\pm0.028$
\\
 &  GJ
 &  $-0.033\pm0.055\pm0.032$
 &  $-0.023\pm0.037\pm0.029$
\\[ 1.5mm]
\hline
\\[-2.5mm]
 \multirow{3}{*}{~ 6$\,-$~ 8}
 &  HX
 &  $-0.112\pm0.061\pm0.030$
 &  $-0.056\pm0.042\pm0.023$
\\
 &  CS
 &  $-0.100\pm0.061\pm0.024$
 &  $-0.048\pm0.042\pm0.022$
\\
 &  GJ
 &  $-0.093\pm0.062\pm0.027$
 &  $-0.045\pm0.042\pm0.024$
\\[ 1.5mm]
\hline
\\[-2.5mm]
 \multirow{3}{*}{~ 8$\,-$~10}
 &  HX
 &  $-0.061\pm0.069\pm0.026$
 &  $\pz0.016\pm0.047\pm0.023$
\\
 &  CS
 &  $-0.059\pm0.069\pm0.026$
 &  $\pz0.021\pm0.048\pm0.024$
\\
 &  GJ
 &  $-0.060\pm0.069\pm0.029$
 &  $\pz0.023\pm0.048\pm0.025$
\\[ 1.5mm]
\hline
\\[-2.5mm]
 \multirow{3}{*}{~10$\,-$~15}
 &  HX
 &  $\pz0.063\pm0.067\pm0.028$
 &  $\pz0.087\pm0.044\pm0.026$
\\
 &  CS
 &  $\pz0.062\pm0.067\pm0.031$
 &  $\pz0.085\pm0.044\pm0.025$
\\
 &  GJ
 &  $\pz0.063\pm0.067\pm0.028$
 &  $\pz0.085\pm0.045\pm0.023$
\\[ 1.5mm]
\hline
\\[-2.5mm]
 \multirow{3}{*}{~15$\,-$~20}
 &  HX
 &  $\pz0.046\pm0.135\pm0.085$
 &  $\pz0.009\pm0.087\pm0.061$
\\
 &  CS
 &  $\pz0.051\pm0.136\pm0.101$
 &  $\pz0.007\pm0.087\pm0.065$
\\
 &  GJ
 &  $\pz0.056\pm0.136\pm0.077$
 &  $\pz0.005\pm0.087\pm0.061$
\\[ 1.5mm]
\hline
\\[-2.5mm]
 \multirow{3}{*}{~20$\,-$~30}
 &  HX
 &  $\pz0.474\pm0.286\pm0.245$
 &  $\pz0.083\pm0.152\pm0.145$
\\
 &  CS
 &  $\pz0.484\pm0.287\pm0.323$
 &  $\pz0.082\pm0.153\pm0.225$
\\
 &  GJ
 &  $\pz0.495\pm0.288\pm0.248$
 &  $\pz0.081\pm0.152\pm0.176$
\end{tabular*}
\end{footnotesize}
\end{center} 
\end{table}  

\clearpage 
{\boldmath{\section{Polarization results for the \threes~state}\label{sec:upsthreespol}}}

Values of the polarization parameters $\pmb{\uplambda}$ and $\tilde\uplambda$
for the \threes~meson
are presented 
in Tables~\ref{tab:Y3S_Results_HX_Slices_7TeV} and~\ref{tab:Y3S_Results_HX_Slices_8TeV}
for the~HX~frame,
in Tables~\ref{tab:Y3S_Results_CS_Slices_7TeV} and~\ref{tab:Y3S_Results_CS_Slices_8TeV}
for the~CS~frame and
in Tables~\ref{tab:Y3S_Results_GJ_Slices_7TeV} and~\ref{tab:Y3S_Results_GJ_Slices_8TeV}
for the~GJ~frame for $\sqs=7$~and~8\tev, respectively.
The~polarization parameters $\pmb{\uplambda}$ 
measured in the~wide rapidity bin $2.2<\yy<4.5$ are presented 
in Tables~\ref{tab:Y3S_Results_HX_CS_GJ_2.2y4.5_7TeV} 
      and~\ref{tab:Y3S_Results_HX_CS_GJ_2.2y4.5_8TeV}, while 
the~parameters $\tilde\uplambda$ are listed 
in Table~\ref{tab:Y3S_lamT_Results_HX_CS_GJ_2.2y4.5}.
 
\begin{table}[htb] 
\begin{center} 
\caption{\small Values of
  $\uplambda_{\theta}$, $\uplambda_{\theta\phi}$, $\uplambda_{\phi}$
  and $\tilde \uplambda$
  measured in the HX frame for the \threes produced
  at $\sqrt{s}=7\,\,{\mathrm{TeV}}$.
  The first uncertainty is statistical and the second systematic.
}\label{tab:Y3S_Results_HX_Slices_7TeV}
\vspace*{2mm}
\begin{footnotesize}
\begin{tabular*}{0.99\textwidth}{@{\hspace{1mm}}c@{\extracolsep{\fill}}cccc@{\hspace{1mm}}}
 $\pty~\left[\!\gevc\right]$ 
 & $\uplambda$ 
 & $2.2<y<3.0$
 & $3.0<y<3.5$ 
 & $3.5<y<4.5$ 
\\[ 1.5mm]
\hline
\\[-2.5mm]
 \multirow{4}{*}{$0-2$}
 &  $\uplambda_{\theta}$
 &  $\pz0.479\pm0.259\pm0.104$
 &  $-0.052\pm0.125\pm0.074$
 &  $-0.277\pm0.154\pm0.101$
\\
 &  $\uplambda_{\theta\phi}$
 &  $-0.026\pm0.064\pm0.015$
 &  $\pz0.068\pm0.043\pm0.012$
 &  $-0.047\pm0.057\pm0.020$
\\
 &  $\uplambda_{\phi}$
 &  $-0.024\pm0.034\pm0.007$
 &  $-0.050\pm0.035\pm0.008$
 &  $\pz0.006\pm0.035\pm0.008$
\\
 &  $\tilde \uplambda$
 &  $\pz0.396\pm0.276\pm0.108$
 &  $-0.193\pm0.151\pm0.074$
 &  $-0.261\pm0.188\pm0.106$
\\[ 1.5mm]
\hline
\\[-2.5mm]
 \multirow{4}{*}{$2-4$}
 &  $\uplambda_{\theta}$
 &  $\pz0.051\pm0.165\pm0.065$
 &  $\pz0.021\pm0.095\pm0.058$
 &  $-0.451\pm0.089\pm0.058$
\\
 &  $\uplambda_{\theta\phi}$
 &  $-0.098\pm0.048\pm0.017$
 &  $\pz0.056\pm0.040\pm0.015$
 &  $\pz0.129\pm0.054\pm0.026$
\\
 &  $\uplambda_{\phi}$
 &  $\pz0.029\pm0.023\pm0.005$
 &  $\pz0.062\pm0.027\pm0.007$
 &  $\pz0.043\pm0.028\pm0.010$
\\
 &  $\tilde \uplambda$
 &  $\pz0.141\pm0.192\pm0.075$
 &  $\pz0.218\pm0.148\pm0.079$
 &  $-0.338\pm0.134\pm0.083$
\\[ 1.5mm]
\hline
\\[-2.5mm]
 \multirow{4}{*}{$4-6$}
 &  $\uplambda_{\theta}$
 &  $\pz0.111\pm0.150\pm0.059$
 &  $-0.039\pm0.085\pm0.045$
 &  $-0.030\pm0.089\pm0.061$
\\
 &  $\uplambda_{\theta\phi}$
 &  $-0.072\pm0.060\pm0.024$
 &  $\pz0.062\pm0.048\pm0.018$
 &  $\pz0.088\pm0.068\pm0.038$
\\
 &  $\uplambda_{\phi}$
 &  $\pz0.061\pm0.027\pm0.009$
 &  $\pz0.029\pm0.032\pm0.010$
 &  $\pz0.022\pm0.040\pm0.019$
\\
 &  $\tilde \uplambda$
 &  $\pz0.314\pm0.204\pm0.084$
 &  $\pz0.048\pm0.146\pm0.070$
 &  $\pz0.038\pm0.167\pm0.115$
\\[ 1.5mm]
\hline
\\[-2.5mm]
 \multirow{4}{*}{$6-8$}
 &  $\uplambda_{\theta}$
 &  $-0.244\pm0.133\pm0.050$
 &  $\pz0.011\pm0.090\pm0.036$
 &  $-0.035\pm0.090\pm0.049$
\\
 &  $\uplambda_{\theta\phi}$
 &  $\pz0.029\pm0.068\pm0.034$
 &  $\pz0.111\pm0.053\pm0.019$
 &  $\pz0.238\pm0.079\pm0.045$
\\
 &  $\uplambda_{\phi}$
 &  $\pz0.100\pm0.037\pm0.017$
 &  $\pz0.031\pm0.045\pm0.019$
 &  $-0.041\pm0.060\pm0.036$
\\
 &  $\tilde \uplambda$
 &  $\pz0.063\pm0.209\pm0.089$
 &  $\pz0.107\pm0.167\pm0.070$
 &  $-0.151\pm0.175\pm0.094$
\\[ 1.5mm]
\hline
\\[-2.5mm]
 \multirow{4}{*}{$8-10$}
 &  $\uplambda_{\theta}$
 &  $-0.087\pm0.144\pm0.070$
 &  $\pz0.126\pm0.124\pm0.070$
 &  $\pz0.015\pm0.122\pm0.087$
\\
 &  $\uplambda_{\theta\phi}$
 &  $-0.068\pm0.074\pm0.039$
 &  $\pz0.184\pm0.053\pm0.014$
 &  $\pz0.209\pm0.077\pm0.040$
\\
 &  $\uplambda_{\phi}$
 &  $\pz0.124\pm0.050\pm0.026$
 &  $-0.071\pm0.063\pm0.028$
 &  $-0.050\pm0.084\pm0.060$
\\
 &  $\tilde \uplambda$
 &  $\pz0.326\pm0.242\pm0.116$
 &  $-0.080\pm0.163\pm0.046$
 &  $-0.129\pm0.198\pm0.099$
\\[ 1.5mm]
\hline
\\[-2.5mm]
 \multirow{4}{*}{$10-15$}
 &  $\uplambda_{\theta}$
 &  $\pz0.192\pm0.134\pm0.109$
 &  $\pz0.322\pm0.121\pm0.103$
 &  $\pz0.053\pm0.118\pm0.121$
\\
 &  $\uplambda_{\theta\phi}$
 &  $\pz0.126\pm0.056\pm0.026$
 &  $\pz0.091\pm0.046\pm0.017$
 &  $\pz0.106\pm0.055\pm0.019$
\\
 &  $\uplambda_{\phi}$
 &  $\pz0.081\pm0.045\pm0.028$
 &  $-0.147\pm0.062\pm0.039$
 &  $-0.026\pm0.076\pm0.068$
\\
 &  $\tilde \uplambda$
 &  $\pz0.474\pm0.181\pm0.077$
 &  $-0.103\pm0.138\pm0.043$
 &  $-0.025\pm0.185\pm0.099$
\\[ 1.5mm]
\hline
\\[-2.5mm]
 \multirow{4}{*}{$15-20$}
 &  $\uplambda_{\theta}$
 &  $\pz0.181\pm0.203\pm0.198$
 &  $\pz0.156\pm0.161\pm0.125$
 &  $\pz0.521\pm0.292\pm0.253$
\\
 &  $\uplambda_{\theta\phi}$
 &  $\pz0.084\pm0.093\pm0.053$
 &  $\pz0.121\pm0.093\pm0.060$
 &  $\pz0.265\pm0.147\pm0.080$
\\
 &  $\uplambda_{\phi}$
 &  $\pz0.137\pm0.067\pm0.047$
 &  $\pz0.086\pm0.082\pm0.052$
 &  $\pz0.178\pm0.122\pm0.087$
\\
 &  $\tilde \uplambda$
 &  $\pz0.685\pm0.335\pm0.204$
 &  $\pz0.453\pm0.333\pm0.171$
 &  $\pz1.282\pm0.644\pm0.294$
\end{tabular*}
\end{footnotesize} 
\end{center} 
\end{table}  
%
%
%
%
%
\begin{table}[t] 
\begin{center} 
\caption{\small Values of
 $\uplambda_{\theta}$, $\uplambda_{\theta\phi}$, $\uplambda_{\phi}$
 and $\tilde \uplambda$
 measured in the HX frame for the \threes produced
 at $\sqrt{s}=8\,\,{\mathrm{TeV}}$.
 The first uncertainty is statistical and the second systematic.
}\label{tab:Y3S_Results_HX_Slices_8TeV}
\vspace*{2mm}
\begin{footnotesize}
\begin{tabular*}{0.99\textwidth}{@{\hspace{1mm}}c@{\extracolsep{\fill}}cccc@{\hspace{1mm}}}
 $\pty~\left[\!\gevc\right]$ 
 & $\uplambda$ 
 & $2.2<y<3.0$ 
 & $3.0<y<3.5$ 
 & $3.5<y<4.5$ 
\\[ 1.5mm]
\hline
\\[-2.5mm]
 \multirow{4}{*}{$0-2$}
 &  $\uplambda_{\theta}$
 &  $\pz0.212\pm0.165\pm0.083$
 &  $-0.073\pm0.089\pm0.078$
 &  $-0.494\pm0.101\pm0.080$
\\
 &  $\uplambda_{\theta\phi}$
 &  $-0.060\pm0.042\pm0.011$
 &  $\pz0.015\pm0.031\pm0.009$
 &  $\pz0.021\pm0.039\pm0.015$
\\
 &  $\uplambda_{\phi}$
 &  $-0.004\pm0.022\pm0.005$
 &  $-0.029\pm0.025\pm0.006$
 &  $-0.019\pm0.024\pm0.006$
\\
 &  $\tilde \uplambda$
 &  $\pz0.198\pm0.180\pm0.086$
 &  $-0.155\pm0.112\pm0.080$
 &  $-0.540\pm0.119\pm0.081$
\\[ 1.5mm]
\hline
\\[-2.5mm]
 \multirow{4}{*}{$2-4$}
 &  $\uplambda_{\theta}$
 &  $\pz0.233\pm0.113\pm0.045$
 &  $-0.008\pm0.063\pm0.057$
 &  $-0.147\pm0.068\pm0.072$
\\
 &  $\uplambda_{\theta\phi}$
 &  $-0.108\pm0.033\pm0.011$
 &  $\pz0.016\pm0.028\pm0.013$
 &  $\pz0.014\pm0.038\pm0.024$
\\
 &  $\uplambda_{\phi}$
 &  $\pz0.040\pm0.016\pm0.005$
 &  $\pz0.008\pm0.019\pm0.006$
 &  $\pz0.027\pm0.020\pm0.009$
\\
 &  $\tilde \uplambda$
 &  $\pz0.367\pm0.135\pm0.055$
 &  $\pz0.015\pm0.091\pm0.071$
 &  $-0.068\pm0.101\pm0.092$
\\[ 1.5mm]
\hline
\\[-2.5mm]
 \multirow{4}{*}{$4-6$}
 &  $\uplambda_{\theta}$
 &  $\pz0.106\pm0.102\pm0.042$
 &  $\pz0.096\pm0.061\pm0.045$
 &  $-0.183\pm0.062\pm0.057$
\\
 &  $\uplambda_{\theta\phi}$
 &  $-0.078\pm0.041\pm0.017$
 &  $\pz0.059\pm0.033\pm0.020$
 &  $\pz0.267\pm0.050\pm0.034$
\\
 &  $\uplambda_{\phi}$
 &  $\pz0.075\pm0.019\pm0.008$
 &  $\pz0.037\pm0.022\pm0.011$
 &  $-0.047\pm0.028\pm0.017$
\\
 &  $\tilde \uplambda$
 &  $\pz0.357\pm0.141\pm0.064$
 &  $\pz0.215\pm0.105\pm0.077$
 &  $-0.309\pm0.104\pm0.091$
\\[ 1.5mm]
\hline
\\[-2.5mm]
 \multirow{4}{*}{$6-8$}
 &  $\uplambda_{\theta}$
 &  $-0.150\pm0.095\pm0.042$
 &  $-0.019\pm0.062\pm0.033$
 &  $-0.088\pm0.062\pm0.044$
\\
 &  $\uplambda_{\theta\phi}$
 &  $\pz0.024\pm0.048\pm0.023$
 &  $\pz0.273\pm0.038\pm0.017$
 &  $\pz0.406\pm0.065\pm0.042$
\\
 &  $\uplambda_{\phi}$
 &  $\pz0.080\pm0.026\pm0.012$
 &  $-0.068\pm0.032\pm0.016$
 &  $-0.146\pm0.047\pm0.029$
\\
 &  $\tilde \uplambda$
 &  $\pz0.097\pm0.145\pm0.067$
 &  $-0.210\pm0.098\pm0.058$
 &  $-0.459\pm0.116\pm0.080$
\\[ 1.5mm]
\hline
\\[-2.5mm]
 \multirow{4}{*}{$8-10$}
 &  $\uplambda_{\theta}$
 &  $-0.100\pm0.103\pm0.057$
 &  $\pz0.109\pm0.078\pm0.052$
 &  $\pz0.064\pm0.081\pm0.063$
\\
 &  $\uplambda_{\theta\phi}$
 &  $\pz0.096\pm0.052\pm0.030$
 &  $\pz0.193\pm0.035\pm0.014$
 &  $\pz0.175\pm0.050\pm0.029$
\\
 &  $\uplambda_{\phi}$
 &  $\pz0.092\pm0.036\pm0.019$
 &  $-0.017\pm0.041\pm0.023$
 &  $-0.049\pm0.055\pm0.035$
\\
 &  $\tilde \uplambda$
 &  $\pz0.194\pm0.159\pm0.088$
 &  $\pz0.056\pm0.116\pm0.042$
 &  $-0.079\pm0.132\pm0.079$
\\[ 1.5mm]
\hline
\\[-2.5mm]
 \multirow{4}{*}{$10-15$}
 &  $\uplambda_{\theta}$
 &  $\pz0.222\pm0.093\pm0.067$
 &  $\pz0.254\pm0.075\pm0.067$
 &  $\pz0.186\pm0.086\pm0.072$
\\
 &  $\uplambda_{\theta\phi}$
 &  $\pz0.054\pm0.038\pm0.021$
 &  $\pz0.129\pm0.029\pm0.012$
 &  $\pz0.142\pm0.037\pm0.015$
\\
 &  $\uplambda_{\phi}$
 &  $-0.017\pm0.033\pm0.022$
 &  $-0.057\pm0.037\pm0.025$
 &  $-0.091\pm0.051\pm0.033$
\\
 &  $\tilde \uplambda$
 &  $\pz0.168\pm0.108\pm0.056$
 &  $\pz0.080\pm0.099\pm0.046$
 &  $-0.079\pm0.112\pm0.056$
\\[ 1.5mm]
\hline
\\[-2.5mm]
 \multirow{4}{*}{$15-20$}
 &  $\uplambda_{\theta}$
 &  $\pz0.512\pm0.153\pm0.143$
 &  $\pz0.238\pm0.120\pm0.128$
 &  $\pz0.313\pm0.176\pm0.160$
\\
 &  $\uplambda_{\theta\phi}$
 &  $-0.064\pm0.073\pm0.049$
 &  $\pz0.032\pm0.067\pm0.056$
 &  $\pz0.176\pm0.096\pm0.071$
\\
 &  $\uplambda_{\phi}$
 &  $-0.123\pm0.054\pm0.045$
 &  $-0.061\pm0.060\pm0.058$
 &  $-0.203\pm0.097\pm0.093$
\\
 &  $\tilde \uplambda$
 &  $\pz0.126\pm0.168\pm0.093$
 &  $\pz0.052\pm0.184\pm0.135$
 &  $-0.246\pm0.223\pm0.178$
\end{tabular*}
\end{footnotesize}
\end{center} 
\end{table}  
%
%
%
%
%
\begin{table}[t] 
\begin{center} 
\caption{\small Values of
 $\uplambda_{\theta}$, $\uplambda_{\theta\phi}$, $\uplambda_{\phi}$
 and $\tilde \uplambda$
 measured in the CS frame for the \threes produced
 at $\sqrt{s}=7\,\,{\mathrm{TeV}}$.
 The first uncertainty is statistical and the second systematic.
}\label{tab:Y3S_Results_CS_Slices_7TeV}
\vspace*{2mm}
\begin{footnotesize}
\begin{tabular*}{0.99\textwidth}{@{\hspace{1mm}}c@{\extracolsep{\fill}}cccc@{\hspace{1mm}}}
 $\pty~\left[\!\gevc\right]$ 
 & $\uplambda$ 
 & 2.2~$<y<$~3.0 
 & 3.0~$<y<$~3.5 
 & 3.5~$<y<$~4.5 
\\[ 1.5mm]
\hline
\\[-2.5mm]
 \multirow{4}{*}{$0-2$}
 &  $\uplambda_{\theta}$
 &  $\pz0.457\pm0.257\pm0.104$
 &  $-0.065\pm0.125\pm0.072$
 &  $-0.189\pm0.167\pm0.098$
\\
 &  $\uplambda_{\theta\phi}$
 &  $\pz0.089\pm0.065\pm0.014$
 &  $\pz0.081\pm0.042\pm0.011$
 &  $-0.047\pm0.053\pm0.016$
\\
 &  $\uplambda_{\phi}$
 &  $-0.020\pm0.033\pm0.008$
 &  $-0.042\pm0.035\pm0.009$
 &  $-0.008\pm0.034\pm0.009$
\\
 &  $\tilde \uplambda$
 &  $\pz0.388\pm0.277\pm0.102$
 &  $-0.185\pm0.152\pm0.071$
 &  $-0.210\pm0.192\pm0.102$
\\[ 1.5mm]
\hline
\\[-2.5mm]
 \multirow{4}{*}{$2-4$}
 &  $\uplambda_{\theta}$
 &  $\pz0.115\pm0.173\pm0.065$
 &  $\pz0.023\pm0.102\pm0.060$
 &  $-0.344\pm0.111\pm0.063$
\\
 &  $\uplambda_{\theta\phi}$
 &  $\pz0.029\pm0.047\pm0.012$
 &  $\pz0.064\pm0.035\pm0.007$
 &  $\pz0.010\pm0.041\pm0.013$
\\
 &  $\uplambda_{\phi}$
 &  $\pz0.015\pm0.023\pm0.005$
 &  $\pz0.072\pm0.026\pm0.006$
 &  $\pz0.045\pm0.025\pm0.007$
\\
 &  $\tilde \uplambda$
 &  $\pz0.161\pm0.195\pm0.070$
 &  $\pz0.257\pm0.151\pm0.079$
 &  $-0.218\pm0.140\pm0.080$
\\[ 1.5mm]
\hline
\\[-2.5mm]
 \multirow{4}{*}{$4-6$}
 &  $\uplambda_{\theta}$
 &  $\pz0.240\pm0.172\pm0.052$
 &  $-0.043\pm0.103\pm0.047$
 &  $\pz0.047\pm0.130\pm0.070$
\\
 &  $\uplambda_{\theta\phi}$
 &  $\pz0.106\pm0.052\pm0.012$
 &  $\pz0.033\pm0.039\pm0.008$
 &  $\pz0.065\pm0.050\pm0.013$
\\
 &  $\uplambda_{\phi}$
 &  $\pz0.052\pm0.023\pm0.006$
 &  $\pz0.043\pm0.027\pm0.007$
 &  $\pz0.039\pm0.028\pm0.010$
\\
 &  $\tilde \uplambda$
 &  $\pz0.419\pm0.213\pm0.064$
 &  $\pz0.091\pm0.149\pm0.064$
 &  $\pz0.170\pm0.174\pm0.099$
\\[ 1.5mm]
\hline
\\[-2.5mm]
 \multirow{4}{*}{$6-8$}
 &  $\uplambda_{\theta}$
 &  $-0.080\pm0.161\pm0.047$
 &  $-0.092\pm0.096\pm0.041$
 &  $-0.224\pm0.118\pm0.049$
\\
 &  $\uplambda_{\theta\phi}$
 &  $-0.036\pm0.063\pm0.017$
 &  $\pz0.041\pm0.051\pm0.014$
 &  $\pz0.109\pm0.061\pm0.018$
\\
 &  $\uplambda_{\phi}$
 &  $\pz0.074\pm0.027\pm0.007$
 &  $\pz0.074\pm0.032\pm0.013$
 &  $\pz0.055\pm0.033\pm0.011$
\\
 &  $\tilde \uplambda$
 &  $\pz0.153\pm0.216\pm0.066$
 &  $\pz0.141\pm0.169\pm0.078$
 &  $-0.063\pm0.179\pm0.074$
\\[ 1.5mm]
\hline
\\[-2.5mm]
 \multirow{4}{*}{$8-10$}
 &  $\uplambda_{\theta}$
 &  $\pz0.201\pm0.176\pm0.071$
 &  $-0.219\pm0.081\pm0.031$
 &  $-0.238\pm0.115\pm0.046$
\\
 &  $\uplambda_{\theta\phi}$
 &  $\pz0.010\pm0.086\pm0.023$
 &  $\pz0.097\pm0.068\pm0.016$
 &  $\pz0.076\pm0.084\pm0.025$
\\
 &  $\uplambda_{\phi}$
 &  $\pz0.056\pm0.035\pm0.012$
 &  $\pz0.053\pm0.040\pm0.011$
 &  $\pz0.049\pm0.045\pm0.013$
\\
 &  $\tilde \uplambda$
 &  $\pz0.393\pm0.248\pm0.087$
 &  $-0.063\pm0.164\pm0.053$
 &  $-0.096\pm0.200\pm0.068$
\\[ 1.5mm]
\hline
\\[-2.5mm]
 \multirow{4}{*}{$10-15$}
 &  $\uplambda_{\theta}$
 &  $-0.080\pm0.091\pm0.035$
 &  $-0.184\pm0.060\pm0.028$
 &  $-0.137\pm0.090\pm0.044$
\\
 &  $\uplambda_{\theta\phi}$
 &  $\pz0.025\pm0.073\pm0.021$
 &  $\pz0.204\pm0.063\pm0.021$
 &  $\pz0.027\pm0.079\pm0.033$
\\
 &  $\uplambda_{\phi}$
 &  $\pz0.159\pm0.032\pm0.014$
 &  $\pz0.023\pm0.042\pm0.016$
 &  $\pz0.038\pm0.050\pm0.022$
\\
 &  $\tilde \uplambda$
 &  $\pz0.471\pm0.180\pm0.076$
 &  $-0.118\pm0.138\pm0.047$
 &  $-0.024\pm0.187\pm0.073$
\\[ 1.5mm]
\hline
\\[-2.5mm]
 \multirow{4}{*}{$15-20$}
 &  $\uplambda_{\theta}$
 &  $\pz0.010\pm0.117\pm0.090$
 &  $-0.074\pm0.106\pm0.094$
 &  $-0.153\pm0.146\pm0.141$
\\
 &  $\uplambda_{\theta\phi}$
 &  $-0.023\pm0.115\pm0.067$
 &  $-0.014\pm0.095\pm0.041$
 &  $\pz0.008\pm0.139\pm0.067$
\\
 &  $\uplambda_{\phi}$
 &  $\pz0.184\pm0.072\pm0.052$
 &  $\pz0.151\pm0.083\pm0.053$
 &  $\pz0.337\pm0.103\pm0.066$
\\
 &  $\tilde \uplambda$
 &  $\pz0.689\pm0.336\pm0.207$
 &  $\pz0.447\pm0.333\pm0.178$
 &  $\pz1.295\pm0.652\pm0.324$
\end{tabular*}
\end{footnotesize}
\end{center} 
\end{table}  
%
%
%
%
%
\begin{table}[t] 
\begin{center} 
\caption{\small Values of
 $\uplambda_{\theta}$, $\uplambda_{\theta\phi}$, $\uplambda_{\phi}$
 and $\tilde \uplambda$
 measured in the CS frame for the \threes produced
 at $\sqrt{s}=8\,\,{\mathrm{TeV}}$.
 The first uncertainty is statistical and the second systematic.
}\label{tab:Y3S_Results_CS_Slices_8TeV}
\vspace*{2mm}
\begin{footnotesize}
\begin{tabular*}{0.99\textwidth}{@{\hspace{1mm}}c@{\extracolsep{\fill}}cccc@{\hspace{1mm}}}
 $\pty~\left[\!\gevc\right]$ 
 & $\uplambda$ 
 & $2.2<y<3.0$ 
 & $3.0<y<3.5$
 & $3.5<y<4.5$ 
\\[ 1.5mm]
\hline
\\[-2.5mm]
 \multirow{4}{*}{$0-2$}
 &  $\uplambda_{\theta}$
 &  $\pz0.231\pm0.166\pm0.072$
 &  $-0.072\pm0.090\pm0.076$
 &  $-0.437\pm0.108\pm0.081$
\\
 &  $\uplambda_{\theta\phi}$
 &  $\pz0.033\pm0.043\pm0.013$
 &  $\pz0.022\pm0.030\pm0.008$
 &  $-0.002\pm0.036\pm0.010$
\\
 &  $\uplambda_{\phi}$
 &  $-0.007\pm0.022\pm0.005$
 &  $-0.028\pm0.025\pm0.007$
 &  $-0.024\pm0.024\pm0.007$
\\
 &  $\tilde \uplambda$
 &  $\pz0.209\pm0.181\pm0.077$
 &  $-0.151\pm0.112\pm0.079$
 &  $-0.499\pm0.122\pm0.083$
\\[ 1.5mm]
\hline
\\[-2.5mm]
 \multirow{4}{*}{$2-4$}
 &  $\uplambda_{\theta}$
 &  $\pz0.282\pm0.119\pm0.052$
 &  $\pz0.023\pm0.070\pm0.058$
 &  $\pz0.000\pm0.086\pm0.075$
\\
 &  $\uplambda_{\theta\phi}$
 &  $\pz0.070\pm0.032\pm0.010$
 &  $\pz0.036\pm0.024\pm0.007$
 &  $\pz0.003\pm0.030\pm0.009$
\\
 &  $\uplambda_{\phi}$
 &  $\pz0.030\pm0.016\pm0.004$
 &  $\pz0.009\pm0.018\pm0.005$
 &  $\pz0.013\pm0.018\pm0.005$
\\
 &  $\tilde \uplambda$
 &  $\pz0.382\pm0.138\pm0.061$
 &  $\pz0.051\pm0.093\pm0.069$
 &  $\pz0.039\pm0.105\pm0.086$
\\[ 1.5mm]
\hline
\\[-2.5mm]
 \multirow{4}{*}{$4-6$}
 &  $\uplambda_{\theta}$
 &  $\pz0.184\pm0.116\pm0.044$
 &  $\pz0.060\pm0.071\pm0.052$
 &  $-0.238\pm0.084\pm0.064$
\\
 &  $\uplambda_{\theta\phi}$
 &  $\pz0.094\pm0.035\pm0.011$
 &  $\pz0.082\pm0.027\pm0.008$
 &  $\pz0.131\pm0.032\pm0.011$
\\
 &  $\uplambda_{\phi}$
 &  $\pz0.061\pm0.016\pm0.005$
 &  $\pz0.061\pm0.018\pm0.007$
 &  $\pz0.030\pm0.018\pm0.007$
\\
 &  $\tilde \uplambda$
 &  $\pz0.391\pm0.145\pm0.056$
 &  $\pz0.259\pm0.107\pm0.075$
 &  $-0.154\pm0.108\pm0.083$
\\[ 1.5mm]
\hline
\\[-2.5mm]
 \multirow{4}{*}{$6-8$}
 &  $\uplambda_{\theta}$
 &  $-0.044\pm0.114\pm0.041$
 &  $-0.330\pm0.059\pm0.033$
 &  $-0.404\pm0.080\pm0.044$
\\
 &  $\uplambda_{\theta\phi}$
 &  $\pz0.040\pm0.043\pm0.011$
 &  $\pz0.119\pm0.033\pm0.009$
 &  $\pz0.155\pm0.039\pm0.015$
\\
 &  $\uplambda_{\phi}$
 &  $\pz0.068\pm0.018\pm0.006$
 &  $\pz0.056\pm0.021\pm0.008$
 &  $\pz0.037\pm0.022\pm0.010$
\\
 &  $\tilde \uplambda$
 &  $\pz0.172\pm0.150\pm0.056$
 &  $-0.170\pm0.099\pm0.055$
 &  $-0.302\pm0.114\pm0.068$
\\[ 1.5mm]
\hline
\\[-2.5mm]
 \multirow{4}{*}{$8-10$}
 &  $\uplambda_{\theta}$
 &  $-0.093\pm0.108\pm0.043$
 &  $-0.213\pm0.054\pm0.023$
 &  $-0.177\pm0.078\pm0.036$
\\
 &  $\uplambda_{\theta\phi}$
 &  $\pz0.031\pm0.057\pm0.017$
 &  $\pz0.078\pm0.044\pm0.012$
 &  $\pz0.085\pm0.054\pm0.022$
\\
 &  $\uplambda_{\phi}$
 &  $\pz0.104\pm0.023\pm0.010$
 &  $\pz0.092\pm0.027\pm0.009$
 &  $\pz0.046\pm0.030\pm0.015$
\\
 &  $\tilde \uplambda$
 &  $\pz0.246\pm0.163\pm0.067$
 &  $\pz0.068\pm0.117\pm0.044$
 &  $-0.041\pm0.134\pm0.074$
\\[ 1.5mm]
\hline
\\[-2.5mm]
 \multirow{4}{*}{$10-15$}
 &  $\uplambda_{\theta}$
 &  $-0.038\pm0.063\pm0.027$
 &  $-0.168\pm0.039\pm0.021$
 &  $-0.213\pm0.053\pm0.028$
\\
 &  $\uplambda_{\theta\phi}$
 &  $\pz0.110\pm0.050\pm0.017$
 &  $\pz0.115\pm0.040\pm0.016$
 &  $\pz0.122\pm0.051\pm0.024$
\\
 &  $\uplambda_{\phi}$
 &  $\pz0.068\pm0.022\pm0.012$
 &  $\pz0.078\pm0.027\pm0.013$
 &  $\pz0.040\pm0.032\pm0.016$
\\
 &  $\tilde \uplambda$
 &  $\pz0.177\pm0.108\pm0.049$
 &  $\pz0.073\pm0.099\pm0.043$
 &  $-0.097\pm0.113\pm0.051$
\\[ 1.5mm]
\hline
\\[-2.5mm]
 \multirow{4}{*}{$15-20$}
 &  $\uplambda_{\theta}$
 &  $-0.083\pm0.075\pm0.049$
 &  $-0.109\pm0.072\pm0.070$
 &  $-0.368\pm0.082\pm0.081$
\\
 &  $\uplambda_{\theta\phi}$
 &  $\pz0.242\pm0.074\pm0.040$
 &  $\pz0.116\pm0.068\pm0.035$
 &  $\pz0.107\pm0.088\pm0.050$
\\
 &  $\uplambda_{\phi}$
 &  $\pz0.064\pm0.048\pm0.037$
 &  $\pz0.049\pm0.058\pm0.048$
 &  $\pz0.044\pm0.079\pm0.077$
\\
 &  $\tilde \uplambda$
 &  $\pz0.117\pm0.168\pm0.114$
 &  $\pz0.041\pm0.183\pm0.112$
 &  $-0.246\pm0.224\pm0.182$
\end{tabular*}
\end{footnotesize}
\end{center} 
\end{table}  
%
%
%
%
%
\begin{table}[t] 
\begin{center} 
\caption{\small Values of
 $\uplambda_{\theta}$, $\uplambda_{\theta\phi}$, $\uplambda_{\phi}$
 and $\tilde \uplambda$
 measured in the GJ frame for the \threes produced
 at $\sqrt{s}=7\,\,{\mathrm{TeV}}$.
 The first uncertainty is statistical and the second systematic.
}\label{tab:Y3S_Results_GJ_Slices_7TeV}
\vspace*{2mm}
\begin{footnotesize}
\begin{tabular*}{0.99\textwidth}{@{\hspace{1mm}}c@{\extracolsep{\fill}}cccc@{\hspace{1mm}}}
 $\pty~\left[\!\gevc\right]$ 
 & $\uplambda$ 
 & $2.2<y<3.0$ 
 & $3.0<y<3.5$ 
 & $3.5<y<4.5$ 
\\[ 1.5mm]
\hline
\\[-2.5mm]
 \multirow{4}{*}{$0-2$}
 &  $\uplambda_{\theta}$
 &  $\pz0.317\pm0.233\pm0.116$
 &  $-0.092\pm0.120\pm0.070$
 &  $-0.126\pm0.170\pm0.095$
\\
 &  $\uplambda_{\theta\phi}$
 &  $\pz0.175\pm0.074\pm0.030$
 &  $\pz0.091\pm0.044\pm0.016$
 &  $-0.024\pm0.053\pm0.017$
\\
 &  $\uplambda_{\phi}$
 &  $\pz0.003\pm0.034\pm0.008$
 &  $-0.031\pm0.035\pm0.007$
 &  $-0.015\pm0.034\pm0.007$
\\
 &  $\tilde \uplambda$
 &  $\pz0.326\pm0.270\pm0.132$
 &  $-0.178\pm0.152\pm0.070$
 &  $-0.168\pm0.194\pm0.102$
\\[ 1.5mm]
\hline
\\[-2.5mm]
 \multirow{4}{*}{$2-4$}
 &  $\uplambda_{\theta}$
 &  $-0.088\pm0.132\pm0.074$
 &  $-0.016\pm0.093\pm0.053$
 &  $-0.227\pm0.117\pm0.053$
\\
 &  $\uplambda_{\theta\phi}$
 &  $\pz0.091\pm0.066\pm0.041$
 &  $\pz0.073\pm0.039\pm0.018$
 &  $-0.025\pm0.038\pm0.018$
\\
 &  $\uplambda_{\phi}$
 &  $\pz0.056\pm0.028\pm0.012$
 &  $\pz0.092\pm0.027\pm0.008$
 &  $\pz0.038\pm0.025\pm0.008$
\\
 &  $\tilde \uplambda$
 &  $\pz0.085\pm0.190\pm0.111$
 &  $\pz0.288\pm0.152\pm0.083$
 &  $-0.119\pm0.143\pm0.071$
\\[ 1.5mm]
\hline
\\[-2.5mm]
 \multirow{4}{*}{$4-6$}
 &  $\uplambda_{\theta}$
 &  $-0.099\pm0.104\pm0.046$
 &  $-0.052\pm0.091\pm0.043$
 &  $\pz0.006\pm0.126\pm0.059$
\\
 &  $\uplambda_{\theta\phi}$
 &  $\pz0.160\pm0.071\pm0.042$
 &  $\pz0.017\pm0.045\pm0.019$
 &  $\pz0.118\pm0.048\pm0.029$
\\
 &  $\uplambda_{\phi}$
 &  $\pz0.142\pm0.034\pm0.019$
 &  $\pz0.057\pm0.029\pm0.010$
 &  $\pz0.077\pm0.028\pm0.016$
\\
 &  $\tilde \uplambda$
 &  $\pz0.382\pm0.212\pm0.116$
 &  $\pz0.127\pm0.151\pm0.069$
 &  $\pz0.258\pm0.178\pm0.105$
\\[ 1.5mm]
\hline
\\[-2.5mm]
 \multirow{4}{*}{$6-8$}
 &  $\uplambda_{\theta}$
 &  $-0.110\pm0.103\pm0.042$
 &  $-0.071\pm0.095\pm0.037$
 &  $-0.250\pm0.122\pm0.052$
\\
 &  $\uplambda_{\theta\phi}$
 &  $-0.053\pm0.083\pm0.053$
 &  $-0.039\pm0.049\pm0.017$
 &  $-0.026\pm0.052\pm0.021$
\\
 &  $\uplambda_{\phi}$
 &  $\pz0.087\pm0.051\pm0.032$
 &  $\pz0.076\pm0.038\pm0.017$
 &  $\pz0.081\pm0.035\pm0.017$
\\
 &  $\tilde \uplambda$
 &  $\pz0.165\pm0.217\pm0.121$
 &  $\pz0.171\pm0.171\pm0.072$
 &  $-0.008\pm0.182\pm0.083$
\\[ 1.5mm]
\hline
\\[-2.5mm]
 \multirow{4}{*}{$8-10$}
 &  $\uplambda_{\theta}$
 &  $-0.076\pm0.115\pm0.066$
 &  $-0.171\pm0.108\pm0.054$
 &  $-0.210\pm0.152\pm0.068$
\\
 &  $\uplambda_{\theta\phi}$
 &  $\pz0.083\pm0.077\pm0.050$
 &  $-0.111\pm0.048\pm0.015$
 &  $-0.092\pm0.059\pm0.019$
\\
 &  $\uplambda_{\phi}$
 &  $\pz0.142\pm0.061\pm0.044$
 &  $\pz0.042\pm0.049\pm0.020$
 &  $\pz0.043\pm0.048\pm0.018$
\\
 &  $\tilde \uplambda$
 &  $\pz0.409\pm0.249\pm0.140$
 &  $-0.047\pm0.165\pm0.050$
 &  $-0.085\pm0.202\pm0.085$
\\[ 1.5mm]
\hline
\\[-2.5mm]
 \multirow{4}{*}{$10-15$}
 &  $\uplambda_{\theta}$
 &  $\pz0.105\pm0.119\pm0.111$
 &  $-0.315\pm0.092\pm0.054$
 &  $-0.047\pm0.145\pm0.084$
\\
 &  $\uplambda_{\theta\phi}$
 &  $-0.128\pm0.051\pm0.027$
 &  $-0.123\pm0.037\pm0.011$
 &  $-0.078\pm0.055\pm0.019$
\\
 &  $\uplambda_{\phi}$
 &  $\pz0.105\pm0.055\pm0.050$
 &  $\pz0.064\pm0.039\pm0.017$
 &  $\pz0.006\pm0.047\pm0.021$
\\
 &  $\tilde \uplambda$
 &  $\pz0.468\pm0.178\pm0.091$
 &  $-0.130\pm0.137\pm0.049$
 &  $-0.029\pm0.189\pm0.075$
\\[ 1.5mm]
\hline
\\[-2.5mm]
 \multirow{4}{*}{$15-20$}
 &  $\uplambda_{\theta}$
 &  $\pz0.261\pm0.198\pm0.277$
 &  $\pz0.188\pm0.203\pm0.190$
 &  $\pz0.470\pm0.331\pm0.247$
\\
 &  $\uplambda_{\theta\phi}$
 &  $-0.076\pm0.088\pm0.065$
 &  $-0.104\pm0.084\pm0.044$
 &  $-0.249\pm0.137\pm0.051$
\\
 &  $\uplambda_{\phi}$
 &  $\pz0.117\pm0.086\pm0.105$
 &  $\pz0.073\pm0.070\pm0.041$
 &  $\pz0.193\pm0.101\pm0.057$
\\
 &  $\tilde \uplambda$
 &  $\pz0.693\pm0.334\pm0.238$
 &  $\pz0.437\pm0.332\pm0.225$
 &  $\pz1.300\pm0.649\pm0.346$
\end{tabular*}
\end{footnotesize}
\end{center} 
\end{table}  
%
%
%
%
%
\begin{table}[t] 
\begin{center} 
\caption{\small Values of
 $\uplambda_{\theta}$, $\uplambda_{\theta\phi}$, $\uplambda_{\phi}$
 and $\tilde \uplambda$
 measured in the GJ frame for the \threes produced
 at $\sqrt{s}=8\,\,{\mathrm{TeV}}$.
 The first uncertainty is statistical and the second systematic.
}\label{tab:Y3S_Results_GJ_Slices_8TeV}
\vspace*{2mm}
\begin{footnotesize}
\begin{tabular*}{0.99\textwidth}{@{\hspace{1mm}}c@{\extracolsep{\fill}}cccc@{\hspace{1mm}}}
 $\pty~\left[\!\gevc\right]$ 
 & $\uplambda$ 
 & $2.2<y<3.0$ 
 & $3.0<y<3.5$ 
 & $3.5<y<4.5$
\\[ 1.5mm]
\hline
\\[-2.5mm]
 \multirow{4}{*}{$0-2$}
 &  $\uplambda_{\theta}$
 &  $\pz0.147\pm0.152\pm0.072$
 &  $-0.081\pm0.087\pm0.072$
 &  $-0.385\pm0.111\pm0.074$
\\
 &  $\uplambda_{\theta\phi}$
 &  $\pz0.110\pm0.049\pm0.019$
 &  $\pz0.030\pm0.031\pm0.013$
 &  $-0.009\pm0.036\pm0.014$
\\
 &  $\uplambda_{\phi}$
 &  $\pz0.009\pm0.023\pm0.005$
 &  $-0.024\pm0.025\pm0.006$
 &  $-0.027\pm0.024\pm0.006$
\\
 &  $\tilde \uplambda$
 &  $\pz0.177\pm0.177\pm0.079$
 &  $-0.148\pm0.112\pm0.076$
 &  $-0.453\pm0.123\pm0.078$
\\[ 1.5mm]
\hline
\\[-2.5mm]
 \multirow{4}{*}{$2-4$}
 &  $\uplambda_{\theta}$
 &  $\pz0.008\pm0.089\pm0.043$
 &  $-0.002\pm0.065\pm0.051$
 &  $\pz0.054\pm0.088\pm0.065$
\\
 &  $\uplambda_{\theta\phi}$
 &  $\pz0.156\pm0.044\pm0.022$
 &  $\pz0.064\pm0.026\pm0.017$
 &  $\pz0.070\pm0.029\pm0.022$
\\
 &  $\uplambda_{\phi}$
 &  $\pz0.084\pm0.019\pm0.007$
 &  $\pz0.025\pm0.018\pm0.007$
 &  $\pz0.019\pm0.018\pm0.008$
\\
 &  $\tilde \uplambda$
 &  $\pz0.283\pm0.134\pm0.067$
 &  $\pz0.075\pm0.093\pm0.073$
 &  $\pz0.113\pm0.108\pm0.085$
\\[ 1.5mm]
\hline
\\[-2.5mm]
 \multirow{4}{*}{$4-6$}
 &  $\uplambda_{\theta}$
 &  $-0.166\pm0.071\pm0.049$
 &  $-0.038\pm0.060\pm0.037$
 &  $-0.247\pm0.081\pm0.045$
\\
 &  $\uplambda_{\theta\phi}$
 &  $\pz0.114\pm0.048\pm0.043$
 &  $\pz0.079\pm0.030\pm0.019$
 &  $\pz0.058\pm0.031\pm0.024$
\\
 &  $\uplambda_{\phi}$
 &  $\pz0.138\pm0.023\pm0.017$
 &  $\pz0.099\pm0.020\pm0.011$
 &  $\pz0.068\pm0.018\pm0.013$
\\
 &  $\tilde \uplambda$
 &  $\pz0.288\pm0.142\pm0.118$
 &  $\pz0.289\pm0.108\pm0.076$
 &  $-0.044\pm0.110\pm0.082$
\\[ 1.5mm]
\hline
\\[-2.5mm]
 \multirow{4}{*}{$6-8$}
 &  $\uplambda_{\theta}$
 &  $-0.223\pm0.066\pm0.031$
 &  $-0.306\pm0.056\pm0.024$
 &  $-0.375\pm0.080\pm0.032$
\\
 &  $\uplambda_{\theta\phi}$
 &  $-0.011\pm0.055\pm0.043$
 &  $-0.106\pm0.032\pm0.016$
 &  $-0.074\pm0.036\pm0.020$
\\
 &  $\uplambda_{\phi}$
 &  $\pz0.119\pm0.033\pm0.024$
 &  $\pz0.062\pm0.025\pm0.014$
 &  $\pz0.060\pm0.023\pm0.016$
\\
 &  $\tilde \uplambda$
 &  $\pz0.153\pm0.150\pm0.110$
 &  $-0.129\pm0.100\pm0.055$
 &  $-0.207\pm0.116\pm0.064$
\\[ 1.5mm]
\hline
\\[-2.5mm]
 \multirow{4}{*}{$8-10$}
 &  $\uplambda_{\theta}$
 &  $-0.157\pm0.077\pm0.046$
 &  $-0.134\pm0.074\pm0.039$
 &  $-0.183\pm0.098\pm0.049$
\\
 &  $\uplambda_{\theta\phi}$
 &  $-0.078\pm0.056\pm0.048$
 &  $-0.132\pm0.032\pm0.011$
 &  $-0.060\pm0.039\pm0.016$
\\
 &  $\uplambda_{\phi}$
 &  $\pz0.128\pm0.043\pm0.041$
 &  $\pz0.069\pm0.031\pm0.014$
 &  $\pz0.054\pm0.031\pm0.021$
\\
 &  $\tilde \uplambda$
 &  $\pz0.259\pm0.163\pm0.122$
 &  $\pz0.079\pm0.117\pm0.044$
 &  $-0.024\pm0.136\pm0.074$
\\[ 1.5mm]
\hline
\\[-2.5mm]
 \multirow{4}{*}{$10-15$}
 &  $\uplambda_{\theta}$
 &  $-0.114\pm0.069\pm0.072$
 &  $-0.126\pm0.067\pm0.051$
 &  $-0.218\pm0.086\pm0.049$
\\
 &  $\uplambda_{\theta\phi}$
 &  $-0.058\pm0.031\pm0.019$
 &  $-0.133\pm0.025\pm0.010$
 &  $-0.130\pm0.033\pm0.011$
\\
 &  $\uplambda_{\phi}$
 &  $\pz0.092\pm0.035\pm0.036$
 &  $\pz0.063\pm0.027\pm0.015$
 &  $\pz0.035\pm0.029\pm0.015$
\\
 &  $\tilde \uplambda$
 &  $\pz0.178\pm0.107\pm0.068$
 &  $\pz0.068\pm0.099\pm0.047$
 &  $-0.117\pm0.113\pm0.054$
\\[ 1.5mm]
\hline
\\[-2.5mm]
 \multirow{4}{*}{$15-20$}
 &  $\uplambda_{\theta}$
 &  $-0.213\pm0.099\pm0.126$
 &  $-0.115\pm0.118\pm0.121$
 &  $-0.066\pm0.181\pm0.164$
\\
 &  $\uplambda_{\theta\phi}$
 &  $-0.159\pm0.047\pm0.029$
 &  $-0.121\pm0.051\pm0.025$
 &  $-0.239\pm0.077\pm0.046$
\\
 &  $\uplambda_{\phi}$
 &  $\pz0.102\pm0.050\pm0.054$
 &  $\pz0.047\pm0.046\pm0.028$
 &  $-0.068\pm0.063\pm0.036$
\\
 &  $\tilde \uplambda$
 &  $\pz0.105\pm0.167\pm0.115$
 &  $\pz0.028\pm0.182\pm0.126$
 &  $-0.252\pm0.223\pm0.148$
\end{tabular*}
\end{footnotesize}
\end{center} 
\end{table}  

\begin{table}[t] 
\begin{center} 
\caption{\small Values of
 $\uplambda_{\theta}$, $\uplambda_{\theta\phi}$ and $\uplambda_{\phi}$
 measured in the HX, CS and GJ frames for the \threes produced
 at $\sqrt{s}=7\,\,{\mathrm{TeV}}$ in the rapidity range
 $2.2<\yy<4.5$.
 The first uncertainty is statistical and the second systematic.
}\label{tab:Y3S_Results_HX_CS_GJ_2.2y4.5_7TeV}
\vspace*{2mm}
\begin{footnotesize}
  \begin{tabular*}{0.99\textwidth}{@{\hspace{1mm}}c@{\extracolsep{\fill}}cccc@{\hspace{1mm}}}
 $\pty~~\left[\!\gevc\right]$ 
 & 
 & $\uplambda_{\theta}$ 
 & $\uplambda_{\theta\phi}$ 
 & $\uplambda_{\phi}$ 
\\[ 1.5mm]
\hline
\\[-2.5mm]
 \multirow{3}{*}{~ 0$\,-$~ 2}
 &  HX
 &  $-0.054\pm0.087\pm0.069$
 &  $\pz0.012\pm0.030\pm0.009$
 &  $-0.026\pm0.020\pm0.006$
\\
 &  CS
 &  $-0.039\pm0.089\pm0.069$
 &  $\pz0.038\pm0.029\pm0.007$
 &  $-0.026\pm0.020\pm0.004$
\\
 &  GJ
 &  $-0.054\pm0.086\pm0.068$
 &  $\pz0.065\pm0.030\pm0.012$
 &  $-0.018\pm0.020\pm0.004$
\\[ 1.5mm]
\hline
\\[-2.5mm]
 \multirow{3}{*}{~ 2$\,-$~ 4}
 &  HX
 &  $-0.141\pm0.057\pm0.046$
 &  $\pz0.022\pm0.026\pm0.013$
 &  $\pz0.043\pm0.015\pm0.005$
\\
 &  CS
 &  $-0.097\pm0.064\pm0.052$
 &  $\pz0.023\pm0.022\pm0.006$
 &  $\pz0.040\pm0.014\pm0.004$
\\
 &  GJ
 &  $-0.123\pm0.059\pm0.042$
 &  $\pz0.034\pm0.024\pm0.014$
 &  $\pz0.055\pm0.015\pm0.006$
\\[ 1.5mm]
\hline
\\[-2.5mm]
 \multirow{3}{*}{~ 4$\,-$~ 6}
 &  HX
 &  $-0.004\pm0.054\pm0.037$
 &  $\pz0.027\pm0.032\pm0.016$
 &  $\pz0.040\pm0.018\pm0.010$
\\
 &  CS
 &  $\pz0.028\pm0.069\pm0.043$
 &  $\pz0.053\pm0.025\pm0.007$
 &  $\pz0.044\pm0.015\pm0.007$
\\
 &  GJ
 &  $-0.075\pm0.057\pm0.031$
 &  $\pz0.068\pm0.028\pm0.016$
 &  $\pz0.080\pm0.017\pm0.011$
\\[ 1.5mm]
\hline
\\[-2.5mm]
 \multirow{3}{*}{~ 6$\,-$~ 8}
 &  HX
 &  $-0.056\pm0.053\pm0.027$
 &  $\pz0.098\pm0.035\pm0.017$
 &  $\pz0.049\pm0.025\pm0.015$
\\
 &  CS
 &  $-0.101\pm0.064\pm0.028$
 &  $\pz0.032\pm0.031\pm0.009$
 &  $\pz0.074\pm0.017\pm0.009$
\\
 &  GJ
 &  $-0.115\pm0.059\pm0.025$
 &  $-0.038\pm0.031\pm0.015$
 &  $\pz0.083\pm0.021\pm0.014$
\\[ 1.5mm]
\hline
\\[-2.5mm]
 \multirow{3}{*}{~ 8$\,-$~10}
 &  HX
 &  $-0.008\pm0.067\pm0.039$
 &  $\pz0.120\pm0.035\pm0.015$
 &  $\pz0.001\pm0.035\pm0.022$
\\
 &  CS
 &  $-0.148\pm0.061\pm0.023$
 &  $\pz0.048\pm0.042\pm0.012$
 &  $\pz0.051\pm0.022\pm0.013$
\\
 &  GJ
 &  $-0.148\pm0.069\pm0.032$
 &  $-0.074\pm0.032\pm0.010$
 &  $\pz0.052\pm0.028\pm0.016$
\\[ 1.5mm]
\hline
\\[-2.5mm]
 \multirow{3}{*}{~10$\,-$~15}
 &  HX
 &  $\pz0.176\pm0.067\pm0.064$
 &  $\pz0.126\pm0.028\pm0.012$
 &  $-0.010\pm0.033\pm0.026$
\\
 &  CS
 &  $-0.147\pm0.042\pm0.019$
 &  $\pz0.073\pm0.038\pm0.012$
 &  $\pz0.091\pm0.022\pm0.013$
\\
 &  GJ
 &  $-0.066\pm0.064\pm0.037$
 &  $-0.128\pm0.026\pm0.010$
 &  $\pz0.064\pm0.025\pm0.013$
\\[ 1.5mm]
\hline
\\[-2.5mm]
 \multirow{3}{*}{~15$\,-$~20}
 &  HX
 &  $\pz0.227\pm0.112\pm0.108$
 &  $\pz0.138\pm0.058\pm0.034$
 &  $\pz0.125\pm0.048\pm0.040$
\\
 &  CS
 &  $-0.062\pm0.068\pm0.068$
 &  $-0.018\pm0.062\pm0.031$
 &  $\pz0.204\pm0.048\pm0.041$
\\
 &  GJ
 &  $\pz0.281\pm0.128\pm0.108$
 &  $-0.122\pm0.055\pm0.033$
 &  $\pz0.111\pm0.046\pm0.028$
\\[ 1.5mm]
\hline
\\[-2.5mm]
 \multirow{3}{*}{~20$\,-$~30}
 &  HX
 &  $\pz0.570\pm0.221\pm0.232$
 &  $\pz0.149\pm0.113\pm0.102$
 &  $\pz0.017\pm0.074\pm0.069$
\\
 &  CS
 &  $-0.226\pm0.099\pm0.134$
 &  $\pz0.083\pm0.082\pm0.048$
 &  $\pz0.239\pm0.083\pm0.094$
\\
 &  GJ
 &  $\pz0.313\pm0.212\pm0.231$
 &  $-0.274\pm0.096\pm0.089$
 &  $\pz0.093\pm0.069\pm0.060$
\end{tabular*}
\end{footnotesize}
\end{center} 
\end{table}  
\begin{table}[t]
\begin{center}
\caption{\small Values of
 $\uplambda_{\theta}$, $\uplambda_{\theta\phi}$ and $\uplambda_{\phi}$
 measured in the HX, CS and GJ frames for the \threes produced
 at $\sqrt{s}=8\,\,{\mathrm{TeV}}$ in the rapidity range
 $2.2<\yy<4.5$.
 The first uncertainty is statistical and the second systematic.
}\label{tab:Y3S_Results_HX_CS_GJ_2.2y4.5_8TeV}
\vspace*{2mm}
\begin{footnotesize}
\begin{tabular*}{0.99\textwidth}{@{\hspace{1mm}}c@{\extracolsep{\fill}}cccc@{\hspace{1mm}}}
 $\pty~~\left[\!\gevc\right]$ 
 & 
 & $\uplambda_{\theta}$ 
 & $\uplambda_{\theta\phi}$ 
 & $\uplambda_{\phi}$ 
\\[ 1.5mm]
\hline
\\[-2.5mm]
 \multirow{3}{*}{~ 0$\,-$~ 2}
 &  HX
 &  $-0.141\pm0.059\pm0.063$
 &  $-0.010\pm0.021\pm0.009$
 &  $-0.016\pm0.014\pm0.003$
\\
 &  CS
 &  $-0.119\pm0.061\pm0.067$
 &  $\pz0.007\pm0.020\pm0.006$
 &  $-0.019\pm0.014\pm0.003$
\\
 &  GJ
 &  $-0.122\pm0.059\pm0.063$
 &  $\pz0.027\pm0.021\pm0.011$
 &  $-0.015\pm0.014\pm0.003$
\\[ 1.5mm]
\hline
\\[-2.5mm]
 \multirow{3}{*}{~ 2$\,-$~ 4}
 &  HX
 &  $-0.014\pm0.041\pm0.048$
 &  $-0.022\pm0.018\pm0.013$
 &  $\pz0.026\pm0.010\pm0.005$
\\
 &  CS
 &  $\pz0.044\pm0.046\pm0.053$
 &  $\pz0.025\pm0.015\pm0.006$
 &  $\pz0.016\pm0.010\pm0.004$
\\
 &  GJ
 &  $-0.017\pm0.042\pm0.045$
 &  $\pz0.076\pm0.017\pm0.015$
 &  $\pz0.038\pm0.010\pm0.006$
\\[ 1.5mm]
\hline
\\[-2.5mm]
 \multirow{3}{*}{~ 4$\,-$~ 6}
 &  HX
 &  $\pz0.008\pm0.037\pm0.037$
 &  $\pz0.057\pm0.022\pm0.017$
 &  $\pz0.036\pm0.013\pm0.010$
\\
 &  CS
 &  $-0.009\pm0.046\pm0.045$
 &  $\pz0.079\pm0.017\pm0.007$
 &  $\pz0.051\pm0.010\pm0.007$
\\
 &  GJ
 &  $-0.139\pm0.038\pm0.030$
 &  $\pz0.065\pm0.019\pm0.016$
 &  $\pz0.093\pm0.011\pm0.010$
\\[ 1.5mm]
\hline
\\[-2.5mm]
 \multirow{3}{*}{~ 6$\,-$~ 8}
 &  HX
 &  $-0.076\pm0.036\pm0.027$
 &  $\pz0.205\pm0.025\pm0.014$
 &  $-0.019\pm0.018\pm0.014$
\\
 &  CS
 &  $-0.270\pm0.042\pm0.027$
 &  $\pz0.091\pm0.020\pm0.007$
 &  $\pz0.059\pm0.011\pm0.009$
\\
 &  GJ
 &  $-0.291\pm0.037\pm0.020$
 &  $-0.083\pm0.021\pm0.012$
 &  $\pz0.072\pm0.014\pm0.014$
\\[ 1.5mm]
\hline
\\[-2.5mm]
 \multirow{3}{*}{~ 8$\,-$~10}
 &  HX
 &  $\pz0.033\pm0.045\pm0.029$
 &  $\pz0.154\pm0.024\pm0.011$
 &  $\pz0.017\pm0.024\pm0.014$
\\
 &  CS
 &  $-0.172\pm0.039\pm0.017$
 &  $\pz0.059\pm0.027\pm0.010$
 &  $\pz0.086\pm0.015\pm0.010$
\\
 &  GJ
 &  $-0.146\pm0.046\pm0.029$
 &  $-0.104\pm0.021\pm0.009$
 &  $\pz0.077\pm0.018\pm0.014$
\\[ 1.5mm]
\hline
\\[-2.5mm]
 \multirow{3}{*}{~10$\,-$~15}
 &  HX
 &  $\pz0.198\pm0.045\pm0.039$
 &  $\pz0.114\pm0.018\pm0.012$
 &  $-0.049\pm0.022\pm0.016$
\\
 &  CS
 &  $-0.150\pm0.027\pm0.014$
 &  $\pz0.101\pm0.025\pm0.011$
 &  $\pz0.063\pm0.015\pm0.013$
\\
 &  GJ
 &  $-0.129\pm0.041\pm0.026$
 &  $-0.112\pm0.016\pm0.010$
 &  $\pz0.055\pm0.016\pm0.013$
\\[ 1.5mm]
\hline
\\[-2.5mm]
 \multirow{3}{*}{~15$\,-$~20}
 &  HX
 &  $\pz0.307\pm0.080\pm0.075$
 &  $\pz0.042\pm0.042\pm0.029$
 &  $-0.111\pm0.037\pm0.030$
\\
 &  CS
 &  $-0.169\pm0.043\pm0.038$
 &  $\pz0.140\pm0.042\pm0.026$
 &  $\pz0.048\pm0.034\pm0.030$
\\
 &  GJ
 &  $-0.123\pm0.070\pm0.077$
 &  $-0.156\pm0.032\pm0.018$
 &  $\pz0.030\pm0.029\pm0.022$
\\[ 1.5mm]
\hline
\\[-2.5mm]
 \multirow{3}{*}{~20$\,-$~30}
 &  HX
 &  $\pz0.060\pm0.114\pm0.168$
 &  $\pz0.087\pm0.066\pm0.060$
 &  $\pz0.004\pm0.046\pm0.040$
\\
 &  CS
 &  $-0.086\pm0.077\pm0.133$
 &  $-0.021\pm0.059\pm0.040$
 &  $\pz0.046\pm0.064\pm0.113$
\\
 &  GJ
 &  $\pz0.069\pm0.128\pm0.198$
 &  $-0.037\pm0.060\pm0.060$
 &  $-0.011\pm0.047\pm0.044$
\end{tabular*}
\end{footnotesize} 
\end{center} 
\end{table}  
\begin{table}[t]
\begin{center}
\caption{\small Values of
 $\tilde \uplambda$
 measured in the HX, CS and GJ frames for the \threes produced
 at $\sqrt{s}=7$ and $8\,\,{\mathrm{TeV}}$ in the rapidity range
 $2.2<\yy<4.5$.
 The first uncertainty is statistical and the second systematic.
}\label{tab:Y3S_lamT_Results_HX_CS_GJ_2.2y4.5}
\vspace*{7mm}
\begin{footnotesize}
\begin{tabular*}{0.90\textwidth}{@{\hspace{1mm}}c@{\extracolsep{\fill}}ccc@{\hspace{1mm}}}
 $\pty~~~\left[\!\gevc\right]$ & $\tilde \uplambda$ & $\sqrt{s}=7\tev$ & $\sqrt{s}=8\tev$ 
\\[ 1.5mm]
\hline
\\[-2.5mm]
 \multirow{3}{*}{~ 0$\,-$~ 2}
 &  HX
 &  $-0.129\pm0.102\pm0.070$
 &  $-0.187\pm0.071\pm0.066$
\\
 &  CS
 &  $-0.113\pm0.103\pm0.069$
 &  $-0.174\pm0.071\pm0.070$
\\
 &  GJ
 &  $-0.105\pm0.102\pm0.072$
 &  $-0.165\pm0.071\pm0.067$
\\[ 1.5mm]
\hline
\\[-2.5mm]
 \multirow{3}{*}{~ 2$\,-$~ 4}
 &  HX
 &  $-0.012\pm0.080\pm0.063$
 &  $\pz0.065\pm0.056\pm0.064$
\\
 &  CS
 &  $\pz0.024\pm0.082\pm0.063$
 &  $\pz0.095\pm0.057\pm0.064$
\\
 &  GJ
 &  $\pz0.043\pm0.082\pm0.061$
 &  $\pz0.100\pm0.057\pm0.064$
\\[ 1.5mm]
\hline
\\[-2.5mm]
 \multirow{3}{*}{~ 4$\,-$~ 6}
 &  HX
 &  $\pz0.121\pm0.089\pm0.068$
 &  $\pz0.120\pm0.061\pm0.070$
\\
 &  CS
 &  $\pz0.167\pm0.091\pm0.066$
 &  $\pz0.153\pm0.062\pm0.068$
\\
 &  GJ
 &  $\pz0.181\pm0.091\pm0.067$
 &  $\pz0.155\pm0.062\pm0.068$
\\[ 1.5mm]
\hline
\\[-2.5mm]
 \multirow{3}{*}{~ 6$\,-$~ 8}
 &  HX
 &  $\pz0.097\pm0.097\pm0.063$
 &  $-0.129\pm0.061\pm0.059$
\\
 &  CS
 &  $\pz0.130\pm0.099\pm0.056$
 &  $-0.099\pm0.062\pm0.055$
\\
 &  GJ
 &  $\pz0.145\pm0.099\pm0.059$
 &  $-0.081\pm0.062\pm0.056$
\\[ 1.5mm]
\hline
\\[-2.5mm]
 \multirow{3}{*}{~ 8$\,-$~10}
 &  HX
 &  $-0.004\pm0.102\pm0.055$
 &  $\pz0.086\pm0.071\pm0.048$
\\
 &  CS
 &  $\pz0.007\pm0.103\pm0.056$
 &  $\pz0.093\pm0.071\pm0.049$
\\
 &  GJ
 &  $\pz0.009\pm0.103\pm0.052$
 &  $\pz0.093\pm0.071\pm0.049$
\\[ 1.5mm]
\hline
\\[-2.5mm]
 \multirow{3}{*}{~10$\,-$~15}
 &  HX
 &  $\pz0.143\pm0.091\pm0.042$
 &  $\pz0.047\pm0.057\pm0.042$
\\
 &  CS
 &  $\pz0.140\pm0.091\pm0.043$
 &  $\pz0.043\pm0.057\pm0.043$
\\
 &  GJ
 &  $\pz0.136\pm0.091\pm0.037$
 &  $\pz0.039\pm0.057\pm0.042$
\\[ 1.5mm]
\hline
\\[-2.5mm]
 \multirow{3}{*}{~15$\,-$~20}
 &  HX
 &  $\pz0.689\pm0.218\pm0.134$
 &  $-0.023\pm0.104\pm0.069$
\\
 &  CS
 &  $\pz0.692\pm0.218\pm0.142$
 &  $-0.027\pm0.104\pm0.076$
\\
 &  GJ
 &  $\pz0.692\pm0.218\pm0.108$
 &  $-0.033\pm0.104\pm0.058$
\\[ 1.5mm]
\hline
\\[-2.5mm]
 \multirow{3}{*}{~20$\,-$~30}
 &  HX
 &  $\pz0.631\pm0.360\pm0.327$
 &  $\pz0.073\pm0.187\pm0.215$
\\
 &  CS
 &  $\pz0.646\pm0.365\pm0.313$
 &  $\pz0.056\pm0.187\pm0.276$
\\
 &  GJ
 &  $\pz0.654\pm0.366\pm0.335$
 &  $\pz0.035\pm0.186\pm0.201$
\end{tabular*}
\end{footnotesize}
\end{center} 
\end{table}  


\clearpage
\addcontentsline{toc}{section}{References}
\setboolean{inbibliography}{true}
\bibliographystyle{LHCb}
\bibliography{local,main,LHCb-PAPER,LHCb-CONF,LHCb-DP,LHCb-TDR}

\newpage
\centerline{\large\bf LHCb collaboration}
\begin{flushleft}
\small
R.~Aaij$^{40}$,
B.~Adeva$^{39}$,
M.~Adinolfi$^{48}$,
Z.~Ajaltouni$^{5}$,
S.~Akar$^{59}$,
J.~Albrecht$^{10}$,
F.~Alessio$^{40}$,
M.~Alexander$^{53}$,
A.~Alfonso~Albero$^{38}$,
S.~Ali$^{43}$
% ...

{\footnotesize \it
% ...
$ ^{5}$Clermont Universit{\'e}, Universit{\'e} Blaise Pascal, CNRS/IN2P3, LPC, Clermont-Ferrand, France\\
% ...
$ ^{10}$Fakult{\"a}t Physik, Technische Universit{\"a}t Dortmund, Dortmund, Germany\\
% ...
$ ^{38}$ICCUB, Universitat de Barcelona, Barcelona, Spain\\
$ ^{39}$Universidad de Santiago de Compostela, Santiago de Compostela, Spain\\
$ ^{40}$European Organization for Nuclear Research (CERN), Geneva, Switzerland\\
% ...
$ ^{43}$Nikhef National Institute for Subatomic Physics, Amsterdam, The Netherlands\\
% ...
$ ^{48}$H.H. Wills Physics Laboratory, University of Bristol, Bristol, United Kingdom\\
% ...
$ ^{53}$School of Physics and Astronomy, University of Glasgow, Glasgow, United Kingdom\\
% ...
$ ^{59}$University of Cincinnati, Cincinnati, OH, United States\\
% ...
}
\end{flushleft}

\end{document}